\documentclass[%
 reprint,
superscriptaddress,
 amsmath,amssymb,
 aps,
pra,
]{revtex4-2}

\usepackage[english]{babel}

\usepackage{graphicx}
\usepackage{dcolumn}
\usepackage{bm}
\usepackage{blindtext}
\usepackage[super]{nth}
\usepackage{xspace, color}

\newcommand{\fm}[1]{\ifmmode#1\else$#1$\fi}
\newcommand{\Ca}{\fm{\text{Ca}^{+}}\xspace}
\newcommand{\Al}{\fm{\text{Al}^{+}}\xspace}

\newcommand{\Alts}{\fm{^{27}\Al}\xspace}
\newcommand{\Sres}{\fm{^{87}\text{Sr}}\xspace}

\def\ssz{\fm{{}^1\mathrm{S}_0}\xspace}

\def\tpo{\fm{{}^3\mathrm{P}_1}\xspace}

\def\ssztpo{\ssz\fm{\leftrightarrow} \tpo}

\begin{document}

\preprint{APS/123-QED}

\title{A high-stability optical clock based on a continuously ground-state cooled Al$^+$ ion without compromising its accuracy}

\author{Fabian Dawel}
 \affiliation{Physikalisch-Technische Bundesanstalt, Bundesallee 100, 38116 Braunschweig, Germany}
 \affiliation{Institut für Quantenoptik, Leibniz Universität Hannover, Welfengarten 1, 30167 Hannover, Germany}
\author{Lennart Pelzer}%
\affiliation{Physikalisch-Technische Bundesanstalt, Bundesallee 100, 38116 Braunschweig, Germany}
\author{Kai Dietze}%
 \affiliation{Physikalisch-Technische Bundesanstalt, Bundesallee 100, 38116 Braunschweig, Germany}
 \affiliation{Institut für Quantenoptik, Leibniz Universität Hannover, Welfengarten 1, 30167 Hannover, Germany}
\author{Johannes Kramer}%
 \affiliation{Physikalisch-Technische Bundesanstalt, Bundesallee 100, 38116 Braunschweig, Germany}
  \affiliation{Institut für Quantenoptik, Leibniz Universität Hannover, Welfengarten 1, 30167 Hannover, Germany}
\author{Marek Hild}%
 \affiliation{Physikalisch-Technische Bundesanstalt, Bundesallee 100, 38116 Braunschweig, Germany}
\author{Steven A.~King}%
 \affiliation{Physikalisch-Technische Bundesanstalt, Bundesallee 100, 38116 Braunschweig, Germany}
 \affiliation{current address: Oxford Ionics, Oxford, OX5 1GN, UK}
\author{Nicolas C.H.~Spethmann}%
 \affiliation{Physikalisch-Technische Bundesanstalt, Bundesallee 100, 38116 Braunschweig, Germany}
 \author{Joshua Klose}%
 \affiliation{Physikalisch-Technische Bundesanstalt, Bundesallee 100, 38116 Braunschweig, Germany}
\author{Kilian Stahl}%
 \affiliation{Physikalisch-Technische Bundesanstalt, Bundesallee 100, 38116 Braunschweig, Germany}
\author{Sören Dörscher}%
 \affiliation{Physikalisch-Technische Bundesanstalt, Bundesallee 100, 38116 Braunschweig, Germany}
\author{Erik Benkler}%
 \affiliation{Physikalisch-Technische Bundesanstalt, Bundesallee 100, 38116 Braunschweig, Germany}
\author{Christian Lisdat}%
 \affiliation{Physikalisch-Technische Bundesanstalt, Bundesallee 100, 38116 Braunschweig, Germany}
\author{Sergey G. Porsev}
\affiliation{Department of Physics and Astronomy, University of Delaware, Newark, Delaware 19716, USA}
\author{Marianna S. Safronova}
\affiliation{Department of Physics and Astronomy, University of Delaware, Newark, Delaware 19716, USA}
\author{Piet O.~Schmidt}%
 \affiliation{Physikalisch-Technische Bundesanstalt, Bundesallee 100, 38116 Braunschweig, Germany}
 \affiliation{Institut für Quantenoptik, Leibniz Universität Hannover, Welfengarten 1, 30167 Hannover, Germany}
 
\date{\today}

\begin{abstract}

Single ion optical clocks have shown systematic frequency uncertainties below $10^{-18}$, but typically require more than one week of averaging to achieve a corresponding statistical uncertainty. This time can be reduced with longer probe times, but comes at the cost of a higher time-dilation shift due to motional heating of the ions in the trap. We show that sympathetic ground-state cooling using electromagnetically-induced transparency (EIT) of an \Al clock ion via a co-trapped \Ca ion during clock interrogation suppresses the heating of the ions. \Al can be kept close to the motional ground state, independent from the chosen interrogation time, at a relative time dilation shift of $(-1.69\pm0.20)\times10^{-18}$. The \Ca cooling light introduces an additional light shift on the \Al clock transition of $(-9.27\pm 1.03)\times10^{-18}$. We project that the uncertainty of this light shift can be further reduced by nearly an order of magnitude. This sympathetic cooling enables seconds of interrogation time with $10^{-19}$ motional and cooling laser-induced uncertainties for \Al and can be employed in other ion clocks as well.

\end{abstract}

\maketitle

Optical clocks are the most accurate measurement devices, reaching estimated relative systematic frequency uncertainties in the $10^{-19}$ regime \cite{brewer_27al_2019, aeppli_clock_2024, marshall_highstability_2025, arnold_validating_2024}. Frequency ratios between optical clocks have been measured down to an uncertainty level of a few $10^{-18}$  \cite{mcgrew_atomic_2018, sanner_optical_2019, takamoto_test_2020,hausser_115in+172yb_2025, amy-klein_international_2024, beloy_frequency_2021}. This level of precision can be used for measuring centimeter height differences in relativistic geodesy \cite{mehlstäubler_atomic_2018, grotti_longdistance_2024, grotti_geodesy_2018, yuan_demonstration_2024}, to set new bounds on variation of fundamental constants and dark matter candidates \cite{godun_frequency_2014, safronova_search_2019, sherrill_analysis_2023, filzinger_ultralight_2023, roberts_search_2020, filzinger_improved_2023, beloy_frequency_2021}, or test relativity \cite{yeh_robust_2023, sanner_optical_2019, takamoto_test_2020}. 
All these applications benefit from fast averaging times to achieve a certain precision: in relativistic geodesy and tests of relativity, dynamic changes could be resolved as the Earth is rotating; in the search for dark matter, the mass range of dark matter candidates can be extended towards heavier masses.

While neutral atom lattice clocks probe hundreds to thousands of atoms at a time, ion clocks are currently restricted to one or a few ions, limiting their signal-to-noise ratio \cite{ludlow_optical_2015} and thus requiring long averaging times.

The stability of clocks is fundamentally limited by quantum-projection noise (QPN) \cite{itano_quantum_1993}.
The statistical uncertainty expressed in the form of an Allan deviation for a clock probing $N_{at}$ atoms using Ramsey interrogation and averaging for a time $\tau$, is given by \cite{peik_laser_2006,riehle_frequency_2004}
\begin{equation}
    \sigma_y(\tau) \approx C 
    \frac{\Delta\nu}{\nu_0}\sqrt{\frac{t_c}{N_{at}\tau}},
    \label{eq:QPN}
\end{equation}
where $C$ is a constant on the order of 1, $\Delta\nu$ is the Ramsey fringe width, $\nu_0$ is the transition frequency, and $t_c$ is the time of one complete clock cycle.
Single ion clocks typically achieve a stability of $10^{-15}/\sqrt{\tau/1\,\textrm{s}}$ \cite{beloy_frequency_2021, sanner_optical_2019}.
To improve the ion clock stability one can use multiple ions to increase the amount of signal per cycle \cite{pyka_highprecision_2014, keller_evaluation_2016, hausser_115in+172yb_2025, tan_suppressing_2019, pelzer_multiion_2024, akerman_operating_2025}, or longer interrogations times to reduce $\Delta\nu$. The maximum interrogation time is ultimately limited by the excited state lifetime of the clock transition, or by the laser coherence time \cite{leroux_online_2017}. The former is a fundamental limitation, while the latter can be improved by better clock lasers \cite{matei_15_2017} or compound clocks \cite{dörscher_dynamical_2020, rosenband_exponential_2013, borregaard_efficient_2013} that use additional atomic ensembles for pre-stabilisation of the laser.

Another limitation to long probe times in ion clocks is the \nth{2}-order Doppler or time-dilation shift \cite{martínez-lahuerta_initio_2022, ludlow_optical_2015} arising from residual thermal motion of the ion. It depends on the average kinetic energy during the probe time, which increases with time due to motional heating of the ion. The responsible heating rates depend on the size and type of the used trap with reported heating rates of $1\dots 10^4$ phonons/s for room temperature setups \cite{brownnutt_iontrap_2015}. The determination of the resulting kinetic energy is model dependent and the heating rate can change with environmental disturbances, complicating the systematic shift analysis \cite{zeng_transportable_2023, huntemann_singleion_2016, brewer_27al_2019, ma_quantumlogicbased_2024, rosenband_frequency_2008}.

Cooling during interrogation ensures that the ion motion is in a steady state. This eliminates the effect of the heating rate on the \nth{2}-order Doppler shift. Several two-ion species experiments implemented Doppler cooling during clock interrogation \cite{chou_frequency_2010, rosenband_frequency_2008, cui_evaluation_2022, marshall_highstability_2025}. The Doppler cooling rate is typically much larger than the heating rates and will keep the ion at the Doppler temperature. Here, we go one step further and use electromagnetically-induced transparency (EIT) cooling, since it can reach the motional ground state \cite{roos_experimental_2000, lechner_electromagneticallyinducedtransparency_2016} while still cooling all modes at once \cite{scharnhorst_experimental_2018, sun_sympathetic_2023}. 
The advantage of EIT cooling over Doppler cooling is the achievable lower mean motional quantum number, resulting in an order of magnitude smaller time dilation shift. This significantly simplifies characterization of this shift, which otherwise requires careful calibration 
\cite{marshall_highstability_2025, chen_sympathetic_2017}.

We demonstrate that the systematic frequency uncertainty associated with the ac-Stark shift induced by the cooling lasers are compatible with clock operation at the $10^{-18}$ level and below.

\begin{figure}[b]
\includegraphics[width=0.48\textwidth]{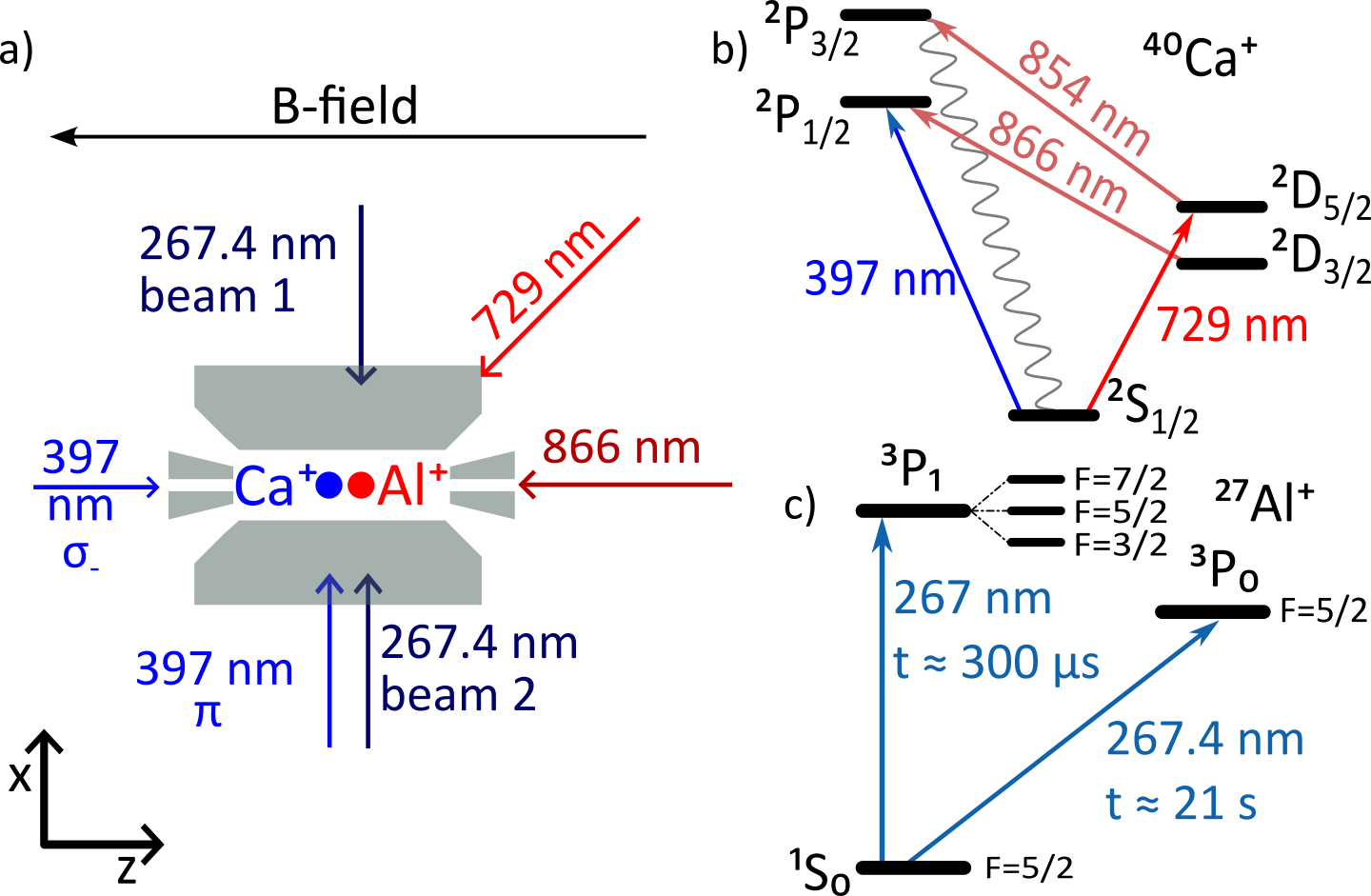}
\caption{\label{fig:lasersetup} a) Laser setup around the trap, showing cooling and clock lasers. Compensation of the first order Doppler shift is realized with two counter propagating clock laser beams. For cooling of the \Al-\Ca crystal we use $397\,$nm $\sigma^-$ and $\pi$  beams and a repump $866\,$nm beam. The $\sigma^-$ and repump beams are aligned along the trap axis. The $397\,$nm $\pi$ beam is perpendicular to the trap axis, as well as the clock laser. For motional sideband spectroscopy via \Ca, the $729\,$nm laser is aligned with an angle of around $45^\circ$ to the trap axis in the $xz$ plane, such that it can address all motional modes. b) and c) show partial level schemes of \Ca and \Al, respectively.}
\end{figure}

Our experimental setup is described in Refs.~\cite{scharnhorst_experimental_2018, kramer_aluminum_2023}. In short, we use a linear Paul trap, consisting of four blades and two endcap electrodes, as well as two additional electrodes for micromotion compensation. The distance between the blades is $1.6\,$mm and the distance between the endcaps is $5\,$mm. We drive the trap blade pairs with near-equal amplitude at an rf drive frequency of $28\,$MHz, resulting in single \Ca trap frequencies of $(\omega_x,\omega_y,\omega_z)=(1.95,1.85,1.1)\,$MHz, where $z$ is the axial direction. Fig. \ref{fig:lasersetup} illustrates the cooling and clock laser directions relative to the trap. The intensity of all cooling lasers involved are measured via photodiodes before the vacuum chamber. These voltages are used for an intensity stabilization of the laser light via a digital sample and hold circuitry. The trap is inside a stainless-steel vacuum chamber at room temperature with a pressure of $11(6)\,$nPa, measured via swapping rates of an \Al-\Ca crystal \cite{hankin_systematic_2019}. The magnetic field of $0.15\,$mT along the trap axis is generated by two external coils. We use an additional set of coils for active magnetic field stabilization to reduce magnetic field drift and noise, mainly at $50\,$Hz. With the magnetic field stabilization we measure on the $S_{1/2}(m=-1/2)\leftrightarrow D_{5/2}(m=-1/2)$ coherence times of around $1\,$ms. 

For EIT cooling we use $397\,$nm $\pi$- and $\sigma^-$-polarized light. Here, $\sigma^-$ is the strong pump beam while $\pi$ is the probe beam. These beams have an effective wavevector which overlaps with all motional modes of the crystal (see also Fig.~\ref{fig:lasersetup}). We use a common blue detuning of $64(4)\,$MHz, limited by the available laser power \cite{morigi_ground_2000}. To cool all modes at once we set the light shift of the $\sigma^-$ EIT beam to a value of $2.3\,$MHz, which maximizes the cooling efficiency of modes around this motional frequency. We use $10\,$ms of precooling before clock interrogation to establish a steady state of the mean motional mode occupation. 

The clock is interrogated by probing the stretched states of the $^1\mathrm{S}_0\leftrightarrow $ $^3\mathrm{P}_0$ transition of \Al, to eliminate the first order Zeeman shift. The Zeeman states are prepared through frequency-addressed optical pumping on the \Al \ssztpo transition. The interrogation order between the states is randomized. The clock state is read out via a repeated quantum non-demolition measurement protocol \cite{hume_highfidelity_2007}.

Long coherence times are achieved by frequency locking the clock laser at its second sub-harmonic at $1069\,$nm to a high finesse cavity \cite{amairi_long_2014} and transfer lock \cite{scharnhorst_highbandwidth_2015} it to a laser with a relative frequency stability noise floor of $4\times10^{-17}$ \cite{matei_15_2017} using the end-to-end topology to minimize differential optical path length fluctuations \cite{benkler_endtoend_2019}. All fiber paths are length-stabilized using acousto-optic modulators \cite{ye_delivery_2003}. We use two second harmonic generation stages with similar path length stabilization in between,
resulting in a stability of below $1\times10^{-16}/\sqrt{\tau/1\,\text{s}}$ directly after the last doubling system \cite{kraus_phasestabilized_2022}. Using length-stabilized UV fibers \cite{colombe_singlemode_2014, marciniak_fully_2017}, we bring the light near the vacuum chamber, leaving an uncompensated free-space path of around $40\,$cm to the ion. For the clock interrogation we use Ramsey spectroscopy with $250\,$ms dark time and $25\,$ms pulse duration per pulse. The dead time is around $40\,$\% of the overall clock cycle. The EIT cooling and repump lasers are on during the complete clock interrogation.

\begin{figure}[b]
\includegraphics[width=0.48\textwidth]{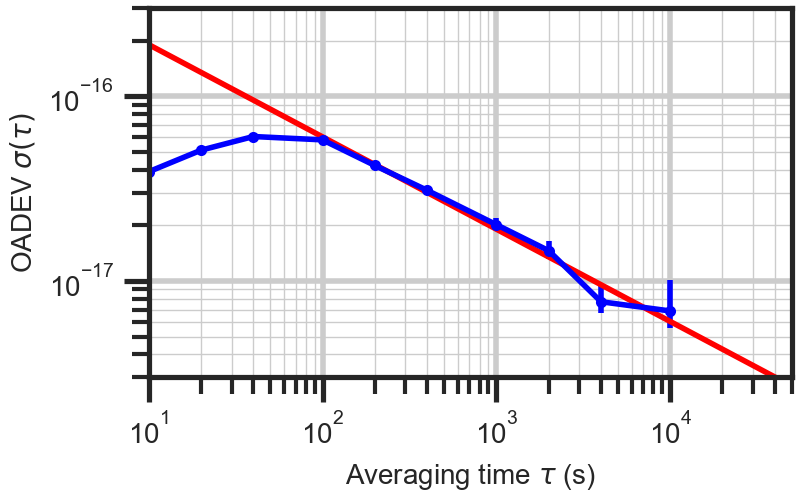}
\caption{\label{fig:Stability_Sr} Stability measurement of the frequency ratio between \Sres and \Alts (blue dots). 
The red line is a fit to the white frequency noise regime with a resulting stability of $6.1\times10^{-16}\sqrt{\tau/1\,\text{s}}$. This measurement data is concatenated from multiple shorter runs during one week.}
\end{figure}

We measured the stability of the frequency ratio of the \Alts clock against a \Sres-optical lattice clock \cite{hausser_115in+172yb_2025,schwarz_cryogenic_2022}, as shown in Fig.~\ref{fig:Stability_Sr}. When the statistical uncertainty of the Sr-lattice clock of $1.1\times10^{-16}/\sqrt{\tau/1\,\text{s}}$ is subtracted from the fitted value, we obtain a frequency stability of $6.0\times10^{-16}/\sqrt{\tau/1\,\text{s}}$ for \Al. This is in agreement with a quantum projection noise-limited measurement with a contrast of $93$\%. 
The result confirms that statistical fluctuations from the presence of the cooling laser are absent down to a level of $6\times 10^{-18}$ fractional frequency uncertainty.

\begin{figure}[b]
\includegraphics[width=0.45\textwidth]{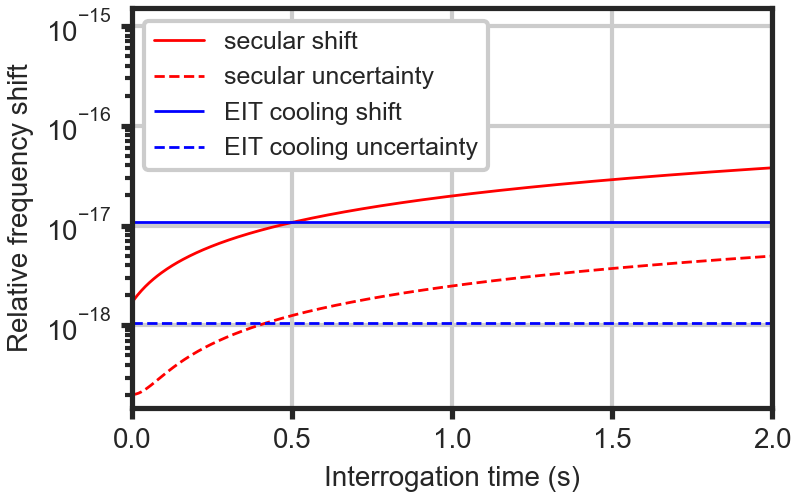}
\caption{\label{fig:interrogation_time} Frequency shift versus interrogation time. The frequency shift includes the measured time dilation shift and the ac-Stark shift of the cooling lasers. With EIT cooling the shift is stationary, while the shift increases without cooling due to motional heating. For a lifetime-limited interrogation of $21\,$s, the shift increases up to $(3.77\pm0.51)\times 10^{-16}$.}
\end{figure}

\begin{table*}[t!]
\caption{\label{tab:secularfreq} Motional mode parameters of an \Al-\Ca crystal. Here $\omega$ is the motional mode frequency, $b_i$ are the normalized eigenvector components of aluminum, $\eta_{729}$ the Lamb-Dicke parameter of the $729\,$nm laser on \Ca (which includes the projection of the laser on the mode), $S$ is the multiplicative contribution due to intrinsic micromotion. The mean motional state after Doppler and EIT cooling are given by $\bar{n}_\text{Dop}$ (calculated) and $\bar{n}_\text{EIT}$ (measured), respectively. The heating rate is $\dot{\bar{n}}$. The time-dilation shift (TDS) with cooling column shows the resulting frequency shift of each mode for the measured $\bar{n}_\text{EIT}$. The radial (rad) in-phase (ip) mode and the axial (ax) out-of-phase (op) mode have the largest weight for the secular motional shift on \Al, which is shown in the next-to-last column. The last column shows the shift per time when no cooling is applied during the interrogation. In the last two columns we neglect the zero-point energy.}
\begin{ruledtabular}
\begin{tabular}{l|cccccccccc}
Mode & $\omega$  & $b_m$ & $S$ &$\eta_{729}$ & $\bar{n}_\text{Dop}$ & $\bar{n}_\text{EIT}$ & TDS with &$\dot{\bar{n}}$ & shift/quantum  & shift/time \\
 & (MHz) &  & & &  &  &  cooling $(10^{-18})$& (quantum/s) & ($10^{-19}$) & ($10^{-18}$t/s)\\
\hline
ax ip & 1.25& 0.562 & 0 & 0.051& 9.9 & 1.1(4) & 0.054(15) &56(6) & 0.33& 0.97(10)\\
ax op & 2.26& 0.827 & 0 & 0.026& 5.5& 0.11(5) & 0.077(6)&3.7(1.0)& 1.27 & 0.31(6)\\
rad x op & 1.759& 0.157& 3.17 & 0.039& 7.0& 0.35(8) & 0.00125(12)&37(4)& 0.15& 0.28(3)\\
rad y op & 1.766& 0.157 & 3.17 & 0.030& 7.0& 0.18(4) &0.00101(7)&7.8(1.3)& 0.15& 0.07(1)\\
rad x ip & 2.868& 0.988 & 1.21 & 0.005 & 4.3& 0.78(19) &0.66(10) &53(8)& 5.2& 14.4(2.1)\\
rad y ip & 2.912& 0.988 & 1.21 & 0.004& 4.3& 1.18(34) & 0.88(18)&11(5)& 5.2 & 3.7(1.3)
 
\end{tabular}
\end{ruledtabular}
\end{table*}
While simultaneous cooling allows long probe times and therefore improve the statistical frequency uncertainty, we need to evaluate the systematic frequency shifts from residual secular motion and the cooling laser-induced ac-Stark shift on the \Al clock transition (see Fig.~\ref{fig:interrogation_time}). The secular kinetic energy is determined from the mean motional state occupation and the normal mode frequencies (for details see Supplemental Material \cite{dawel_supplement_2025}). Secular frequencies are measured via the S$_{1/2}\to \mathrm{D}_{5/2}$ transition in \Ca in the resolved sideband regime. Measuring the mean motional state occupation relies on knowing the underlying state distribution, which changes depending on the cooling technique. Resolved sideband cooling generates a state distribution that can be approximated by a double-thermal distribution \cite{chen_sympathetic_2017}, while Doppler cooling \cite{Rasmusson_optimized_2021} and EIT cooling yields a thermal distribution \cite{lechner_electromagneticallyinducedtransparency_2016}. We employ the sideband thermometry method to obtain the mean motional state, predicated on the presence of the expected thermal distribution \cite{Rasmusson_optimized_2021}. 

We measured the excitation probability ratio between blue and red sidebands for all axial and radial out of phase modes for times of up to $250\,$ms of cooling, confirming that the mean motional state occupation is stable over the probe time. The ratio for the radial in-phase modes is inferred from frequency scans over the respective sideband for up to 80\,ms. This is necessary due to weak coupling (see Tab.~\ref{tab:secularfreq}) and a frequency drift of the sidebands. The drift is presumably due to thermal changes of the helical resonator driving the trap rf electrodes, which slightly change the resonance frequency. We also confirmed that the cooling performance is independent of the ion crystal ordering.

The advantage of EIT over Doppler cooling becomes evident from the measured mean motional quantum numbers, $\bar{n}$, shown in Tab.~\ref{tab:secularfreq}: all modes are cooled well below the Doppler limit, resulting in an overall motional shift and associated uncertainty of $(-1.69\pm0.20)\times10^{-18}$, which is smaller compared to Dopper cooling \cite{marshall_highstability_2025}.  The axial out-of-phase mode is most efficiently cooled, since it is closest to the optimal EIT cooling condition.

A downside of cooling during interrogation is the differential ac-Stark shift of the cooling lasers on the clock transition. The three light beams which cause a shift are the $397\,$nm $\pi$ and $\sigma^-$ beams and the $866\,$nm repump beam. To estimate the light shift on the clock transition $\Delta \nu_c$ we need to know the differential polarizability of the clock transition of \Al, $\Delta \alpha$, and the electric field $E(t)=E_0\cos(\omega t)$ of the cooling beams. The former is extrapolated from polarizability measurements in the infra-red wavelength regime, while the latter is derived from the ac-Stark shift of the lasers on the \Ca levels and its known polarizabilities.

In general the ac-Stark shift of an atomic state can be described by three different polarizability contributions, namely the scalar ($\alpha_S$), vector ($\alpha_V$) and tensor ($\alpha_T$) polarizability \cite{steck_quantum_2007, bonin_electricdipole_1997, dounas-frazer_measurement_2010}:
\begin{widetext}
\begin{equation}
    \Delta \nu =-\frac{1}{4h}\left(\alpha_S(\omega)|E|^2+\alpha_V(\omega)(|E_{\sigma^+}|^2-|E_{\sigma^-}|^2)\frac{m_J}{J}+\alpha_T(\omega)\frac{(3|E_{\pi}|^2-|E|^2)}{2}\frac{3m_J^2-J(J+1)}{J(2J-1)}\right).
    \label{eq:ac_stark_shift}
\end{equation}
\end{widetext}
Where $E_{\pi},E_{\sigma^+},E_{\sigma^-}$ is the $\pi,\sigma^+,\sigma^-$ polarized electrical field, respectively, $|E|^2=|E_{\pi}|^2+|E_{\sigma^+}|^2+|E_{\sigma^-}|^2$ is the overall electric field, $m_J$ is the magnetic quantum number, $J$ is the total angular momentum quantum number of the atomic state and $h$ is the Planck constant. The overall light shift is then calculated by the difference of the frequency shift in the excited and ground state of the driven transition.

The tensor and the vector light shift for \Al can be neglected, since its fine structure in the ground and excited state have zero angular momentum. Contributions from hyperfine interaction to the ac-Stark shift are small and can be neglected as well \cite{shi_polarizabilities_2015}. The methods for calculating the differential scalar polarizability of \Al are given in the supplement material \cite{dawel_supplement_2025}.

The electric field strength of the lasers are determined by measuring the frequency shift on the $S_{1/2}\leftrightarrow D_{5/2}$ transition of \Ca as a function of applied laser intensity. 
Intensity difference between the \Ca and \Al ions, axially separated by $5.2\,\mu$m, adds only a small uncertainty on the ac-Stark shift determination (see Supplemental Material \cite{dawel_supplement_2025}).

To measure the frequency shift on the \Ca $S_{1/2}\leftrightarrow D_{5/2}$ transition, we first pump the ion with a $\pi$ pulse into the $D_{5/2} (m_{\pm 1/2})$ state.
Then the $D_{5/2}\rightarrow S_{1/2}$ ($\Delta m=0$) transition frequency is probed for different $397\,$nm $\pi$ ($\sigma^-$) laser powers. The frequency difference between  $397\,$nm laser on or off is a measure of the ac-Stark shift. From the known transition matrix elements \cite{hettrich_measurement_2015, UDportal} one can determine the frequency-dependent polarizabilities. With these and the measured frequency shift $\Delta\nu$ we can determine the electric field via Eq.~\eqref{eq:ac_stark_shift} (see Supplemental Material \cite{dawel_supplement_2025}).
Due to the small detuning from the $S_{1/2} \leftrightarrow  P_{1/2}$ transition, the dominant frequency shift arises from coupling of the laser to the $S_{1/2}$ state, while the frequency shift of the D state is eight orders of magnitude smaller. 
The main uncertainty for the intensity of the 397\,nm laser comes from its $4\,$MHz frequency uncertainty and the strongly detuning-dependent polarizability.

The frequency shift on \Al from the $397\,$nm beams are $\Delta\nu_{Al,397\sigma}=(-7.06\pm0.99)\times10^{-18}$
and $\Delta\nu_{Al,397\pi}=(-0.84\pm0.12)\times10^{-18}$ 
where the main error is due to the uncertainty in the \Al polarizability. One can improve the light shift measurement uncertainty by measuring the differential polarizability at $397\,$nm, e.g., using the method described in Ref.~\cite{wei_improved_2024}. The uncertainty of the total ac-Stark shift on \Al would reduce to $3.7\times10^{-19}$ if the uncertainty of the differential polarizability of \Al at $397\,$nm was known at the $1\,$\% level. This could be further reduced to the $2\times10^{-19}$ by stabilizing the 397\,nm laser frequency to below 1\,MHz. 

\begin{figure}[b]
\includegraphics[width=0.45\textwidth]{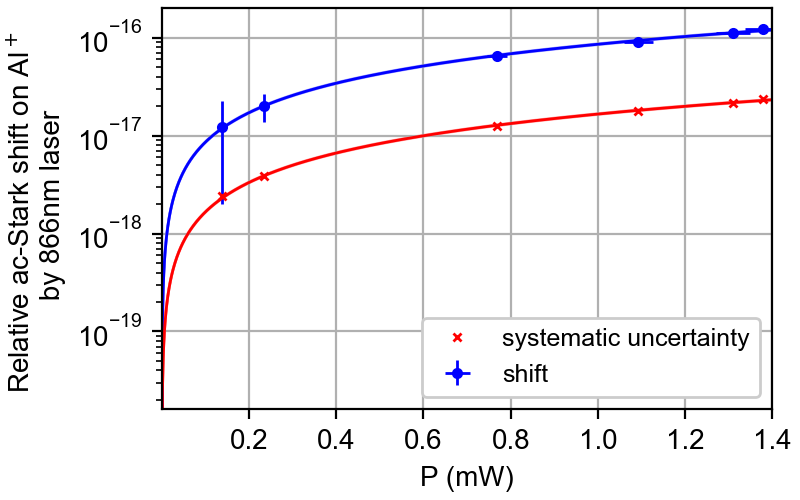}
\caption{\label{fig:AC_stark_866} Ac-Stark shift of the $866\,$nm laser on \Al. The blue line shows the calculated shift, while the red line shows the uncertainty. Since we expect a linear behaviour, we can extrapolate to lower powers, to estimate the value for the laser power of $17\,\mu$W used in the experiment. Here the error bars represent the statistical error of the shift.}
\end{figure}

The intensity of the repump laser can be obtained from its shift of the  S$_{1/2}\leftrightarrow$D$_{5/2}$ transition, dominated by its coupling to the D$_{5/2}$ state. The large detuning of 4.9\,THz results in a small ac-Stark shift. We use the quantum lockin-amplifier technique \cite{kotler_singleion_2011} and probe on the S$_{1/2}, m_{\pm 1/2}\leftrightarrow \mathrm{D}_{5/2}, m_{\pm 5/2}$ transitions, where the $866\,$nm laser was switched such that the light shifts add up, and the magnetic field noise is dynamically decoupled. For this measurement we also used different powers of the $866\,$nm laser to extrapolate to the low power of 17\,$\mu$W used during EIT cooling. For our calculations we assumed that we have no $\pi$ polarization (since the beam propagates along the magnetic field direction). Averaging the transition frequency measurements involving the D, $m_{\pm5/2}$ states eliminates the dependence on vector light shifts and the electric field can be determined from Eq.~\eqref{eq:ac_stark_shift} (see also Supplemental Material \cite{dawel_supplement_2025}). The conversion of laser power to shift is shown in Fig.~\ref{fig:AC_stark_866}. 
The systematic uncertainty for the light shift on \Al is the quadratic sum of the uncertainties in the \Al and \Ca polarizabilities and the statistical uncertainty from the shift measurement. For small laser powers, the uncertainty is dominated by the shift measurement uncertainty, while for large powers it is dominated by the uncertainty of the \Al polarizability.
For the laser power used for cooling we get a shift of $\Delta\nu_{Al,866}=(-1.37\pm0.27)\times10^{-18}$. 
This means we have an overall frequency shift due to the cooling and repump lasers of $\Delta\nu=(-9.27\pm1.03)\times10^{-18}$ 
, where the main uncertainty is due to the strong $397\,$nm $\sigma^-$ pump beam. This shift is added as a constant shift on the now time independent second order Doppler shift in Fig.~\ref{fig:interrogation_time}. As a result, for interrogation times longer than $420\,$ms, continuous cooling exhibits smaller uncertainties. This threshold is dictated by the motional heating rates specific to our experimental setup and the uncertainty of the \Al polarizability at 397\,nm.

In conclusion, we have shown that simultaneous EIT cooling during clock interrogation significantly reduces time dilation shifts and its uncertainty. The added ac-Stark shift from the cooling lasers on the clock transition has been measured to an uncertainty level of $1.0\times10^{-18}$, limited by the knowledge of the \Al polarizability at 397\,nm. This allows for long interrogation times without degrading the clock's systematic frequency uncertainty. Improved calculations or measurements of the polarizability of the \Al clock transition for all used wavelengths with an uncertainty of $1\%$ \cite{wei_improved_2024} together with improved intensity and frequency control would reduce the shift uncertainty by an order of magnitude, making it the method of choice even for interrogation times below several hundred milliseconds. The combined time dilation and ac-Stark shifts can be further tuned by changing the EIT cooling parameters towards an optimal trade-off between steady-state $\bar{n}$ for the most relevant modes and required laser power. 
Using EIT cooling during clock interrogation, we demonstrated Ramsey dark times of up to 250\,ms, resulting in an instability of $6.1\times 10^{-16}/\sqrt{\tau/1\,\mathrm{s}}$ in a comparison to a Sr lattice clock, which can be extended to even longer times by phase stabilization of the probe light all the way to the ion. 
This paves the way for time-efficient frequency ratio measurements between different clocks at $5\times 10^{-18}$ frequency uncertainty and below as required by the roadmap for a redefinition of the SI second \cite{dimarcq_roadmap_2024}.

\section*{Acknowledgements}
We thank Daniele Nicolodi, Thomas Legero and Uwe Sterr for providing the ultra-stable Si cavity as a laser reference. We like to thank Vicent Barbé for comments on the manuscript. The project was supported by the Physikalisch-Technische Bundesanstalt, the Max-Planck Society, the Max-Planck–Riken–PTB–Center for Time, Constants and Fundamental Symmetries, and the Deutsche Forschungsgemeinschaft (DFG, German Research Foundation) under Germany’s Excellence Strategy – EXC-2123 QuantumFrontiers – 390837967 and Project-ID 274200144 – SFB 1227, projects B03 and B02. This joint research project was financially supported by the State of Lower Saxony, Hanover, Germany, through Nieders\"achsisches Vorab. This project also received funding from the European Metrology Programme for Innovation and Research (EMPIR) and the European Partnership on Metrology, cofinanced by the five participating States and from the European Union’s Horizon 2020 and Horizon Europe research and innovation programme (Projects No. 20FUN01 TSCAC and 23FUN03 HIOC). This project further received funding from the European Research Council (ERC) under the European Union’s Horizon 2020 research and innovation programme (grant agreement No 101019987). This project has been supported by the German Federal Ministry of Education and Research within the funding program „Clusters4Future“, contract number 03ZU1209EE (QVLS-iLabs) and by the State of Lower Saxony, Hannover, Germany, through Niedersächsisches Vorab.
The theoretical work has been supported in part by the US Office of Naval Research Grant N000142512105  and by the European Research Council (ERC) under the Horizon 2020 Research and Innovation Program of the European Union (Grant Agreement No. 856415).

\section{appendix}

\subsection{Time dilation shift}
The movement of the ion in the trap causes a time dilation frequency shift \cite{ludlow_optical_2015}: $\frac{\Delta\nu}{\nu}=-\frac{\langle E_{kin}\rangle}{mc^2}$. It depends on the kinetic energy $E_{kin}$ of the ion, the mass $m$ and the speed of light $c$. The kinetic energy is given by the mean harmonic oscillator populations $\langle \bar{n}\rangle_i$ of the normal modes of motion $i$. For a two-ion crystal the kinetic energy is then given by:
\begin{equation}
     E_{kin}=0.5\left(\sum_{i} b_i^2\hbar\omega_i(\langle \bar{n}\rangle_i+0.5)(1+S)\right),
     \label{eq:Ekin}
\end{equation}
where we sum over all secular modes $i$ and $b_i$ is the motional mode amplitude, $\omega_i$ is the mode frequency, 
and $S$ is the influence of intrinsic micromotion \cite{wübbena_sympathetic_2012, berkeland_minimization_1998}. 
Without cooling, the motional state will increase over time due to coupling to electric field noise, which is described via the heating rate $\dot{\bar{n}}$ \cite{brownnutt_iontrap_2015}. Assuming a linear increase in mean motional quanta, the motional phonon occupation is described by
\begin{equation}
    \langle \bar{n}\rangle = \frac{1}{T}\int^T _0 \bar{n}_0+\dot{\bar{n}}t\,dt=\bar{n}_0+\frac{\dot{\bar{n}}T}{2}.
    \label{eq:heating_rate}
\end{equation}
Here $\bar{n}_0$ is the mean motional mode occupation after cooling. 
To minimize the frequency shift due to motion one can cool the ion before the clock interrogation to the Doppler limit \cite{huntemann_singleion_2016,cui_evaluation_2022}, or to the motional ground state \cite{brewer_27al_2019, chen_sympathetic_2017}. Using sympathetic cooling during the clock interrogation will counteract the anomalous heating of the trap, and $\dot{\bar{n}}=0$ in Eq.~\eqref{eq:heating_rate}. Therefore, the motional occupation depends on the cooling method used.

\subsection{Intensity uncertainties due to ion positioning}
We measure the intensities of the \Ca cooling lasers on a single \Ca ion and assume the same intensities on the \Al to estimate their frequency shift on the clock transition. Here we bound intensity differences between the two ions due to the beam geometry.
In a two-ion crystal the intensity of a laser beam on one of the two ions depends on the ordering of the ions, the spatial mode of out of a fiber (here we assume a Gaussian TEM$_{00}$ mode) and the position and orientation of the laser beam with respect to the ion crystal's symmetry axis.
Given a beam waist of $37\,\mu$m for the radial 397\,nm~$\pi$ beam, an ion-ion separation of $5.2\,\mu$m, derived from the axial secular frequencies, and a $8.2\,\mu$m maximal axial misalignment of the beam from the center position of the two-ion crystal, we obtain an intensity difference between the two ions of 13\,\%, which we take as the uncertainty.

For the axial 397\,nm~$\sigma^-$ and 866\,nm beams, an angle between laser beams and ion crystal axis can lead to a differential intensity between the two ions. From geometric constraints of the beam and ion crystal tilt, we estimate a maximum angle of $9^\circ$, corresponding to an intensity difference of $3\,\%$ for a $25\,\mu$m beam waist for the 397\, nm $\sigma^-$ and 0.7\,\% for a $110\,\mu$m waist for the 866\,nm beam.
We estimate that for our parameters a worst case displacement of the focus along the laser beam direction by the Rayleigh length leads to a difference in intensity of $<0.1\,\%$.

\subsection{AC-Stark shift calculations for Calcium}

\begin{table*}
\caption{\label{tab:polarizabilities} Calculated polarizabilities for the $S_{1/2}$ and $D_{5/2}$ state of Calcium for $866\,$nm light and $397\,$nm light with a detuning of $+60$\,MHz from the $\mathrm{S}_{1/2}\leftrightarrow \mathrm{P}_{1/2}$ transition.}
\begin{ruledtabular}
\begin{tabular}{l|ccc}
Laser & $\alpha_S$  & $\alpha_V$ & $\alpha_T$ \\
\hline

$397\,$nm  $S_{1/2}$ a.u. & $-1.55(10)\times10^{8}$ & $1.55(10)\times10^{8}$ & 0 \\
$397\,$nm  $D_{5/2}$ a.u.& $4.8(1.5)$ &$22.87(40)$ & $3.90(51)$ \\ 
$866\,$nm  $S_{1/2}$ a.u. & $94.8(2)$& $-0.30(6)$ & 0   \\ 
$866\,$nm $D_{5/2}$ a.u. & $825.9(3.4)$ & $-1205.7(4.4)$  & $-817.9(3.1)$ 
\end{tabular}
\end{ruledtabular}
\end{table*}

The scalar, vector, and tensor polarizabilities used in Eq. \eqref{eq:ac_stark_shift} can be calculated from the following expressions:
\begin{widetext}
\begin{align}
        \alpha_S(\omega) &= \frac{1}{2J+1} \sum_{J'}\frac{2\omega_{J'J}|\langle J\|\hat{d}\|J'\rangle|^2}{3\hbar(\omega_{J'J}^2-\omega^2-i\omega\Gamma+\Gamma_{J'}^2/4)} \label{eq:scal_pol} \\
        \alpha_V(\omega) &= \frac{1}{2J+1} \sum_{J'}(-1)^{J+J'+1}\sqrt{\frac{6J(2J+1)}{J+1}}        \begin{Bmatrix}
            1 & 1 & 1 \\
            J & J & J' 
            \end{Bmatrix}
        \frac{(\omega+\frac{i}{2}\Gamma_{J'})|\langle J\|\hat{d}\|J'\rangle|^2}{\hbar(\omega_{J'J}^2-\omega^2-i\omega\Gamma+\Gamma_{J'}^2/4)} \label{eq:vec_pol}\\ 
        \alpha_T(\omega) &=\frac{1}{2J+1} \sum_{J'}(-1)^{J+J'}\sqrt{\frac{40J(2J+1)(2J-1)}{3(J+1)(2J+3)}}        \begin{Bmatrix}
            1 & 1 & 2 \\
            J & J & J' 
            \end{Bmatrix}
        \frac{\omega_{J'J}|\langle J\|\hat{d}\|J'\rangle|^2}{\hbar(\omega_{J'J}^2-\omega^2-i\omega\Gamma+\Gamma_{J'}^2/4)}
        \label{eq:tensor_pol}
\end{align}
\end{widetext}
Here, $\{\}$ are the Wigner-$6j$ symbols, $\omega_{JJ'}$ is the transition frequency between the state $J$ and $J'$, and $\langle J\|\hat{d}\|J'\rangle$ is the reduced dipole matrix element. We included an additional $\frac{1}{2J+1}$ in front to match the definition provided in Refs.~\cite{UDportal,bonin_electricdipole_1997}, which accounts for the degeneracy of the state. $\Gamma_{J'}$ is the natural linewidth of the state $J'$ and $\hbar$ is the reduced Planck constant. Although polarizabilities are generally complex numbers, our interest lies solely in the real part. The real part is  called the dispersive part, which corresponds to a frequency shift. The imaginary part is the absorptive part and can be used to describe the photoabsorption cross section. For large detunings ($|\omega_{J'J}-\omega|\gg\Gamma/2$) the imaginary part gets small and can be neglected. In our case, we neglect the $\Gamma$ terms for all transitions apart from the $S_{1/2}\leftrightarrow P_{1/2}$ where we include it to first order.
The calculated polarizabilities for \Ca, used to calibrate the laser intensities in the experiment, are listed in Tab. \ref{tab:polarizabilities}. The Wigner-$6j$ symbols dictate a vanishing vector shift for a total angular momentum of $J = 0$, while the tensor shift is zero for $J<1$.

The calculation of Ca$^+$  polarizabilities can be separated into the calculations of the ionic core
contribution and a valence contribution for the scalar case. The small scalar core polarizability is essentially independent of the wavelength for the wavelengths of interest, and a static value is used. It is calculated using the random phase
approximation (RPA) \cite{Ca2007}. The core polarizability is zero for the vector and tensor cases, since the total angular momentum is zero for the core. Valence polarizabilities are calculated using Eq.~\eqref{eq:scal_pol}-\eqref{eq:tensor_pol}. 

We use the B-spline method to reduce infinite sums in Eqs.~\eqref{eq:scal_pol}-\eqref{eq:tensor_pol} to a finite number of terms. We separated the calculation of polarizability into two parts, the main term
containing the first contributions from the matrix elements given in the online portal \cite{Portal} and the remaining tail part. The online portal values are computed using the coupled cluster approach and their uncertainties are estimated as described in Ref.~\cite{Portal}. The remaining part is computed in the Dirac-Hartree-Fock (DHF) approach and corrected for RPA. The same set of B-spline orbitals is used to compute the tail as in the online portal computations to ensure consistency of the separation to the main part and the remainder term. Tail uncertainties are conservatively based on the accuracy of the DHF approach, which is on the order of 50\%. The contributions to the scalar and vector polarizabilities of the $4s$ state at 866.45~nm in a.u. are listed in Table~\ref{tab:4s_state_Ca}. Contributions to scalar, vector, and tensor polarizabilities of the $5D_{3/2}$ state at 397.0~nm and 866.45~nm in a.u. are listed in Table~\ref{tab:4d_state_Ca}.

\begin{table}[h!]
\caption{Contributions to scalar and vector polarizabilities of the $4s$ state at 866.45~nm in a.u.}
\label{tab:4s_state_Ca}
\begin{ruledtabular}
\begin{tabular}{lcccc}
State & Matrix el. & Energy& $\alpha_S$&$\alpha_V$\\
\hline
$4P_{1/2}$       & 2.8928(43)  &  25191.51  &  30.77(9)&    -14.09(4)  \\
$4P_{3/2}$       & 4.0920(60)  &  25414.4   & 60.74(18)&    13.79(4)    \\
$(5-12)P_{1/2}$  &             &            &   0.02     &  -0.004       \\
$(5-12)P_{3/2}$  &             &            &  0.04      &   0.003     \\
$(>12)P_{1/2}$     &             &            &  0.03(2)   &  -0.002(2)  \\
$(>12)P_{3/2}$     &             &            &  0.06(3)   &  0.001(1)   \\
Core             &             &            &  3.13(9)   &            \\
Total            &             &            &   94.8(2)  &    -0.30(6) \\
\end{tabular}
\end{ruledtabular}
\end{table}

\begin{table*}[h!]
\caption{Contributions to scalar, vector, and tensor polarizabilities of the $5D_{5/2}$ state in \Ca at 397.0~nm and 866.45~nm in a.u.}
\label{tab:4d_state_Ca}
\begin{ruledtabular}
\begin{tabular}{lcccccccc}
Contr.& Matrix el. & Energy& $\alpha_S$&$\alpha_V$& $\alpha_T$& $\alpha_S$&$\alpha_V$& $\alpha_T$\\
\cline{4-6}\cline{7-9}
      &            &       & \multicolumn{3}{c}{397.0~nm} & \multicolumn{3}{c}{866.45~nm} \\
\hline
$4P_{3/2}$  &3.283(6) &  11704&  -6.18(2)  & 19.96(7) &  6.18(2) & 815.9(3.0)&   -1207(4)&  -815.9(3.0)   \\
$(>4P)_{3/2}$ &         &       &   0.03(2)  & -0.02    & -0.03(2) & 0.01(1)   &     0.0   &   0.0          \\ [0.5pc]
$4F_{5/2}$  &0.516(6) &  54346&   0.15     & -0.03    &   0.17   & 0.13      &     0.0   &   0.1          \\
$5F_{5/2}$  &0.319(2) &  64324&   0.05     & -0.01    &   0.05   & 0.04      &     0.0   &   0.0          \\ 
6F$_{5/2}$  &0.224(4) &  69747&   0.02     & 0.00     &   0.02   & 0.02      &     0.0   &   0.0          \\
$(>6)F_{5/2}$ &         &       &   0.15(7)  & -0.02(1) &   0.18(9)& 0.14      &     0.0   &   0.2(1)       \\ [0.5pc]
$4F_{7/2}$  &2.309(25)& 54346 &   3.05(7)  & 1.51(3)  &  -1.09(2)& 2.51(5)   &     0.6   &   -0.9         \\ 
$5F_{7/2}$  &1.428(8) & 64324 &   0.91(1)  & 0.38     &   -0.33  & 0.80(1)   &     0.2   &   -0.3         \\ 
$6F_{7/2}$  &1.000(19)& 69747 &   0.40(2)  & 0.16(1)  &  -0.14(1)& 0.36(1)   &     0.1   &   -0.1         \\ 
$7F_{7/2}$  &0.755(13)& 73016 &   0.22(1)  & 0.08     &   -0.08  & 0.20(1)   &     0.0   &   -0.1         \\ 
$8F_{7/2}$  &0.598(9) & 75136 &   0.13     & 0.05     &   -0.05  & 0.12      &     0.0   &   0.0            \\  
$9F_{7/2}$  &0.491(7) & 76589 &   0.09     & 0.03     &   -0.03  & 0.08      &     0.0   &   0.0            \\
$10F_{7/2}$ &0.412(6) & 77627 &   0.06     & 0.02     &   -0.02  & 0.05      &     0.0   &   0.0            \\
$11F_{7/2}$ &0.354(6) & 78395 &   0.04     & 0.02     &   -0.02  & 0.04      &     0.0   &   0.0            \\
$(>11)F_{7/2}$&         &       &   2.6(1.4) & 0.7(4)   &   -0.9(5)& 2.5(1.3)  &     0.3(2)&   -0.9(4)        \\[0.5pc]
 Core       &         &       &   3.02(9)   &          & &  3.02(9)           &           &                 \\   
 Total      &         &       &   4.8(1.4) & 22.9(4)  &  3.9(5)  & 825.9(3.4)&  -1206(4) &   -817.9(3.1)    \\
\end{tabular}
\end{ruledtabular}
\end{table*}

We determine the intensity of the cooling and repumper lasers on the \Ca ion by determining their light shifts on the S-D transition. Using Eq. \eqref{eq:ac_stark_shift} with the polarizability component $X$ of the state $Y$, $\alpha_{X,Y}$, we can calculate the light shifts for the cooling lasers on the $\mathrm{S}_{1/2}, m_J=\pm 1/2 \leftrightarrow \mathrm{D}_{5/2}, m_J'=\pm 1/2$ ($m_J-m_J'=0$) transition and the repump laser on the $\mathrm{S}_{1/2}, m_J=\pm 1/2 \leftrightarrow \mathrm{D}_{5/2}, m_J'=\pm 5/2$ ($m_J-m_J'=\pm 2$) transition:
\begin{widetext}
\begin{equation}
\begin{split}
    \Delta\nu_{397_\pi} =&-\frac{1}{4h}\left(\alpha_{S,D}|E|^2+\frac{m_J}{J}\alpha_{V,D}(|E_{\sigma^+}|^2-|E_{\sigma^-}|^2)-\frac{4}{10}(3|E_{\pi}|^2-|E|^2)\alpha_{T,D}\right) \\
     &-\frac{1}{4h}\left(-\alpha_{S,S}|E|^2-\frac{m_J'}{J'}\alpha_{V,S}(|E_{\sigma^+}|^2-|E_{\sigma^-}|^2)\right) \\
    \Delta\nu_{397_\sigma} &=-\frac{1}{4h}\left(\alpha_{S,D}-\frac{1}{5}\alpha_{V,D}+\frac{2}{5}\alpha_{T,D}-\alpha_{S,S}+\alpha_{V,S}\right)|E_{\sigma^-}|^2 \\
    \Delta\nu_{866} &=-\frac{1}{4h}\left(\alpha_{S,D}|E|^2+\frac{m_J}{J}\alpha_{V,D}(|E_{\sigma^+}|^2-|E_{\sigma^-}|^2)-\alpha_{T,D}\frac{|E|^2}{2}-\alpha_{S,S}|E|^2-\frac{m_J'}{J'}\alpha_{V,S}(|E_{\sigma^+}|^2-|E_{\sigma^-}|^2)\right)
\end{split}
\end{equation}
\end{widetext}
Here we assume a linear polarization of the $866\,$nm laser oriented perpendicular to the quantization axis ($E_\pi=0$). We obtain the overall electric field by averaging the light shifts over the two measured transitions  and inverting the above equations. Note that for $\Delta\nu_{866}$ and $\Delta\nu_{397_\pi}$ the averaging removes the dependence on the vector light shifts. The tensor shift component $\alpha_{T,D}$ for $\Delta\nu_{397_\pi}$ is eight orders of magnitude smaller than $\alpha_{S,S}$ (see Tab.~\ref{tab:polarizabilities}) and its effect is negligible.

\subsection{AC-Stark shift calculations for Aluminium}
\subsubsection{Method of calculation}
We consider Al$^+$ as a divalent ion with the core $[1s^2, 2s^2, 2p^6]$ and two valence electrons above it. The initial Dirac-Hartree-Fock (DHF) self-consistency procedure included the Breit interaction and was performed for the core electrons. Then, the DHF $s,p$, and $d$ orbitals with the principal quantum number $n=3$ and 4 were constructed in the frozen core potential. The remaining virtual orbitals were formed using a recurrent procedure described in~\cite{KozPorFla96,KozPorSaf15}. In total, the basis set includes six partial waves ($l_{\rm max} = 5$) and orbitals with $n$ up to 25.

In an approach combining configuration interaction (CI) and a method allowing us to include core-valence correlations~\cite{DzuFlaKoz96,SafKozJoh09}, the wave functions and energy levels of the valence electrons were found by solving the multiparticle relativistic equation~\cite{DzuFlaKoz96},
\begin{equation}
H_{\rm eff}(E_n) \Phi_n = E_n \Phi_n,
\label{Heff}
\end{equation}
where the effective Hamiltonian is defined as
\begin{equation}
H_{\rm eff}(E) = H_{\rm FC} + \Sigma(E),
\label{Heff1}
\end{equation}
with $H_{\rm FC}$ being the Hamiltonian in the frozen-core approximation.
The energy-dependent operator $\Sigma(E)$ accounts for the virtual excitations of the core electrons.
We constructed it in two ways: using (i) the second-order many-body perturbation theory (MBPT) over the residual Coulomb interaction~\cite{DzuFlaKoz96}
and (ii) the linearized coupled cluster single-double (LCCSD) method~\cite{SafKozJoh09}.
In the following, we refer to these approaches as CI+MBPT and CI+all-order methods.
\subsubsection{Polarizabilities}
We calculated the dynamic polarizabilities of the clock $3s^2\,\,^1\!S_0$ and $3s3p\,\,^3\!P_0$ states and the differential polarizabilities
$\Delta \alpha = \alpha(^3\!P_0) -\alpha(^1\!S_0)$ at wavelengths $\lambda = 397$ and $866$ nm.

We carried out calculations in the framework of the CI+MBPT and CI+all-order methods.
In both cases, the random phase approximation (RPA) corrections were included.
We present respective results (in $a_{\rm B}^3$, where $a_{\rm B}$ is the Bohr radius) in the \nth{3} and \nth{4} columns of Table~\ref{Tab:Polar}.
\begin{table}[t]
\caption{The ac polarizabilities of the $^1\!S_0$ and $^3\!P_0$ states of \Al and the differential polarizabilities $\Delta \alpha$ (in $a_{\rm B}^3$), calculated at the wavelengths $\lambda = 397$ and $866$ nm in the CI+MBPT, CI+all-order (labeled as ``CI+All''), and CI+all-order+AC (labeled as ``CI+All+AC'')  approximations, are presented. The uncertainties are given in parentheses.}
\label{Tab:Polar}%
\begin{ruledtabular}
\begin{tabular}{lccccc}
         &                    &  CI+MBPT   &  CI+All    &  CI+All+AC &   Final        \\
\hline \\ [-0.6pc]
 397 nm  & $\alpha(^1\!S_0)$  &  29.177    &   29.198   &   29.082   &  $29.08(12)$   \\[0.2pc]
         & $\alpha(^3\!P_0)$  &  30.285    &   30.309   &   30.232   &  $30.23(8)$    \\[0.2pc]
         & $\Delta \alpha$    &   1.108    &    1.111   &    1.150   &   $1.15(14)$   \\[0.5pc]
 866 nm  & $\alpha(^1\!S_0)$  &  24.997    &   25.013   &   24.913   &  $24.91(10)$   \\[0.2pc]
         & $\alpha(^3\!P_0)$  &  25.578    &   25.594   &   25.529   &  $25.53(7)$    \\[0.2pc]
         & $\Delta \alpha$    &   0.581    &    0.581   &    0.616   &   $0.62(12)$
\end{tabular}
\end{ruledtabular}
\end{table}
We also took into account other corrections to the electric-dipole operator: the two-particle (2P) and core Brueckner ($\sigma$)~\cite{DzuKozPor98},
the structural radiation (SR)~\cite{DzuFlaSil87,BlaJohLiu89}, and the normalization (Norm.) corrections. The results of this calculation, labeled as ``CI+All+AC''
(where the abbreviation ``AC'' means all corrections beyond RPA), are displayed in the \nth{5} column. We consider these results to be final.
Based on the difference between the CI+all-order+AC and CI+all-order values, we determined the uncertainties of
$\alpha(^1\!S_0)$ and $\alpha(^3\!P_0)$. The uncertainties of the final values are given in parentheses.

We also calculated the polarizabilities of the $^1\!S_0$ and $^3\!P_0$ states
at the zero frequency (static) and at $\lambda = 1068\,\,{\rm nm}$. The results obtained in the CI+all-order+AC approximation are given in Table~\ref{Tab:Polar2} and compared to the experimental values.
\begin{table}[t]
\caption{The CI+all-order+AC values of the static and ac ($\lambda = 1068$ nm) polarizabilities of the $^1\!S_0$ and $^3\!P_0$ states of \Al and the differential polarizabilities $\Delta \alpha$ (in $a_{\rm B}^3$) are presented.}
\label{Tab:Polar2}
\begin{ruledtabular}
\begin{tabular}{lccc}
         &                    &  CI+All+AC &   Experiment                      \\
\hline \\ [-0.6pc]
 Static  & $\alpha(^1\!S_0)$  &   24.00    &                                   \\[0.2pc]
         & $\alpha(^3\!P_0)$  &   24.52    &                                   \\[0.2pc]
         & $\Delta \alpha$    &    0.52    &   $0.43(6)$~\cite{BreCheHan19}    \\[0.2pc]
         &                    &            &   $0.416(14)$~\cite{WeiChaCui24}  \\[0.5pc]

1068 nm  & $\alpha(^1\!S_0)$  &   24.60    &                                   \\[0.2pc]
         & $\alpha(^3\!P_0)$  &   25.18    &                                   \\[0.2pc]
         & $\Delta \alpha$    &    0.58    &   $0.476(14)$~\cite{WeiChaCui24}
\end{tabular}
\end{ruledtabular}
\end{table}

We see that our results for $\Delta \alpha$ differ by $\sim 20\%$ from the central values of the experimental results of
Wei {\it et al}.~\cite{WeiChaCui24}. The reason for such a discrepancy is not quite clear to us.
At the same time, we note that the polarizabilities of the $^1\!S_0$ and $^3\!P_0$ states are very close to each other. In differential polarizabilities, they cancel each other out at the level of 96\% or more.
As a result, $\Delta \alpha$ can be rather sensitive to different small corrections still not included in our calculation.

In particular, we disregarded the quantum electro-dynamical (QED) corrections.
We did not take these corrections into account for either the wave functions nor for the electric dipole operator. Furthermore, valence and core triple excitations were not included, and the 2P, $\sigma$, and SR corrections were calculated in the second order of perturbation theory only.

Based on this, we assume that our usual method for determining the uncertainty, as a difference between
CI+all-order+AC and CI+all-order (or CI+MBPT) values can underestimate it for differential polarizabilities.
A more conservative estimate of the absolute uncertainty of the differential polarizability as
\begin{equation}
\Delta(\Delta \alpha) = \sqrt{(\Delta \alpha(^1\!S_0))^2
                            + (\Delta \alpha(^3\!P_0))^2},
\end{equation}
looks more reliable in the present case.

\subsection{Estimated frequency uncertainty of the ac-Stark shift}

\begin{table*}
\caption{\label{tab:uncertainty_light_shift} Source of the uncertainties for all cooling and repumping lasers involved.}
\begin{ruledtabular}
\begin{tabular}{l|ccc}
Effect & uncertainty   & uncertainty  & uncertainty  \\
 & $397\,$nm $\sigma$ ($10^{-18}$)  & $397\,$nm $\pi$  ($10^{-18}$) &  $866\,$nm   ($10^{-18}$) \\
\hline
$\Delta\alpha_{Al}$ & 0.89 & 0.10 & 0.27\\ 
$\alpha_{S,S-Ca}$ & 0.24 & 0.03 & 0.0002\\ 
$\alpha_{V,S-Ca}$ & 0.24 & - & -\\
$\alpha_{S,D-Ca}$ & $\approx10^{-9}$& $\approx10^{-9}$ & 0.004\\ 
$\alpha_{V,D-Ca}$ & $\approx10^{-10}$& - & -\\
$\alpha_{T,D-Ca}$ &$\approx10^{-10}$& $\approx10^{-9}$ & 0.002\\
extrapolation & 0.09& 0.008 & 0.015\\ 
I(Ca)/I(Al) & 0.2& 0.11& 0.01\\
\hline
Total uncertainty & 0.99 & 0.16 &  0.27 

\end{tabular}
\end{ruledtabular}
\end{table*}

Table~\ref{tab:uncertainty_light_shift} shows the different contributions to the uncertainty of the light shifts. The differential polarizability of \Al has the most significant impact on the uncertainty, with the highest contribution coming from the $397\,$nm light, because it has the largest intensity. To reduce the uncertainty, a direct measurement of the differential polarizability at the cooling wavelengths should be conducted. 
The intensity difference between the ions is another uncertainty source due to their distance. Aligning the laser on a single ion ensures it is well-centered for the illumination of a two-ion crystal. For $397\,$nm one large remaining shift is the frequency-dependent polarizability of \Ca. Here, the error is dominated by laser frequency fluctuations, which can be overcome with a more stable frequency lock.

\bibliography{EQM_Master,supplement,AlII,ca}

\begin{thebibliography}{84}%
\makeatletter
\providecommand \@ifxundefined [1]{%
 \@ifx{#1\undefined}
}%
\providecommand \@ifnum [1]{%
 \ifnum #1\expandafter \@firstoftwo
 \else \expandafter \@secondoftwo
 \fi
}%
\providecommand \@ifx [1]{%
 \ifx #1\expandafter \@firstoftwo
 \else \expandafter \@secondoftwo
 \fi
}%
\providecommand \natexlab [1]{#1}%
\providecommand \enquote  [1]{``#1''}%
\providecommand \bibnamefont  [1]{#1}%
\providecommand \bibfnamefont [1]{#1}%
\providecommand \citenamefont [1]{#1}%
\providecommand \href@noop [0]{\@secondoftwo}%
\providecommand \href [0]{\begingroup \@sanitize@url \@href}%
\providecommand \@href[1]{\@@startlink{#1}\@@href}%
\providecommand \@@href[1]{\endgroup#1\@@endlink}%
\providecommand \@sanitize@url [0]{\catcode `\\12\catcode `\$12\catcode `\&12\catcode `\#12\catcode `\^12\catcode `\_12\catcode `\%12\relax}%
\providecommand \@@startlink[1]{}%
\providecommand \@@endlink[0]{}%
\providecommand \url  [0]{\begingroup\@sanitize@url \@url }%
\providecommand \@url [1]{\endgroup\@href {#1}{\urlprefix }}%
\providecommand \urlprefix  [0]{URL }%
\providecommand \Eprint [0]{\href }%
\providecommand \doibase [0]{https://doi.org/}%
\providecommand \selectlanguage [0]{\@gobble}%
\providecommand \bibinfo  [0]{\@secondoftwo}%
\providecommand \bibfield  [0]{\@secondoftwo}%
\providecommand \translation [1]{[#1]}%
\providecommand \BibitemOpen [0]{}%
\providecommand \bibitemStop [0]{}%
\providecommand \bibitemNoStop [0]{.\EOS\space}%
\providecommand \EOS [0]{\spacefactor3000\relax}%
\providecommand \BibitemShut  [1]{\csname bibitem#1\endcsname}%
\let\auto@bib@innerbib\@empty
\bibitem [{\citenamefont {Brewer}\ \emph {et~al.}(2019{\natexlab{a}})\citenamefont {Brewer}, \citenamefont {Chen}, \citenamefont {Hankin}, \citenamefont {Clements}, \citenamefont {Chou}, \citenamefont {Wineland}, \citenamefont {Hume},\ and\ \citenamefont {Leibrandt}}]{brewer_27al_2019}%
  \BibitemOpen
  \bibfield  {author} {\bibinfo {author} {\bibfnamefont {S.~M.}\ \bibnamefont {Brewer}}, \bibinfo {author} {\bibfnamefont {J.-S.}\ \bibnamefont {Chen}}, \bibinfo {author} {\bibfnamefont {A.~M.}\ \bibnamefont {Hankin}}, \bibinfo {author} {\bibfnamefont {E.~R.}\ \bibnamefont {Clements}}, \bibinfo {author} {\bibfnamefont {C.~W.}\ \bibnamefont {Chou}}, \bibinfo {author} {\bibfnamefont {D.~J.}\ \bibnamefont {Wineland}}, \bibinfo {author} {\bibfnamefont {D.~B.}\ \bibnamefont {Hume}},\ and\ \bibinfo {author} {\bibfnamefont {D.~R.}\ \bibnamefont {Leibrandt}},\ }\bibfield  {title} {\bibinfo {title} {{\textsuperscript{27}}{{Al}}{\textsuperscript{+}} {{Quantum-Logic Clock}} with a {{Systematic Uncertainty}} below 10{\textsuperscript{-18}}},\ }\href {https://doi.org/10.1103/PhysRevLett.123.033201} {\bibfield  {journal} {\bibinfo  {journal} {Physical Review Letters}\ }\textbf {\bibinfo {volume} {123}},\ \bibinfo {pages} {033201} (\bibinfo {year} {2019}{\natexlab{a}})}\BibitemShut {NoStop}%
\bibitem [{\citenamefont {Aeppli}\ \emph {et~al.}(2024)\citenamefont {Aeppli}, \citenamefont {Kim}, \citenamefont {Warfield}, \citenamefont {Safronova},\ and\ \citenamefont {Ye}}]{aeppli_clock_2024}%
  \BibitemOpen
  \bibfield  {author} {\bibinfo {author} {\bibfnamefont {A.}~\bibnamefont {Aeppli}}, \bibinfo {author} {\bibfnamefont {K.}~\bibnamefont {Kim}}, \bibinfo {author} {\bibfnamefont {W.}~\bibnamefont {Warfield}}, \bibinfo {author} {\bibfnamefont {M.~S.}\ \bibnamefont {Safronova}},\ and\ \bibinfo {author} {\bibfnamefont {J.}~\bibnamefont {Ye}},\ }\bibfield  {title} {\bibinfo {title} {A clock with 8{\texttimes}10{\textsuperscript{-19}} systematic uncertainty},\ }\href {https://doi.org/10.1103/PhysRevLett.133.023401} {\bibfield  {journal} {\bibinfo  {journal} {Physical Review Letters}\ }\textbf {\bibinfo {volume} {133}},\ \bibinfo {pages} {023401} (\bibinfo {year} {2024})},\ \Eprint {https://arxiv.org/abs/2403.10664} {arXiv:2403.10664 [physics]} \BibitemShut {NoStop}%
\bibitem [{\citenamefont {Marshall}\ \emph {et~al.}(2025)\citenamefont {Marshall}, \citenamefont {Castillo}, \citenamefont {{Arthur-Dworschack}}, \citenamefont {Aeppli}, \citenamefont {Kim}, \citenamefont {Lee}, \citenamefont {Warfield}, \citenamefont {Hinrichs}, \citenamefont {Nardelli}, \citenamefont {Fortier}, \citenamefont {Ye}, \citenamefont {Leibrandt},\ and\ \citenamefont {Hume}}]{marshall_highstability_2025}%
  \BibitemOpen
  \bibfield  {author} {\bibinfo {author} {\bibfnamefont {M.~C.}\ \bibnamefont {Marshall}}, \bibinfo {author} {\bibfnamefont {D.~A.~R.}\ \bibnamefont {Castillo}}, \bibinfo {author} {\bibfnamefont {W.~J.}\ \bibnamefont {{Arthur-Dworschack}}}, \bibinfo {author} {\bibfnamefont {A.}~\bibnamefont {Aeppli}}, \bibinfo {author} {\bibfnamefont {K.}~\bibnamefont {Kim}}, \bibinfo {author} {\bibfnamefont {D.}~\bibnamefont {Lee}}, \bibinfo {author} {\bibfnamefont {W.}~\bibnamefont {Warfield}}, \bibinfo {author} {\bibfnamefont {J.}~\bibnamefont {Hinrichs}}, \bibinfo {author} {\bibfnamefont {N.~V.}\ \bibnamefont {Nardelli}}, \bibinfo {author} {\bibfnamefont {T.~M.}\ \bibnamefont {Fortier}}, \bibinfo {author} {\bibfnamefont {J.}~\bibnamefont {Ye}}, \bibinfo {author} {\bibfnamefont {D.~R.}\ \bibnamefont {Leibrandt}},\ and\ \bibinfo {author} {\bibfnamefont {D.~B.}\ \bibnamefont {Hume}},\ }\bibfield  {title} {\bibinfo {title} {High-{{Stability Single-Ion Clock}} with
  \$5.5{\textbackslash}ifmmode{\textbackslash}times{\textbackslash}else{\textbackslash}texttimes{\textbackslash}fi\{\}\{10\}{\textasciicircum}\{{\textbackslash}ensuremath\{-\}19\}\$ {{Systematic Uncertainty}}},\ }\href {https://doi.org/10.1103/hb3c-dk28} {\bibfield  {journal} {\bibinfo  {journal} {Physical Review Letters}\ }\textbf {\bibinfo {volume} {135}},\ \bibinfo {pages} {033201} (\bibinfo {year} {2025})}\BibitemShut {NoStop}%
\bibitem [{\citenamefont {Arnold}\ \emph {et~al.}(2024)\citenamefont {Arnold}, \citenamefont {Bustabad}, \citenamefont {Qi}, \citenamefont {Qichen}, \citenamefont {Zhang}, \citenamefont {Zhao},\ and\ \citenamefont {Barrett}}]{arnold_validating_2024}%
  \BibitemOpen
  \bibfield  {author} {\bibinfo {author} {\bibfnamefont {K.~J.}\ \bibnamefont {Arnold}}, \bibinfo {author} {\bibfnamefont {S.}~\bibnamefont {Bustabad}}, \bibinfo {author} {\bibfnamefont {Z.}~\bibnamefont {Qi}}, \bibinfo {author} {\bibfnamefont {Q.}~\bibnamefont {Qichen}}, \bibinfo {author} {\bibfnamefont {Z.}~\bibnamefont {Zhang}}, \bibinfo {author} {\bibfnamefont {Z.}~\bibnamefont {Zhao}},\ and\ \bibinfo {author} {\bibfnamefont {M.~D.}\ \bibnamefont {Barrett}},\ }\bibfield  {title} {\bibinfo {title} {Validating a lutetium frequency reference.},\ }\href {https://doi.org/10.1088/1742-6596/2889/1/012040} {\bibfield  {journal} {\bibinfo  {journal} {Journal of Physics: Conference Series}\ }\textbf {\bibinfo {volume} {2889}},\ \bibinfo {pages} {012040} (\bibinfo {year} {2024})}\BibitemShut {NoStop}%
\bibitem [{\citenamefont {McGrew}\ \emph {et~al.}(2018)\citenamefont {McGrew}, \citenamefont {Zhang}, \citenamefont {Fasano}, \citenamefont {Sch{\"a}ffer}, \citenamefont {Beloy}, \citenamefont {Nicolodi}, \citenamefont {Brown}, \citenamefont {Hinkley}, \citenamefont {Milani}, \citenamefont {Schioppo}, \citenamefont {Yoon},\ and\ \citenamefont {Ludlow}}]{mcgrew_atomic_2018}%
  \BibitemOpen
  \bibfield  {author} {\bibinfo {author} {\bibfnamefont {W.~F.}\ \bibnamefont {McGrew}}, \bibinfo {author} {\bibfnamefont {X.}~\bibnamefont {Zhang}}, \bibinfo {author} {\bibfnamefont {R.~J.}\ \bibnamefont {Fasano}}, \bibinfo {author} {\bibfnamefont {S.~A.}\ \bibnamefont {Sch{\"a}ffer}}, \bibinfo {author} {\bibfnamefont {K.}~\bibnamefont {Beloy}}, \bibinfo {author} {\bibfnamefont {D.}~\bibnamefont {Nicolodi}}, \bibinfo {author} {\bibfnamefont {R.~C.}\ \bibnamefont {Brown}}, \bibinfo {author} {\bibfnamefont {N.}~\bibnamefont {Hinkley}}, \bibinfo {author} {\bibfnamefont {G.}~\bibnamefont {Milani}}, \bibinfo {author} {\bibfnamefont {M.}~\bibnamefont {Schioppo}}, \bibinfo {author} {\bibfnamefont {T.~H.}\ \bibnamefont {Yoon}},\ and\ \bibinfo {author} {\bibfnamefont {A.~D.}\ \bibnamefont {Ludlow}},\ }\bibfield  {title} {\bibinfo {title} {Atomic clock performance enabling geodesy below the centimetre level},\ }\href {https://doi.org/10.1038/s41586-018-0738-2} {\bibfield  {journal} {\bibinfo  {journal} {Nature}\
  }\textbf {\bibinfo {volume} {564}},\ \bibinfo {pages} {87} (\bibinfo {year} {2018})}\BibitemShut {NoStop}%
\bibitem [{\citenamefont {Sanner}\ \emph {et~al.}(2019)\citenamefont {Sanner}, \citenamefont {Huntemann}, \citenamefont {Lange}, \citenamefont {Tamm}, \citenamefont {Peik}, \citenamefont {Safronova},\ and\ \citenamefont {Porsev}}]{sanner_optical_2019}%
  \BibitemOpen
  \bibfield  {author} {\bibinfo {author} {\bibfnamefont {C.}~\bibnamefont {Sanner}}, \bibinfo {author} {\bibfnamefont {N.}~\bibnamefont {Huntemann}}, \bibinfo {author} {\bibfnamefont {R.}~\bibnamefont {Lange}}, \bibinfo {author} {\bibfnamefont {C.}~\bibnamefont {Tamm}}, \bibinfo {author} {\bibfnamefont {E.}~\bibnamefont {Peik}}, \bibinfo {author} {\bibfnamefont {M.~S.}\ \bibnamefont {Safronova}},\ and\ \bibinfo {author} {\bibfnamefont {S.~G.}\ \bibnamefont {Porsev}},\ }\bibfield  {title} {\bibinfo {title} {Optical clock comparison for {{Lorentz}} symmetry testing},\ }\href {https://doi.org/10.1038/s41586-019-0972-2} {\bibfield  {journal} {\bibinfo  {journal} {Nature}\ }\textbf {\bibinfo {volume} {567}},\ \bibinfo {pages} {204} (\bibinfo {year} {2019})}\BibitemShut {NoStop}%
\bibitem [{\citenamefont {Takamoto}\ \emph {et~al.}(2020)\citenamefont {Takamoto}, \citenamefont {Ushijima}, \citenamefont {Ohmae}, \citenamefont {Yahagi}, \citenamefont {Kokado}, \citenamefont {Shinkai},\ and\ \citenamefont {Katori}}]{takamoto_test_2020}%
  \BibitemOpen
  \bibfield  {author} {\bibinfo {author} {\bibfnamefont {M.}~\bibnamefont {Takamoto}}, \bibinfo {author} {\bibfnamefont {I.}~\bibnamefont {Ushijima}}, \bibinfo {author} {\bibfnamefont {N.}~\bibnamefont {Ohmae}}, \bibinfo {author} {\bibfnamefont {T.}~\bibnamefont {Yahagi}}, \bibinfo {author} {\bibfnamefont {K.}~\bibnamefont {Kokado}}, \bibinfo {author} {\bibfnamefont {H.}~\bibnamefont {Shinkai}},\ and\ \bibinfo {author} {\bibfnamefont {H.}~\bibnamefont {Katori}},\ }\bibfield  {title} {\bibinfo {title} {Test of general relativity by a pair of transportable optical lattice clocks},\ }\href {https://doi.org/10.1038/s41566-020-0619-8} {\bibfield  {journal} {\bibinfo  {journal} {Nature Photonics}\ }\textbf {\bibinfo {volume} {14}},\ \bibinfo {pages} {411} (\bibinfo {year} {2020})}\BibitemShut {NoStop}%
\bibitem [{\citenamefont {Hausser}\ \emph {et~al.}(2025)\citenamefont {Hausser}, \citenamefont {Keller}, \citenamefont {Nordmann}, \citenamefont {Bhatt}, \citenamefont {Kiethe}, \citenamefont {Liu}, \citenamefont {Richter}, \citenamefont {{von Boehn}}, \citenamefont {Rahm}, \citenamefont {Weyers}, \citenamefont {Benkler}, \citenamefont {Lipphardt}, \citenamefont {D{\"o}rscher}, \citenamefont {Stahl}, \citenamefont {Klose}, \citenamefont {Lisdat}, \citenamefont {Filzinger}, \citenamefont {Huntemann}, \citenamefont {Peik},\ and\ \citenamefont {Mehlst{\"a}ubler}}]{hausser_115in+172yb_2025}%
  \BibitemOpen
  \bibfield  {author} {\bibinfo {author} {\bibfnamefont {H.~N.}\ \bibnamefont {Hausser}}, \bibinfo {author} {\bibfnamefont {J.}~\bibnamefont {Keller}}, \bibinfo {author} {\bibfnamefont {T.}~\bibnamefont {Nordmann}}, \bibinfo {author} {\bibfnamefont {N.~M.}\ \bibnamefont {Bhatt}}, \bibinfo {author} {\bibfnamefont {J.}~\bibnamefont {Kiethe}}, \bibinfo {author} {\bibfnamefont {H.}~\bibnamefont {Liu}}, \bibinfo {author} {\bibfnamefont {I.~M.}\ \bibnamefont {Richter}}, \bibinfo {author} {\bibfnamefont {M.}~\bibnamefont {{von Boehn}}}, \bibinfo {author} {\bibfnamefont {J.}~\bibnamefont {Rahm}}, \bibinfo {author} {\bibfnamefont {S.}~\bibnamefont {Weyers}}, \bibinfo {author} {\bibfnamefont {E.}~\bibnamefont {Benkler}}, \bibinfo {author} {\bibfnamefont {B.}~\bibnamefont {Lipphardt}}, \bibinfo {author} {\bibfnamefont {S.}~\bibnamefont {D{\"o}rscher}}, \bibinfo {author} {\bibfnamefont {K.}~\bibnamefont {Stahl}}, \bibinfo {author} {\bibfnamefont {J.}~\bibnamefont {Klose}}, \bibinfo {author} {\bibfnamefont
  {C.}~\bibnamefont {Lisdat}}, \bibinfo {author} {\bibfnamefont {M.}~\bibnamefont {Filzinger}}, \bibinfo {author} {\bibfnamefont {N.}~\bibnamefont {Huntemann}}, \bibinfo {author} {\bibfnamefont {E.}~\bibnamefont {Peik}},\ and\ \bibinfo {author} {\bibfnamefont {T.~E.}\ \bibnamefont {Mehlst{\"a}ubler}},\ }\bibfield  {title} {\bibinfo {title} {{\textsuperscript{115}}{{In}}{\textsuperscript{+}}-{\textsuperscript{172}}{{Yb}}{\textsuperscript{+}} {{Coulomb Crystal Clock}} with 2.5{\texttimes}10{\textsuperscript{-18}} {{Systematic Uncertainty}}},\ }\href {https://doi.org/10.1103/PhysRevLett.134.023201} {\bibfield  {journal} {\bibinfo  {journal} {Physical Review Letters}\ }\textbf {\bibinfo {volume} {134}},\ \bibinfo {pages} {023201} (\bibinfo {year} {2025})}\BibitemShut {NoStop}%
\bibitem [{\citenamefont {{Amy-Klein}}\ \emph {et~al.}(2024)\citenamefont {{Amy-Klein}}, \citenamefont {Benkler}, \citenamefont {Blond{\'e}}, \citenamefont {Bongs}, \citenamefont {Cantin}, \citenamefont {Chardonnet}, \citenamefont {Denker}, \citenamefont {D{\"o}rscher}, \citenamefont {Feng}, \citenamefont {Gaudron}, \citenamefont {Gill}, \citenamefont {Hill}, \citenamefont {Huang}, \citenamefont {Johnson}, \citenamefont {Kale}, \citenamefont {Katori}, \citenamefont {Klose}, \citenamefont {Kronj{\"a}ger}, \citenamefont {Kuhl}, \citenamefont {Targat}, \citenamefont {Lisdat}, \citenamefont {Lopez}, \citenamefont {L{\"u}cke}, \citenamefont {Mazouth}, \citenamefont {Mukherjee}, \citenamefont {Nosske}, \citenamefont {Pointard}, \citenamefont {Pottie}, \citenamefont {Schioppo}, \citenamefont {Singh}, \citenamefont {Stahl}, \citenamefont {Takamoto}, \citenamefont {T{\o}nnes}, \citenamefont {Tunesi}, \citenamefont {Ushijima},\ and\ \citenamefont {Vishwakarma}}]{amy-klein_international_2024}%
  \BibitemOpen
  \bibfield  {author} {\bibinfo {author} {\bibfnamefont {A.}~\bibnamefont {{Amy-Klein}}}, \bibinfo {author} {\bibfnamefont {E.}~\bibnamefont {Benkler}}, \bibinfo {author} {\bibfnamefont {P.}~\bibnamefont {Blond{\'e}}}, \bibinfo {author} {\bibfnamefont {K.}~\bibnamefont {Bongs}}, \bibinfo {author} {\bibfnamefont {E.}~\bibnamefont {Cantin}}, \bibinfo {author} {\bibfnamefont {C.}~\bibnamefont {Chardonnet}}, \bibinfo {author} {\bibfnamefont {H.}~\bibnamefont {Denker}}, \bibinfo {author} {\bibfnamefont {S.}~\bibnamefont {D{\"o}rscher}}, \bibinfo {author} {\bibfnamefont {C.-H.}\ \bibnamefont {Feng}}, \bibinfo {author} {\bibfnamefont {J.-O.}\ \bibnamefont {Gaudron}}, \bibinfo {author} {\bibfnamefont {P.}~\bibnamefont {Gill}}, \bibinfo {author} {\bibfnamefont {I.~R.}\ \bibnamefont {Hill}}, \bibinfo {author} {\bibfnamefont {W.}~\bibnamefont {Huang}}, \bibinfo {author} {\bibfnamefont {M.~Y.~H.}\ \bibnamefont {Johnson}}, \bibinfo {author} {\bibfnamefont {Y.~B.}\ \bibnamefont {Kale}}, \bibinfo {author} {\bibfnamefont
  {H.}~\bibnamefont {Katori}}, \bibinfo {author} {\bibfnamefont {J.}~\bibnamefont {Klose}}, \bibinfo {author} {\bibfnamefont {J.}~\bibnamefont {Kronj{\"a}ger}}, \bibinfo {author} {\bibfnamefont {A.}~\bibnamefont {Kuhl}}, \bibinfo {author} {\bibfnamefont {R.~L.}\ \bibnamefont {Targat}}, \bibinfo {author} {\bibfnamefont {C.}~\bibnamefont {Lisdat}}, \bibinfo {author} {\bibfnamefont {O.}~\bibnamefont {Lopez}}, \bibinfo {author} {\bibfnamefont {T.}~\bibnamefont {L{\"u}cke}}, \bibinfo {author} {\bibfnamefont {M.}~\bibnamefont {Mazouth}}, \bibinfo {author} {\bibfnamefont {S.}~\bibnamefont {Mukherjee}}, \bibinfo {author} {\bibfnamefont {I.}~\bibnamefont {Nosske}}, \bibinfo {author} {\bibfnamefont {B.}~\bibnamefont {Pointard}}, \bibinfo {author} {\bibfnamefont {P.-E.}\ \bibnamefont {Pottie}}, \bibinfo {author} {\bibfnamefont {M.}~\bibnamefont {Schioppo}}, \bibinfo {author} {\bibfnamefont {Y.}~\bibnamefont {Singh}}, \bibinfo {author} {\bibfnamefont {K.}~\bibnamefont {Stahl}}, \bibinfo {author} {\bibfnamefont
  {M.}~\bibnamefont {Takamoto}}, \bibinfo {author} {\bibfnamefont {M.}~\bibnamefont {T{\o}nnes}}, \bibinfo {author} {\bibfnamefont {J.}~\bibnamefont {Tunesi}}, \bibinfo {author} {\bibfnamefont {I.}~\bibnamefont {Ushijima}},\ and\ \bibinfo {author} {\bibfnamefont {C.}~\bibnamefont {Vishwakarma}},\ }\href {https://doi.org/10.48550/arXiv.2410.22973} {\bibinfo {title} {International comparison of optical frequencies with transportable optical lattice clocks}} (\bibinfo {year} {2024}),\ \Eprint {https://arxiv.org/abs/2410.22973} {arXiv:2410.22973} \BibitemShut {NoStop}%
\bibitem [{\citenamefont {Beloy}\ \emph {et~al.}(2021)\citenamefont {Beloy}, \citenamefont {Bodine}, \citenamefont {Bothwell}, \citenamefont {Brewer}, \citenamefont {Bromley}, \citenamefont {Chen}, \citenamefont {Desch{\^e}nes}, \citenamefont {Diddams}, \citenamefont {Fasano}, \citenamefont {Fortier}, \citenamefont {Hassan}, \citenamefont {Hume}, \citenamefont {Kedar}, \citenamefont {Kennedy}, \citenamefont {Khader}, \citenamefont {Koepke}, \citenamefont {Leibrandt}, \citenamefont {Leopardi}, \citenamefont {Ludlow}, \citenamefont {McGrew}, \citenamefont {Milner}, \citenamefont {Newbury}, \citenamefont {Nicolodi}, \citenamefont {Oelker}, \citenamefont {Parker}, \citenamefont {Robinson}, \citenamefont {Romisch}, \citenamefont {Sch{\"a}ffer}, \citenamefont {Sherman}, \citenamefont {Sinclair}, \citenamefont {Sonderhouse}, \citenamefont {Swann}, \citenamefont {Yao}, \citenamefont {Ye}, \citenamefont {Zhang},\ and\ \citenamefont {Collaboration}}]{beloy_frequency_2021}%
  \BibitemOpen
  \bibfield  {author} {\bibinfo {author} {\bibfnamefont {K.}~\bibnamefont {Beloy}}, \bibinfo {author} {\bibfnamefont {M.~I.}\ \bibnamefont {Bodine}}, \bibinfo {author} {\bibfnamefont {T.}~\bibnamefont {Bothwell}}, \bibinfo {author} {\bibfnamefont {S.~M.}\ \bibnamefont {Brewer}}, \bibinfo {author} {\bibfnamefont {S.~L.}\ \bibnamefont {Bromley}}, \bibinfo {author} {\bibfnamefont {J.-S.}\ \bibnamefont {Chen}}, \bibinfo {author} {\bibfnamefont {J.-D.}\ \bibnamefont {Desch{\^e}nes}}, \bibinfo {author} {\bibfnamefont {S.~A.}\ \bibnamefont {Diddams}}, \bibinfo {author} {\bibfnamefont {R.~J.}\ \bibnamefont {Fasano}}, \bibinfo {author} {\bibfnamefont {T.~M.}\ \bibnamefont {Fortier}}, \bibinfo {author} {\bibfnamefont {Y.~S.}\ \bibnamefont {Hassan}}, \bibinfo {author} {\bibfnamefont {D.~B.}\ \bibnamefont {Hume}}, \bibinfo {author} {\bibfnamefont {D.}~\bibnamefont {Kedar}}, \bibinfo {author} {\bibfnamefont {C.~J.}\ \bibnamefont {Kennedy}}, \bibinfo {author} {\bibfnamefont {I.}~\bibnamefont {Khader}}, \bibinfo {author}
  {\bibfnamefont {A.}~\bibnamefont {Koepke}}, \bibinfo {author} {\bibfnamefont {D.~R.}\ \bibnamefont {Leibrandt}}, \bibinfo {author} {\bibfnamefont {H.}~\bibnamefont {Leopardi}}, \bibinfo {author} {\bibfnamefont {A.~D.}\ \bibnamefont {Ludlow}}, \bibinfo {author} {\bibfnamefont {W.~F.}\ \bibnamefont {McGrew}}, \bibinfo {author} {\bibfnamefont {W.~R.}\ \bibnamefont {Milner}}, \bibinfo {author} {\bibfnamefont {N.~R.}\ \bibnamefont {Newbury}}, \bibinfo {author} {\bibfnamefont {D.}~\bibnamefont {Nicolodi}}, \bibinfo {author} {\bibfnamefont {E.}~\bibnamefont {Oelker}}, \bibinfo {author} {\bibfnamefont {T.~E.}\ \bibnamefont {Parker}}, \bibinfo {author} {\bibfnamefont {J.~M.}\ \bibnamefont {Robinson}}, \bibinfo {author} {\bibfnamefont {S.}~\bibnamefont {Romisch}}, \bibinfo {author} {\bibfnamefont {S.~A.}\ \bibnamefont {Sch{\"a}ffer}}, \bibinfo {author} {\bibfnamefont {J.~A.}\ \bibnamefont {Sherman}}, \bibinfo {author} {\bibfnamefont {L.~C.}\ \bibnamefont {Sinclair}}, \bibinfo {author} {\bibfnamefont {L.}~\bibnamefont
  {Sonderhouse}}, \bibinfo {author} {\bibfnamefont {W.~C.}\ \bibnamefont {Swann}}, \bibinfo {author} {\bibfnamefont {J.}~\bibnamefont {Yao}}, \bibinfo {author} {\bibfnamefont {J.}~\bibnamefont {Ye}}, \bibinfo {author} {\bibfnamefont {X.}~\bibnamefont {Zhang}},\ and\ \bibinfo {author} {\bibfnamefont {B.~A. C. O. N.~B.}\ \bibnamefont {Collaboration}},\ }\bibfield  {title} {\bibinfo {title} {Frequency ratio measurements at 18-digit accuracy using an optical clock network},\ }\href {https://doi.org/10.1038/s41586-021-03253-4} {\bibfield  {journal} {\bibinfo  {journal} {Nature}\ }\textbf {\bibinfo {volume} {591}},\ \bibinfo {pages} {564} (\bibinfo {year} {2021})}\BibitemShut {NoStop}%
\bibitem [{\citenamefont {Mehlst{\"a}ubler}\ \emph {et~al.}(2018)\citenamefont {Mehlst{\"a}ubler}, \citenamefont {Grosche}, \citenamefont {Lisdat}, \citenamefont {Schmidt},\ and\ \citenamefont {Denker}}]{mehlstäubler_atomic_2018}%
  \BibitemOpen
  \bibfield  {author} {\bibinfo {author} {\bibfnamefont {T.~E.}\ \bibnamefont {Mehlst{\"a}ubler}}, \bibinfo {author} {\bibfnamefont {G.}~\bibnamefont {Grosche}}, \bibinfo {author} {\bibfnamefont {C.}~\bibnamefont {Lisdat}}, \bibinfo {author} {\bibfnamefont {P.~O.}\ \bibnamefont {Schmidt}},\ and\ \bibinfo {author} {\bibfnamefont {H.}~\bibnamefont {Denker}},\ }\bibfield  {title} {\bibinfo {title} {Atomic clocks for geodesy},\ }\href {https://doi.org/10.1088/1361-6633/aab409} {\bibfield  {journal} {\bibinfo  {journal} {Reports on Progress in Physics}\ }\textbf {\bibinfo {volume} {81}},\ \bibinfo {pages} {064401} (\bibinfo {year} {2018})}\BibitemShut {NoStop}%
\bibitem [{\citenamefont {Grotti}\ \emph {et~al.}(2024)\citenamefont {Grotti}, \citenamefont {Nosske}, \citenamefont {Koller}, \citenamefont {Herbers}, \citenamefont {Denker}, \citenamefont {Timmen}, \citenamefont {Vishnyakova}, \citenamefont {Grosche}, \citenamefont {Waterholter}, \citenamefont {Kuhl}, \citenamefont {Koke}, \citenamefont {Benkler}, \citenamefont {Giunta}, \citenamefont {Maisenbacher}, \citenamefont {Matveev}, \citenamefont {D{\"o}rscher}, \citenamefont {Schwarz}, \citenamefont {{Al-Masoudi}}, \citenamefont {H{\"a}nsch}, \citenamefont {Udem}, \citenamefont {Holzwarth},\ and\ \citenamefont {Lisdat}}]{grotti_longdistance_2024}%
  \BibitemOpen
  \bibfield  {author} {\bibinfo {author} {\bibfnamefont {J.}~\bibnamefont {Grotti}}, \bibinfo {author} {\bibfnamefont {I.}~\bibnamefont {Nosske}}, \bibinfo {author} {\bibfnamefont {S.}~\bibnamefont {Koller}}, \bibinfo {author} {\bibfnamefont {S.}~\bibnamefont {Herbers}}, \bibinfo {author} {\bibfnamefont {H.}~\bibnamefont {Denker}}, \bibinfo {author} {\bibfnamefont {L.}~\bibnamefont {Timmen}}, \bibinfo {author} {\bibfnamefont {G.}~\bibnamefont {Vishnyakova}}, \bibinfo {author} {\bibfnamefont {G.}~\bibnamefont {Grosche}}, \bibinfo {author} {\bibfnamefont {T.}~\bibnamefont {Waterholter}}, \bibinfo {author} {\bibfnamefont {A.}~\bibnamefont {Kuhl}}, \bibinfo {author} {\bibfnamefont {S.}~\bibnamefont {Koke}}, \bibinfo {author} {\bibfnamefont {E.}~\bibnamefont {Benkler}}, \bibinfo {author} {\bibfnamefont {M.}~\bibnamefont {Giunta}}, \bibinfo {author} {\bibfnamefont {L.}~\bibnamefont {Maisenbacher}}, \bibinfo {author} {\bibfnamefont {A.}~\bibnamefont {Matveev}}, \bibinfo {author} {\bibfnamefont {S.}~\bibnamefont
  {D{\"o}rscher}}, \bibinfo {author} {\bibfnamefont {R.}~\bibnamefont {Schwarz}}, \bibinfo {author} {\bibfnamefont {A.}~\bibnamefont {{Al-Masoudi}}}, \bibinfo {author} {\bibfnamefont {T.}~\bibnamefont {H{\"a}nsch}}, \bibinfo {author} {\bibfnamefont {{\relax Th}.}~\bibnamefont {Udem}}, \bibinfo {author} {\bibfnamefont {R.}~\bibnamefont {Holzwarth}},\ and\ \bibinfo {author} {\bibfnamefont {C.}~\bibnamefont {Lisdat}},\ }\bibfield  {title} {\bibinfo {title} {Long-distance chronometric leveling with a portable optical clock},\ }\href {https://doi.org/10.1103/PhysRevApplied.21.L061001} {\bibfield  {journal} {\bibinfo  {journal} {Physical Review Applied}\ }\textbf {\bibinfo {volume} {21}},\ \bibinfo {pages} {L061001} (\bibinfo {year} {2024})}\BibitemShut {NoStop}%
\bibitem [{\citenamefont {Grotti}\ \emph {et~al.}(2018)\citenamefont {Grotti}, \citenamefont {Koller}, \citenamefont {Vogt}, \citenamefont {H{\"a}fner}, \citenamefont {Sterr}, \citenamefont {Lisdat}, \citenamefont {Denker}, \citenamefont {Voigt}, \citenamefont {Timmen}, \citenamefont {Rolland}, \citenamefont {Baynes}, \citenamefont {Margolis}, \citenamefont {Zampaolo}, \citenamefont {Thoumany}, \citenamefont {Pizzocaro}, \citenamefont {Rauf}, \citenamefont {Bregolin}, \citenamefont {Tampellini}, \citenamefont {Barbieri}, \citenamefont {Zucco}, \citenamefont {Costanzo}, \citenamefont {Clivati}, \citenamefont {Levi},\ and\ \citenamefont {Calonico}}]{grotti_geodesy_2018}%
  \BibitemOpen
  \bibfield  {author} {\bibinfo {author} {\bibfnamefont {J.}~\bibnamefont {Grotti}}, \bibinfo {author} {\bibfnamefont {S.}~\bibnamefont {Koller}}, \bibinfo {author} {\bibfnamefont {S.}~\bibnamefont {Vogt}}, \bibinfo {author} {\bibfnamefont {S.}~\bibnamefont {H{\"a}fner}}, \bibinfo {author} {\bibfnamefont {U.}~\bibnamefont {Sterr}}, \bibinfo {author} {\bibfnamefont {C.}~\bibnamefont {Lisdat}}, \bibinfo {author} {\bibfnamefont {H.}~\bibnamefont {Denker}}, \bibinfo {author} {\bibfnamefont {C.}~\bibnamefont {Voigt}}, \bibinfo {author} {\bibfnamefont {L.}~\bibnamefont {Timmen}}, \bibinfo {author} {\bibfnamefont {A.}~\bibnamefont {Rolland}}, \bibinfo {author} {\bibfnamefont {F.~N.}\ \bibnamefont {Baynes}}, \bibinfo {author} {\bibfnamefont {H.~S.}\ \bibnamefont {Margolis}}, \bibinfo {author} {\bibfnamefont {M.}~\bibnamefont {Zampaolo}}, \bibinfo {author} {\bibfnamefont {P.}~\bibnamefont {Thoumany}}, \bibinfo {author} {\bibfnamefont {M.}~\bibnamefont {Pizzocaro}}, \bibinfo {author} {\bibfnamefont {B.}~\bibnamefont
  {Rauf}}, \bibinfo {author} {\bibfnamefont {F.}~\bibnamefont {Bregolin}}, \bibinfo {author} {\bibfnamefont {A.}~\bibnamefont {Tampellini}}, \bibinfo {author} {\bibfnamefont {P.}~\bibnamefont {Barbieri}}, \bibinfo {author} {\bibfnamefont {M.}~\bibnamefont {Zucco}}, \bibinfo {author} {\bibfnamefont {G.~A.}\ \bibnamefont {Costanzo}}, \bibinfo {author} {\bibfnamefont {C.}~\bibnamefont {Clivati}}, \bibinfo {author} {\bibfnamefont {F.}~\bibnamefont {Levi}},\ and\ \bibinfo {author} {\bibfnamefont {D.}~\bibnamefont {Calonico}},\ }\bibfield  {title} {\bibinfo {title} {Geodesy and metrology with a transportable optical clock},\ }\href {https://doi.org/10.1038/s41567-017-0042-3} {\bibfield  {journal} {\bibinfo  {journal} {Nature Physics}\ }\textbf {\bibinfo {volume} {14}},\ \bibinfo {pages} {437} (\bibinfo {year} {2018})}\BibitemShut {NoStop}%
\bibitem [{\citenamefont {Yuan}\ \emph {et~al.}(2024)\citenamefont {Yuan}, \citenamefont {Cui}, \citenamefont {Liu}, \citenamefont {Yuan}, \citenamefont {Cao}, \citenamefont {Wang}, \citenamefont {Chao}, \citenamefont {Shu},\ and\ \citenamefont {Huang}}]{yuan_demonstration_2024}%
  \BibitemOpen
  \bibfield  {author} {\bibinfo {author} {\bibfnamefont {Y.}~\bibnamefont {Yuan}}, \bibinfo {author} {\bibfnamefont {K.}~\bibnamefont {Cui}}, \bibinfo {author} {\bibfnamefont {D.}~\bibnamefont {Liu}}, \bibinfo {author} {\bibfnamefont {J.}~\bibnamefont {Yuan}}, \bibinfo {author} {\bibfnamefont {J.}~\bibnamefont {Cao}}, \bibinfo {author} {\bibfnamefont {D.}~\bibnamefont {Wang}}, \bibinfo {author} {\bibfnamefont {S.}~\bibnamefont {Chao}}, \bibinfo {author} {\bibfnamefont {H.}~\bibnamefont {Shu}},\ and\ \bibinfo {author} {\bibfnamefont {X.}~\bibnamefont {Huang}},\ }\bibfield  {title} {\bibinfo {title} {Demonstration of chronometric leveling using transportable optical clocks beyond laser coherence limit},\ }\href {https://doi.org/10.1103/PhysRevApplied.21.044052} {\bibfield  {journal} {\bibinfo  {journal} {Physical Review Applied}\ }\textbf {\bibinfo {volume} {21}},\ \bibinfo {pages} {044052} (\bibinfo {year} {2024})}\BibitemShut {NoStop}%
\bibitem [{\citenamefont {Godun}\ \emph {et~al.}(2014)\citenamefont {Godun}, \citenamefont {{Nisbet-Jones}}, \citenamefont {Jones}, \citenamefont {King}, \citenamefont {Johnson}, \citenamefont {Margolis}, \citenamefont {Szymaniec}, \citenamefont {Lea}, \citenamefont {Bongs},\ and\ \citenamefont {Gill}}]{godun_frequency_2014}%
  \BibitemOpen
  \bibfield  {author} {\bibinfo {author} {\bibfnamefont {R.~M.}\ \bibnamefont {Godun}}, \bibinfo {author} {\bibfnamefont {P.~B.~R.}\ \bibnamefont {{Nisbet-Jones}}}, \bibinfo {author} {\bibfnamefont {J.~M.}\ \bibnamefont {Jones}}, \bibinfo {author} {\bibfnamefont {S.~A.}\ \bibnamefont {King}}, \bibinfo {author} {\bibfnamefont {L.~A.~M.}\ \bibnamefont {Johnson}}, \bibinfo {author} {\bibfnamefont {H.~S.}\ \bibnamefont {Margolis}}, \bibinfo {author} {\bibfnamefont {K.}~\bibnamefont {Szymaniec}}, \bibinfo {author} {\bibfnamefont {S.~N.}\ \bibnamefont {Lea}}, \bibinfo {author} {\bibfnamefont {K.}~\bibnamefont {Bongs}},\ and\ \bibinfo {author} {\bibfnamefont {P.}~\bibnamefont {Gill}},\ }\bibfield  {title} {\bibinfo {title} {Frequency {{Ratio}} of {{Two Optical Clock Transitions}} in {\textsuperscript{171}}{{Yb}}{\textsuperscript{+}} and {{Constraints}} on the {{Time Variation}} of {{Fundamental Constants}}},\ }\href {https://doi.org/10.1103/PhysRevLett.113.210801} {\bibfield  {journal} {\bibinfo  {journal} {Physical
  Review Letters}\ }\textbf {\bibinfo {volume} {113}},\ \bibinfo {pages} {210801} (\bibinfo {year} {2014})}\BibitemShut {NoStop}%
\bibitem [{\citenamefont {Safronova}(2019)}]{safronova_search_2019}%
  \BibitemOpen
  \bibfield  {author} {\bibinfo {author} {\bibfnamefont {M.~S.}\ \bibnamefont {Safronova}},\ }\bibfield  {title} {\bibinfo {title} {The {{Search}} for {{Variation}} of {{Fundamental Constants}} with {{Clocks}}},\ }\href {https://doi.org/10.1002/andp.201800364} {\bibfield  {journal} {\bibinfo  {journal} {Annalen der Physik}\ }\textbf {\bibinfo {volume} {531}},\ \bibinfo {pages} {1800364} (\bibinfo {year} {2019})}\BibitemShut {NoStop}%
\bibitem [{\citenamefont {Sherrill}\ \emph {et~al.}(2023)\citenamefont {Sherrill}, \citenamefont {Parsons}, \citenamefont {Baynham}, \citenamefont {Bowden}, \citenamefont {Curtis}, \citenamefont {Hendricks}, \citenamefont {Hill}, \citenamefont {Hobson}, \citenamefont {Margolis}, \citenamefont {Robertson}, \citenamefont {Schioppo}, \citenamefont {Szymaniec}, \citenamefont {Tofful}, \citenamefont {Tunesi}, \citenamefont {Godun},\ and\ \citenamefont {Calmet}}]{sherrill_analysis_2023}%
  \BibitemOpen
  \bibfield  {author} {\bibinfo {author} {\bibfnamefont {N.}~\bibnamefont {Sherrill}}, \bibinfo {author} {\bibfnamefont {A.~O.}\ \bibnamefont {Parsons}}, \bibinfo {author} {\bibfnamefont {C.~F.~A.}\ \bibnamefont {Baynham}}, \bibinfo {author} {\bibfnamefont {W.}~\bibnamefont {Bowden}}, \bibinfo {author} {\bibfnamefont {E.~A.}\ \bibnamefont {Curtis}}, \bibinfo {author} {\bibfnamefont {R.}~\bibnamefont {Hendricks}}, \bibinfo {author} {\bibfnamefont {I.~R.}\ \bibnamefont {Hill}}, \bibinfo {author} {\bibfnamefont {R.}~\bibnamefont {Hobson}}, \bibinfo {author} {\bibfnamefont {H.~S.}\ \bibnamefont {Margolis}}, \bibinfo {author} {\bibfnamefont {B.~I.}\ \bibnamefont {Robertson}}, \bibinfo {author} {\bibfnamefont {M.}~\bibnamefont {Schioppo}}, \bibinfo {author} {\bibfnamefont {K.}~\bibnamefont {Szymaniec}}, \bibinfo {author} {\bibfnamefont {A.}~\bibnamefont {Tofful}}, \bibinfo {author} {\bibfnamefont {J.}~\bibnamefont {Tunesi}}, \bibinfo {author} {\bibfnamefont {R.~M.}\ \bibnamefont {Godun}},\ and\ \bibinfo {author}
  {\bibfnamefont {X.}~\bibnamefont {Calmet}},\ }\bibfield  {title} {\bibinfo {title} {Analysis of atomic-clock data to constrain variations of fundamental constants},\ }\href {https://doi.org/10.1088/1367-2630/aceff6} {\bibfield  {journal} {\bibinfo  {journal} {New Journal of Physics}\ }\textbf {\bibinfo {volume} {25}},\ \bibinfo {pages} {093012} (\bibinfo {year} {2023})}\BibitemShut {NoStop}%
\bibitem [{\citenamefont {Filzinger}\ \emph {et~al.}(2023{\natexlab{a}})\citenamefont {Filzinger}, \citenamefont {Caddell}, \citenamefont {Jani}, \citenamefont {Steinel}, \citenamefont {Giani}, \citenamefont {Huntemann},\ and\ \citenamefont {Roberts}}]{filzinger_ultralight_2023}%
  \BibitemOpen
  \bibfield  {author} {\bibinfo {author} {\bibfnamefont {M.}~\bibnamefont {Filzinger}}, \bibinfo {author} {\bibfnamefont {A.~R.}\ \bibnamefont {Caddell}}, \bibinfo {author} {\bibfnamefont {D.}~\bibnamefont {Jani}}, \bibinfo {author} {\bibfnamefont {M.}~\bibnamefont {Steinel}}, \bibinfo {author} {\bibfnamefont {L.}~\bibnamefont {Giani}}, \bibinfo {author} {\bibfnamefont {N.}~\bibnamefont {Huntemann}},\ and\ \bibinfo {author} {\bibfnamefont {B.~M.}\ \bibnamefont {Roberts}},\ }\href {https://doi.org/10.48550/arXiv.2312.13723} {\bibinfo {title} {Ultralight {{Dark Matter Search}} with {{Space-Time Separated Atomic Clocks}} and {{Cavities}}}} (\bibinfo {year} {2023}{\natexlab{a}}),\ \Eprint {https://arxiv.org/abs/2312.13723} {arXiv:2312.13723 [astro-ph, physics:hep-ph, physics:physics]} \BibitemShut {NoStop}%
\bibitem [{\citenamefont {Roberts}\ \emph {et~al.}(2020)\citenamefont {Roberts}, \citenamefont {Delva}, \citenamefont {{Al-Masoudi}}, \citenamefont {{Amy-Klein}}, \citenamefont {B{\ae}rentsen}, \citenamefont {Baynham}, \citenamefont {Benkler}, \citenamefont {Bilicki}, \citenamefont {Bize}, \citenamefont {Bowden}, \citenamefont {Calvert}, \citenamefont {Cambier}, \citenamefont {Cantin}, \citenamefont {Curtis}, \citenamefont {D{\"o}rscher}, \citenamefont {Favier}, \citenamefont {Frank}, \citenamefont {Gill}, \citenamefont {Godun}, \citenamefont {Grosche}, \citenamefont {Guo}, \citenamefont {Hees}, \citenamefont {Hill}, \citenamefont {Hobson}, \citenamefont {Huntemann}, \citenamefont {Kronj{\"a}ger}, \citenamefont {Koke}, \citenamefont {Kuhl}, \citenamefont {Lange}, \citenamefont {Legero}, \citenamefont {Lipphardt}, \citenamefont {Lisdat}, \citenamefont {Lodewyck}, \citenamefont {Lopez}, \citenamefont {Margolis}, \citenamefont {{\'A}lvarez-Mart{\'i}nez}, \citenamefont {Meynadier}, \citenamefont {Ozimek},
  \citenamefont {Peik}, \citenamefont {Pottie}, \citenamefont {Quintin}, \citenamefont {Sanner}, \citenamefont {Sarlo}, \citenamefont {Schioppo}, \citenamefont {Schwarz}, \citenamefont {Silva}, \citenamefont {Sterr}, \citenamefont {Tamm}, \citenamefont {Targat}, \citenamefont {Tuckey}, \citenamefont {Vallet}, \citenamefont {Waterholter}, \citenamefont {Xu},\ and\ \citenamefont {Wolf}}]{roberts_search_2020}%
  \BibitemOpen
  \bibfield  {author} {\bibinfo {author} {\bibfnamefont {B.~M.}\ \bibnamefont {Roberts}}, \bibinfo {author} {\bibfnamefont {P.}~\bibnamefont {Delva}}, \bibinfo {author} {\bibfnamefont {A.}~\bibnamefont {{Al-Masoudi}}}, \bibinfo {author} {\bibfnamefont {A.}~\bibnamefont {{Amy-Klein}}}, \bibinfo {author} {\bibfnamefont {C.}~\bibnamefont {B{\ae}rentsen}}, \bibinfo {author} {\bibfnamefont {C.~F.~A.}\ \bibnamefont {Baynham}}, \bibinfo {author} {\bibfnamefont {E.}~\bibnamefont {Benkler}}, \bibinfo {author} {\bibfnamefont {S.}~\bibnamefont {Bilicki}}, \bibinfo {author} {\bibfnamefont {S.}~\bibnamefont {Bize}}, \bibinfo {author} {\bibfnamefont {W.}~\bibnamefont {Bowden}}, \bibinfo {author} {\bibfnamefont {J.}~\bibnamefont {Calvert}}, \bibinfo {author} {\bibfnamefont {V.}~\bibnamefont {Cambier}}, \bibinfo {author} {\bibfnamefont {E.}~\bibnamefont {Cantin}}, \bibinfo {author} {\bibfnamefont {E.~A.}\ \bibnamefont {Curtis}}, \bibinfo {author} {\bibfnamefont {S.}~\bibnamefont {D{\"o}rscher}}, \bibinfo {author}
  {\bibfnamefont {M.}~\bibnamefont {Favier}}, \bibinfo {author} {\bibfnamefont {F.}~\bibnamefont {Frank}}, \bibinfo {author} {\bibfnamefont {P.}~\bibnamefont {Gill}}, \bibinfo {author} {\bibfnamefont {R.~M.}\ \bibnamefont {Godun}}, \bibinfo {author} {\bibfnamefont {G.}~\bibnamefont {Grosche}}, \bibinfo {author} {\bibfnamefont {C.}~\bibnamefont {Guo}}, \bibinfo {author} {\bibfnamefont {A.}~\bibnamefont {Hees}}, \bibinfo {author} {\bibfnamefont {I.~R.}\ \bibnamefont {Hill}}, \bibinfo {author} {\bibfnamefont {R.}~\bibnamefont {Hobson}}, \bibinfo {author} {\bibfnamefont {N.}~\bibnamefont {Huntemann}}, \bibinfo {author} {\bibfnamefont {J.}~\bibnamefont {Kronj{\"a}ger}}, \bibinfo {author} {\bibfnamefont {S.}~\bibnamefont {Koke}}, \bibinfo {author} {\bibfnamefont {A.}~\bibnamefont {Kuhl}}, \bibinfo {author} {\bibfnamefont {R.}~\bibnamefont {Lange}}, \bibinfo {author} {\bibfnamefont {T.}~\bibnamefont {Legero}}, \bibinfo {author} {\bibfnamefont {B.}~\bibnamefont {Lipphardt}}, \bibinfo {author} {\bibfnamefont
  {C.}~\bibnamefont {Lisdat}}, \bibinfo {author} {\bibfnamefont {J.}~\bibnamefont {Lodewyck}}, \bibinfo {author} {\bibfnamefont {O.}~\bibnamefont {Lopez}}, \bibinfo {author} {\bibfnamefont {H.~S.}\ \bibnamefont {Margolis}}, \bibinfo {author} {\bibfnamefont {H.}~\bibnamefont {{\'A}lvarez-Mart{\'i}nez}}, \bibinfo {author} {\bibfnamefont {F.}~\bibnamefont {Meynadier}}, \bibinfo {author} {\bibfnamefont {F.}~\bibnamefont {Ozimek}}, \bibinfo {author} {\bibfnamefont {E.}~\bibnamefont {Peik}}, \bibinfo {author} {\bibfnamefont {P.-E.}\ \bibnamefont {Pottie}}, \bibinfo {author} {\bibfnamefont {N.}~\bibnamefont {Quintin}}, \bibinfo {author} {\bibfnamefont {C.}~\bibnamefont {Sanner}}, \bibinfo {author} {\bibfnamefont {L.~D.}\ \bibnamefont {Sarlo}}, \bibinfo {author} {\bibfnamefont {M.}~\bibnamefont {Schioppo}}, \bibinfo {author} {\bibfnamefont {R.}~\bibnamefont {Schwarz}}, \bibinfo {author} {\bibfnamefont {A.}~\bibnamefont {Silva}}, \bibinfo {author} {\bibfnamefont {U.}~\bibnamefont {Sterr}}, \bibinfo {author}
  {\bibfnamefont {C.}~\bibnamefont {Tamm}}, \bibinfo {author} {\bibfnamefont {R.~L.}\ \bibnamefont {Targat}}, \bibinfo {author} {\bibfnamefont {P.}~\bibnamefont {Tuckey}}, \bibinfo {author} {\bibfnamefont {G.}~\bibnamefont {Vallet}}, \bibinfo {author} {\bibfnamefont {T.}~\bibnamefont {Waterholter}}, \bibinfo {author} {\bibfnamefont {D.}~\bibnamefont {Xu}},\ and\ \bibinfo {author} {\bibfnamefont {P.}~\bibnamefont {Wolf}},\ }\bibfield  {title} {\bibinfo {title} {Search for transient variations of the fine structure constant and dark matter using fiber-linked optical atomic clocks},\ }\href {https://doi.org/10.1088/1367-2630/abaace} {\bibfield  {journal} {\bibinfo  {journal} {New Journal of Physics}\ }\textbf {\bibinfo {volume} {22}},\ \bibinfo {pages} {093010} (\bibinfo {year} {2020})}\BibitemShut {NoStop}%
\bibitem [{\citenamefont {Filzinger}\ \emph {et~al.}(2023{\natexlab{b}})\citenamefont {Filzinger}, \citenamefont {D{\"o}rscher}, \citenamefont {Lange}, \citenamefont {Klose}, \citenamefont {Steinel}, \citenamefont {Benkler}, \citenamefont {Peik}, \citenamefont {Lisdat},\ and\ \citenamefont {Huntemann}}]{filzinger_improved_2023}%
  \BibitemOpen
  \bibfield  {author} {\bibinfo {author} {\bibfnamefont {M.}~\bibnamefont {Filzinger}}, \bibinfo {author} {\bibfnamefont {S.}~\bibnamefont {D{\"o}rscher}}, \bibinfo {author} {\bibfnamefont {R.}~\bibnamefont {Lange}}, \bibinfo {author} {\bibfnamefont {J.}~\bibnamefont {Klose}}, \bibinfo {author} {\bibfnamefont {M.}~\bibnamefont {Steinel}}, \bibinfo {author} {\bibfnamefont {E.}~\bibnamefont {Benkler}}, \bibinfo {author} {\bibfnamefont {E.}~\bibnamefont {Peik}}, \bibinfo {author} {\bibfnamefont {C.}~\bibnamefont {Lisdat}},\ and\ \bibinfo {author} {\bibfnamefont {N.}~\bibnamefont {Huntemann}},\ }\bibfield  {title} {\bibinfo {title} {Improved {{Limits}} on the {{Coupling}} of {{Ultralight Bosonic Dark Matter}} to {{Photons}} from {{Optical Atomic Clock Comparisons}}},\ }\href {https://doi.org/10.1103/PhysRevLett.130.253001} {\bibfield  {journal} {\bibinfo  {journal} {Physical Review Letters}\ }\textbf {\bibinfo {volume} {130}},\ \bibinfo {pages} {253001} (\bibinfo {year} {2023}{\natexlab{b}})}\BibitemShut
  {NoStop}%
\bibitem [{\citenamefont {Yeh}\ \emph {et~al.}(2023)\citenamefont {Yeh}, \citenamefont {Grensemann}, \citenamefont {Dreissen}, \citenamefont {F{\"u}rst},\ and\ \citenamefont {Mehlst{\"a}ubler}}]{yeh_robust_2023}%
  \BibitemOpen
  \bibfield  {author} {\bibinfo {author} {\bibfnamefont {C.-H.}\ \bibnamefont {Yeh}}, \bibinfo {author} {\bibfnamefont {K.~C.}\ \bibnamefont {Grensemann}}, \bibinfo {author} {\bibfnamefont {L.~S.}\ \bibnamefont {Dreissen}}, \bibinfo {author} {\bibfnamefont {H.~A.}\ \bibnamefont {F{\"u}rst}},\ and\ \bibinfo {author} {\bibfnamefont {T.~E.}\ \bibnamefont {Mehlst{\"a}ubler}},\ }\bibfield  {title} {\bibinfo {title} {Robust and scalable rf spectroscopy in first-order magnetic sensitive states at second-long coherence time},\ }\href {https://doi.org/10.1088/1367-2630/acfc14} {\bibfield  {journal} {\bibinfo  {journal} {New Journal of Physics}\ }\textbf {\bibinfo {volume} {25}},\ \bibinfo {pages} {093054} (\bibinfo {year} {2023})}\BibitemShut {NoStop}%
\bibitem [{\citenamefont {Ludlow}\ \emph {et~al.}(2015)\citenamefont {Ludlow}, \citenamefont {Boyd}, \citenamefont {Ye}, \citenamefont {Peik},\ and\ \citenamefont {Schmidt}}]{ludlow_optical_2015}%
  \BibitemOpen
  \bibfield  {author} {\bibinfo {author} {\bibfnamefont {A.~D.}\ \bibnamefont {Ludlow}}, \bibinfo {author} {\bibfnamefont {M.~M.}\ \bibnamefont {Boyd}}, \bibinfo {author} {\bibfnamefont {J.}~\bibnamefont {Ye}}, \bibinfo {author} {\bibfnamefont {E.}~\bibnamefont {Peik}},\ and\ \bibinfo {author} {\bibfnamefont {P.~O.}\ \bibnamefont {Schmidt}},\ }\bibfield  {title} {\bibinfo {title} {Optical atomic clocks},\ }\href {https://doi.org/10.1103/RevModPhys.87.637} {\bibfield  {journal} {\bibinfo  {journal} {Reviews of Modern Physics}\ }\textbf {\bibinfo {volume} {87}},\ \bibinfo {pages} {637} (\bibinfo {year} {2015})}\BibitemShut {NoStop}%
\bibitem [{\citenamefont {Itano}\ \emph {et~al.}(1993)\citenamefont {Itano}, \citenamefont {Bergquist}, \citenamefont {Bollinger}, \citenamefont {Gilligan}, \citenamefont {Heinzen}, \citenamefont {Moore}, \citenamefont {Raizen},\ and\ \citenamefont {Wineland}}]{itano_quantum_1993}%
  \BibitemOpen
  \bibfield  {author} {\bibinfo {author} {\bibfnamefont {W.~M.}\ \bibnamefont {Itano}}, \bibinfo {author} {\bibfnamefont {J.~C.}\ \bibnamefont {Bergquist}}, \bibinfo {author} {\bibfnamefont {J.~J.}\ \bibnamefont {Bollinger}}, \bibinfo {author} {\bibfnamefont {J.~M.}\ \bibnamefont {Gilligan}}, \bibinfo {author} {\bibfnamefont {D.~J.}\ \bibnamefont {Heinzen}}, \bibinfo {author} {\bibfnamefont {F.~L.}\ \bibnamefont {Moore}}, \bibinfo {author} {\bibfnamefont {M.~G.}\ \bibnamefont {Raizen}},\ and\ \bibinfo {author} {\bibfnamefont {D.~J.}\ \bibnamefont {Wineland}},\ }\bibfield  {title} {\bibinfo {title} {Quantum projection noise: {{Population}} fluctuations in two-level systems},\ }\href {https://doi.org/10.1103/PhysRevA.47.3554} {\bibfield  {journal} {\bibinfo  {journal} {Physical Review A}\ }\textbf {\bibinfo {volume} {47}},\ \bibinfo {pages} {3554} (\bibinfo {year} {1993})}\BibitemShut {NoStop}%
\bibitem [{\citenamefont {Peik}\ \emph {et~al.}(2006)\citenamefont {Peik}, \citenamefont {Schneider},\ and\ \citenamefont {Tamm}}]{peik_laser_2006}%
  \BibitemOpen
  \bibfield  {author} {\bibinfo {author} {\bibfnamefont {E.}~\bibnamefont {Peik}}, \bibinfo {author} {\bibfnamefont {T.}~\bibnamefont {Schneider}},\ and\ \bibinfo {author} {\bibfnamefont {C.}~\bibnamefont {Tamm}},\ }\bibfield  {title} {\bibinfo {title} {Laser frequency stabilization to a single ion},\ }\href {https://doi.org/10.1088/0953-4075/39/1/012} {\bibfield  {journal} {\bibinfo  {journal} {Journal of Physics B: Atomic, Molecular and Optical Physics}\ }\textbf {\bibinfo {volume} {39}},\ \bibinfo {pages} {145} (\bibinfo {year} {2006})}\BibitemShut {NoStop}%
\bibitem [{\citenamefont {Riehle}(2004)}]{riehle_frequency_2004}%
  \BibitemOpen
  \bibfield  {author} {\bibinfo {author} {\bibfnamefont {F.}~\bibnamefont {Riehle}},\ }\href@noop {} {\emph {\bibinfo {title} {Frequency Standards: Basics and Applications}}}\ (\bibinfo  {publisher} {Wiley-VCH},\ \bibinfo {address} {Weinheim},\ \bibinfo {year} {2004})\BibitemShut {NoStop}%
\bibitem [{\citenamefont {Pyka}\ \emph {et~al.}(2014)\citenamefont {Pyka}, \citenamefont {Herschbach}, \citenamefont {Keller},\ and\ \citenamefont {Mehlst{\"a}ubler}}]{pyka_highprecision_2014}%
  \BibitemOpen
  \bibfield  {author} {\bibinfo {author} {\bibfnamefont {K.}~\bibnamefont {Pyka}}, \bibinfo {author} {\bibfnamefont {N.}~\bibnamefont {Herschbach}}, \bibinfo {author} {\bibfnamefont {J.}~\bibnamefont {Keller}},\ and\ \bibinfo {author} {\bibfnamefont {T.~E.}\ \bibnamefont {Mehlst{\"a}ubler}},\ }\bibfield  {title} {\bibinfo {title} {A high-precision segmented {{Paul}} trap with minimized micromotion for an optical multiple-ion clock},\ }\href {https://doi.org/10.1007/s00340-013-5580-5} {\bibfield  {journal} {\bibinfo  {journal} {Applied Physics B}\ }\textbf {\bibinfo {volume} {114}},\ \bibinfo {pages} {231} (\bibinfo {year} {2014})}\BibitemShut {NoStop}%
\bibitem [{\citenamefont {Keller}\ \emph {et~al.}(2016)\citenamefont {Keller}, \citenamefont {Burgermeister}, \citenamefont {Kalincev}, \citenamefont {Kiethe},\ and\ \citenamefont {Mehlst{\"a}ubler}}]{keller_evaluation_2016}%
  \BibitemOpen
  \bibfield  {author} {\bibinfo {author} {\bibfnamefont {J.}~\bibnamefont {Keller}}, \bibinfo {author} {\bibfnamefont {T.}~\bibnamefont {Burgermeister}}, \bibinfo {author} {\bibfnamefont {D.}~\bibnamefont {Kalincev}}, \bibinfo {author} {\bibfnamefont {J.}~\bibnamefont {Kiethe}},\ and\ \bibinfo {author} {\bibfnamefont {T.~E.}\ \bibnamefont {Mehlst{\"a}ubler}},\ }\bibfield  {title} {\bibinfo {title} {Evaluation of trap-induced systematic frequency shifts for a multi-ion optical clock at the 10{\textsuperscript{-19}} level},\ }\href@noop {} {\bibfield  {journal} {\bibinfo  {journal} {Journal of Physics Conference Series}\ }\textbf {\bibinfo {volume} {723}},\ \bibinfo {pages} {012027} (\bibinfo {year} {2016})},\ \Eprint {https://arxiv.org/abs/1512.05677} {arXiv:1512.05677} \BibitemShut {NoStop}%
\bibitem [{\citenamefont {Tan}\ \emph {et~al.}(2019)\citenamefont {Tan}, \citenamefont {Kaewuam}, \citenamefont {Arnold}, \citenamefont {Chanu}, \citenamefont {Zhang}, \citenamefont {Safronova},\ and\ \citenamefont {Barrett}}]{tan_suppressing_2019}%
  \BibitemOpen
  \bibfield  {author} {\bibinfo {author} {\bibfnamefont {T.~R.}\ \bibnamefont {Tan}}, \bibinfo {author} {\bibfnamefont {R.}~\bibnamefont {Kaewuam}}, \bibinfo {author} {\bibfnamefont {K.~J.}\ \bibnamefont {Arnold}}, \bibinfo {author} {\bibfnamefont {S.~R.}\ \bibnamefont {Chanu}}, \bibinfo {author} {\bibfnamefont {Z.}~\bibnamefont {Zhang}}, \bibinfo {author} {\bibfnamefont {M.~S.}\ \bibnamefont {Safronova}},\ and\ \bibinfo {author} {\bibfnamefont {M.~D.}\ \bibnamefont {Barrett}},\ }\bibfield  {title} {\bibinfo {title} {Suppressing {{Inhomogeneous Broadening}} in a {{Lutetium Multi-ion Optical Clock}}},\ }\href {https://doi.org/10.1103/PhysRevLett.123.063201} {\bibfield  {journal} {\bibinfo  {journal} {Physical Review Letters}\ }\textbf {\bibinfo {volume} {123}},\ \bibinfo {pages} {063201} (\bibinfo {year} {2019})}\BibitemShut {NoStop}%
\bibitem [{\citenamefont {Pelzer}\ \emph {et~al.}(2024)\citenamefont {Pelzer}, \citenamefont {Dietze}, \citenamefont {{Mart{\'i}nez-Lahuerta}}, \citenamefont {Krinner}, \citenamefont {Kramer}, \citenamefont {Dawel}, \citenamefont {Spethmann}, \citenamefont {Hammerer},\ and\ \citenamefont {Schmidt}}]{pelzer_multiion_2024}%
  \BibitemOpen
  \bibfield  {author} {\bibinfo {author} {\bibfnamefont {L.}~\bibnamefont {Pelzer}}, \bibinfo {author} {\bibfnamefont {K.}~\bibnamefont {Dietze}}, \bibinfo {author} {\bibfnamefont {V.~J.}\ \bibnamefont {{Mart{\'i}nez-Lahuerta}}}, \bibinfo {author} {\bibfnamefont {L.}~\bibnamefont {Krinner}}, \bibinfo {author} {\bibfnamefont {J.}~\bibnamefont {Kramer}}, \bibinfo {author} {\bibfnamefont {F.}~\bibnamefont {Dawel}}, \bibinfo {author} {\bibfnamefont {N.~C.~H.}\ \bibnamefont {Spethmann}}, \bibinfo {author} {\bibfnamefont {K.}~\bibnamefont {Hammerer}},\ and\ \bibinfo {author} {\bibfnamefont {P.~O.}\ \bibnamefont {Schmidt}},\ }\bibfield  {title} {\bibinfo {title} {Multi-ion {{Frequency Reference Using Dynamical Decoupling}}},\ }\href {https://doi.org/10.1103/PhysRevLett.133.033203} {\bibfield  {journal} {\bibinfo  {journal} {Physical Review Letters}\ }\textbf {\bibinfo {volume} {133}},\ \bibinfo {pages} {033203} (\bibinfo {year} {2024})}\BibitemShut {NoStop}%
\bibitem [{\citenamefont {Akerman}\ and\ \citenamefont {Ozeri}(2025)}]{akerman_operating_2025}%
  \BibitemOpen
  \bibfield  {author} {\bibinfo {author} {\bibfnamefont {N.}~\bibnamefont {Akerman}}\ and\ \bibinfo {author} {\bibfnamefont {R.}~\bibnamefont {Ozeri}},\ }\bibfield  {title} {\bibinfo {title} {Operating a {{Multi-Ion Clock}} with {{Dynamical Decoupling}}},\ }\href {https://doi.org/10.1103/PhysRevLett.134.013201} {\bibfield  {journal} {\bibinfo  {journal} {Physical Review Letters}\ }\textbf {\bibinfo {volume} {134}},\ \bibinfo {pages} {013201} (\bibinfo {year} {2025})}\BibitemShut {NoStop}%
\bibitem [{\citenamefont {Leroux}\ \emph {et~al.}(2017)\citenamefont {Leroux}, \citenamefont {Scharnhorst}, \citenamefont {Hannig}, \citenamefont {Kramer}, \citenamefont {Pelzer}, \citenamefont {Stepanova},\ and\ \citenamefont {Schmidt}}]{leroux_online_2017}%
  \BibitemOpen
  \bibfield  {author} {\bibinfo {author} {\bibfnamefont {I.~D.}\ \bibnamefont {Leroux}}, \bibinfo {author} {\bibfnamefont {N.}~\bibnamefont {Scharnhorst}}, \bibinfo {author} {\bibfnamefont {S.}~\bibnamefont {Hannig}}, \bibinfo {author} {\bibfnamefont {J.}~\bibnamefont {Kramer}}, \bibinfo {author} {\bibfnamefont {L.}~\bibnamefont {Pelzer}}, \bibinfo {author} {\bibfnamefont {M.}~\bibnamefont {Stepanova}},\ and\ \bibinfo {author} {\bibfnamefont {P.~O.}\ \bibnamefont {Schmidt}},\ }\bibfield  {title} {\bibinfo {title} {On-line estimation of local oscillator noise and optimisation of servo parameters in atomic clocks},\ }\href {https://doi.org/10.1088/1681-7575/aa66e9} {\bibfield  {journal} {\bibinfo  {journal} {Metrologia}\ }\textbf {\bibinfo {volume} {54}},\ \bibinfo {pages} {307} (\bibinfo {year} {2017})}\BibitemShut {NoStop}%
\bibitem [{\citenamefont {Matei}\ \emph {et~al.}(2017)\citenamefont {Matei}, \citenamefont {Legero}, \citenamefont {H{\"a}fner}, \citenamefont {Grebing}, \citenamefont {Weyrich}, \citenamefont {Zhang}, \citenamefont {Sonderhouse}, \citenamefont {Robinson}, \citenamefont {Ye}, \citenamefont {Riehle},\ and\ \citenamefont {Sterr}}]{matei_15_2017}%
  \BibitemOpen
  \bibfield  {author} {\bibinfo {author} {\bibfnamefont {D.~G.}\ \bibnamefont {Matei}}, \bibinfo {author} {\bibfnamefont {T.}~\bibnamefont {Legero}}, \bibinfo {author} {\bibfnamefont {S.}~\bibnamefont {H{\"a}fner}}, \bibinfo {author} {\bibfnamefont {C.}~\bibnamefont {Grebing}}, \bibinfo {author} {\bibfnamefont {R.}~\bibnamefont {Weyrich}}, \bibinfo {author} {\bibfnamefont {W.}~\bibnamefont {Zhang}}, \bibinfo {author} {\bibfnamefont {L.}~\bibnamefont {Sonderhouse}}, \bibinfo {author} {\bibfnamefont {J.~M.}\ \bibnamefont {Robinson}}, \bibinfo {author} {\bibfnamefont {J.}~\bibnamefont {Ye}}, \bibinfo {author} {\bibfnamefont {F.}~\bibnamefont {Riehle}},\ and\ \bibinfo {author} {\bibfnamefont {U.}~\bibnamefont {Sterr}},\ }\bibfield  {title} {\bibinfo {title} {1.5 {$\mu\mkern1mu$}m {{Lasers}} with {{Sub-10 mHz Linewidth}}},\ }\href {https://doi.org/10.1103/PhysRevLett.118.263202} {\bibfield  {journal} {\bibinfo  {journal} {Physical Review Letters}\ }\textbf {\bibinfo {volume} {118}},\ \bibinfo {pages} {263202}
  (\bibinfo {year} {2017})}\BibitemShut {NoStop}%
\bibitem [{\citenamefont {D{\"o}rscher}\ \emph {et~al.}(2020)\citenamefont {D{\"o}rscher}, \citenamefont {{Al-Masoudi}}, \citenamefont {Bober}, \citenamefont {Schwarz}, \citenamefont {Hobson}, \citenamefont {Sterr},\ and\ \citenamefont {Lisdat}}]{dörscher_dynamical_2020}%
  \BibitemOpen
  \bibfield  {author} {\bibinfo {author} {\bibfnamefont {S.}~\bibnamefont {D{\"o}rscher}}, \bibinfo {author} {\bibfnamefont {A.}~\bibnamefont {{Al-Masoudi}}}, \bibinfo {author} {\bibfnamefont {M.}~\bibnamefont {Bober}}, \bibinfo {author} {\bibfnamefont {R.}~\bibnamefont {Schwarz}}, \bibinfo {author} {\bibfnamefont {R.}~\bibnamefont {Hobson}}, \bibinfo {author} {\bibfnamefont {U.}~\bibnamefont {Sterr}},\ and\ \bibinfo {author} {\bibfnamefont {C.}~\bibnamefont {Lisdat}},\ }\bibfield  {title} {\bibinfo {title} {Dynamical decoupling of laser phase noise in compound atomic clocks},\ }\href {https://doi.org/10.1038/s42005-020-00452-9} {\bibfield  {journal} {\bibinfo  {journal} {Communications Physics}\ }\textbf {\bibinfo {volume} {3}},\ \bibinfo {pages} {1} (\bibinfo {year} {2020})}\BibitemShut {NoStop}%
\bibitem [{\citenamefont {Rosenband}\ and\ \citenamefont {Leibrandt}(2013)}]{rosenband_exponential_2013}%
  \BibitemOpen
  \bibfield  {author} {\bibinfo {author} {\bibfnamefont {T.}~\bibnamefont {Rosenband}}\ and\ \bibinfo {author} {\bibfnamefont {D.~R.}\ \bibnamefont {Leibrandt}},\ }\bibfield  {title} {\bibinfo {title} {Exponential scaling of clock stability with atom number},\ }\href@noop {} {\bibfield  {journal} {\bibinfo  {journal} {arXiv:1303.6357}\ } (\bibinfo {year} {2013})},\ \Eprint {https://arxiv.org/abs/1303.6357} {arXiv:1303.6357} \BibitemShut {NoStop}%
\bibitem [{\citenamefont {Borregaard}\ and\ \citenamefont {S{\o}rensen}(2013)}]{borregaard_efficient_2013}%
  \BibitemOpen
  \bibfield  {author} {\bibinfo {author} {\bibfnamefont {J.}~\bibnamefont {Borregaard}}\ and\ \bibinfo {author} {\bibfnamefont {A.~S.}\ \bibnamefont {S{\o}rensen}},\ }\bibfield  {title} {\bibinfo {title} {Efficient {{Atomic Clocks Operated}} with {{Several Atomic Ensembles}}},\ }\href {https://doi.org/10.1103/PhysRevLett.111.090802} {\bibfield  {journal} {\bibinfo  {journal} {Physical Review Letters}\ }\textbf {\bibinfo {volume} {111}},\ \bibinfo {pages} {090802} (\bibinfo {year} {2013})}\BibitemShut {NoStop}%
\bibitem [{\citenamefont {{Mart{\'i}nez-Lahuerta}}\ \emph {et~al.}(2022)\citenamefont {{Mart{\'i}nez-Lahuerta}}, \citenamefont {Eilers}, \citenamefont {Mehlst{\"a}ubler}, \citenamefont {Schmidt},\ and\ \citenamefont {Hammerer}}]{martínez-lahuerta_initio_2022}%
  \BibitemOpen
  \bibfield  {author} {\bibinfo {author} {\bibfnamefont {V.~J.}\ \bibnamefont {{Mart{\'i}nez-Lahuerta}}}, \bibinfo {author} {\bibfnamefont {S.}~\bibnamefont {Eilers}}, \bibinfo {author} {\bibfnamefont {T.~E.}\ \bibnamefont {Mehlst{\"a}ubler}}, \bibinfo {author} {\bibfnamefont {P.~O.}\ \bibnamefont {Schmidt}},\ and\ \bibinfo {author} {\bibfnamefont {K.}~\bibnamefont {Hammerer}},\ }\bibfield  {title} {\bibinfo {title} {Ab initio quantum theory of mass defect and time dilation in trapped-ion optical clocks},\ }\href {https://doi.org/10.1103/PhysRevA.106.032803} {\bibfield  {journal} {\bibinfo  {journal} {Physical Review A}\ }\textbf {\bibinfo {volume} {106}},\ \bibinfo {pages} {032803} (\bibinfo {year} {2022})}\BibitemShut {NoStop}%
\bibitem [{\citenamefont {Brownnutt}\ \emph {et~al.}(2015)\citenamefont {Brownnutt}, \citenamefont {Kumph}, \citenamefont {Rabl},\ and\ \citenamefont {Blatt}}]{brownnutt_iontrap_2015}%
  \BibitemOpen
  \bibfield  {author} {\bibinfo {author} {\bibfnamefont {M.}~\bibnamefont {Brownnutt}}, \bibinfo {author} {\bibfnamefont {M.}~\bibnamefont {Kumph}}, \bibinfo {author} {\bibfnamefont {P.}~\bibnamefont {Rabl}},\ and\ \bibinfo {author} {\bibfnamefont {R.}~\bibnamefont {Blatt}},\ }\bibfield  {title} {\bibinfo {title} {Ion-trap measurements of electric-field noise near surfaces},\ }\href {https://doi.org/10.1103/RevModPhys.87.1419} {\bibfield  {journal} {\bibinfo  {journal} {Reviews of Modern Physics}\ }\textbf {\bibinfo {volume} {87}},\ \bibinfo {pages} {1419} (\bibinfo {year} {2015})}\BibitemShut {NoStop}%
\bibitem [{\citenamefont {Zeng}\ \emph {et~al.}(2023)\citenamefont {Zeng}, \citenamefont {Huang}, \citenamefont {Zhang}, \citenamefont {Hao}, \citenamefont {Ma}, \citenamefont {Hu}, \citenamefont {Zhang}, \citenamefont {Chen}, \citenamefont {Wang}, \citenamefont {Guan},\ and\ \citenamefont {Gao}}]{zeng_transportable_2023}%
  \BibitemOpen
  \bibfield  {author} {\bibinfo {author} {\bibfnamefont {M.}~\bibnamefont {Zeng}}, \bibinfo {author} {\bibfnamefont {Y.}~\bibnamefont {Huang}}, \bibinfo {author} {\bibfnamefont {B.}~\bibnamefont {Zhang}}, \bibinfo {author} {\bibfnamefont {Y.}~\bibnamefont {Hao}}, \bibinfo {author} {\bibfnamefont {Z.}~\bibnamefont {Ma}}, \bibinfo {author} {\bibfnamefont {R.}~\bibnamefont {Hu}}, \bibinfo {author} {\bibfnamefont {H.}~\bibnamefont {Zhang}}, \bibinfo {author} {\bibfnamefont {Z.}~\bibnamefont {Chen}}, \bibinfo {author} {\bibfnamefont {M.}~\bibnamefont {Wang}}, \bibinfo {author} {\bibfnamefont {H.}~\bibnamefont {Guan}},\ and\ \bibinfo {author} {\bibfnamefont {K.}~\bibnamefont {Gao}},\ }\bibfield  {title} {\bibinfo {title} {Toward a {{Transportable Ca}}{\textsuperscript{+}} {{Optical Clock}} with a {{Systematic Uncertainty}} of 4.8x10{\textsuperscript{-18}}},\ }\href {https://doi.org/10.1103/PhysRevApplied.19.064004} {\bibfield  {journal} {\bibinfo  {journal} {Physical Review Applied}\ }\textbf {\bibinfo {volume}
  {19}},\ \bibinfo {pages} {064004} (\bibinfo {year} {2023})}\BibitemShut {NoStop}%
\bibitem [{\citenamefont {Huntemann}\ \emph {et~al.}(2016)\citenamefont {Huntemann}, \citenamefont {Sanner}, \citenamefont {Lipphardt}, \citenamefont {Tamm},\ and\ \citenamefont {Peik}}]{huntemann_singleion_2016}%
  \BibitemOpen
  \bibfield  {author} {\bibinfo {author} {\bibfnamefont {N.}~\bibnamefont {Huntemann}}, \bibinfo {author} {\bibfnamefont {C.}~\bibnamefont {Sanner}}, \bibinfo {author} {\bibfnamefont {B.}~\bibnamefont {Lipphardt}}, \bibinfo {author} {\bibfnamefont {{\relax Chr}.}~\bibnamefont {Tamm}},\ and\ \bibinfo {author} {\bibfnamefont {E.}~\bibnamefont {Peik}},\ }\bibfield  {title} {\bibinfo {title} {Single-{{Ion Atomic Clock}} with 3{\texttimes}10{\textsuperscript{-18}} {{Systematic Uncertainty}}},\ }\href {https://doi.org/10.1103/PhysRevLett.116.063001} {\bibfield  {journal} {\bibinfo  {journal} {Physical Review Letters}\ }\textbf {\bibinfo {volume} {116}},\ \bibinfo {pages} {063001} (\bibinfo {year} {2016})}\BibitemShut {NoStop}%
\bibitem [{\citenamefont {Ma}\ \emph {et~al.}(2024)\citenamefont {Ma}, \citenamefont {Deng}, \citenamefont {Wang}, \citenamefont {Wei}, \citenamefont {Hao}, \citenamefont {Zhang}, \citenamefont {Pang}, \citenamefont {Wang}, \citenamefont {Wu}, \citenamefont {Liu}, \citenamefont {Yuan}, \citenamefont {Chang}, \citenamefont {Zhang}, \citenamefont {Wu}, \citenamefont {Zhang},\ and\ \citenamefont {Lu}}]{ma_quantumlogicbased_2024}%
  \BibitemOpen
  \bibfield  {author} {\bibinfo {author} {\bibfnamefont {Z.~Y.}\ \bibnamefont {Ma}}, \bibinfo {author} {\bibfnamefont {K.}~\bibnamefont {Deng}}, \bibinfo {author} {\bibfnamefont {Z.~Y.}\ \bibnamefont {Wang}}, \bibinfo {author} {\bibfnamefont {W.~Z.}\ \bibnamefont {Wei}}, \bibinfo {author} {\bibfnamefont {P.}~\bibnamefont {Hao}}, \bibinfo {author} {\bibfnamefont {H.~X.}\ \bibnamefont {Zhang}}, \bibinfo {author} {\bibfnamefont {L.~R.}\ \bibnamefont {Pang}}, \bibinfo {author} {\bibfnamefont {B.}~\bibnamefont {Wang}}, \bibinfo {author} {\bibfnamefont {F.~F.}\ \bibnamefont {Wu}}, \bibinfo {author} {\bibfnamefont {H.~L.}\ \bibnamefont {Liu}}, \bibinfo {author} {\bibfnamefont {W.~H.}\ \bibnamefont {Yuan}}, \bibinfo {author} {\bibfnamefont {J.~L.}\ \bibnamefont {Chang}}, \bibinfo {author} {\bibfnamefont {J.~X.}\ \bibnamefont {Zhang}}, \bibinfo {author} {\bibfnamefont {Q.~Y.}\ \bibnamefont {Wu}}, \bibinfo {author} {\bibfnamefont {J.}~\bibnamefont {Zhang}},\ and\ \bibinfo {author} {\bibfnamefont {Z.~H.}\ \bibnamefont
  {Lu}},\ }\bibfield  {title} {\bibinfo {title} {Quantum-logic-based {\textsuperscript{25}}{{Mg}}{\textsuperscript{+}}-{\textsuperscript{27}}{{Al}}{\textsuperscript{+}} optical frequency standard for the redefinition of the {{SI}} second},\ }\href {https://doi.org/10.1103/PhysRevApplied.21.044017} {\bibfield  {journal} {\bibinfo  {journal} {Physical Review Applied}\ }\textbf {\bibinfo {volume} {21}},\ \bibinfo {pages} {044017} (\bibinfo {year} {2024})}\BibitemShut {NoStop}%
\bibitem [{\citenamefont {Rosenband}\ \emph {et~al.}(2008)\citenamefont {Rosenband}, \citenamefont {Hume}, \citenamefont {Schmidt}, \citenamefont {Chou}, \citenamefont {Brusch}, \citenamefont {Lorini}, \citenamefont {Oskay}, \citenamefont {Drullinger}, \citenamefont {Fortier}, \citenamefont {Stalnaker}, \citenamefont {Diddams}, \citenamefont {Swann}, \citenamefont {Newbury}, \citenamefont {Itano}, \citenamefont {Wineland},\ and\ \citenamefont {Bergquist}}]{rosenband_frequency_2008}%
  \BibitemOpen
  \bibfield  {author} {\bibinfo {author} {\bibfnamefont {T.}~\bibnamefont {Rosenband}}, \bibinfo {author} {\bibfnamefont {D.~B.}\ \bibnamefont {Hume}}, \bibinfo {author} {\bibfnamefont {P.~O.}\ \bibnamefont {Schmidt}}, \bibinfo {author} {\bibfnamefont {C.~W.}\ \bibnamefont {Chou}}, \bibinfo {author} {\bibfnamefont {A.}~\bibnamefont {Brusch}}, \bibinfo {author} {\bibfnamefont {L.}~\bibnamefont {Lorini}}, \bibinfo {author} {\bibfnamefont {W.~H.}\ \bibnamefont {Oskay}}, \bibinfo {author} {\bibfnamefont {R.~E.}\ \bibnamefont {Drullinger}}, \bibinfo {author} {\bibfnamefont {T.~M.}\ \bibnamefont {Fortier}}, \bibinfo {author} {\bibfnamefont {J.~E.}\ \bibnamefont {Stalnaker}}, \bibinfo {author} {\bibfnamefont {S.~A.}\ \bibnamefont {Diddams}}, \bibinfo {author} {\bibfnamefont {W.~C.}\ \bibnamefont {Swann}}, \bibinfo {author} {\bibfnamefont {N.~R.}\ \bibnamefont {Newbury}}, \bibinfo {author} {\bibfnamefont {W.~M.}\ \bibnamefont {Itano}}, \bibinfo {author} {\bibfnamefont {D.~J.}\ \bibnamefont {Wineland}},\ and\ \bibinfo
  {author} {\bibfnamefont {J.~C.}\ \bibnamefont {Bergquist}},\ }\bibfield  {title} {\bibinfo {title} {Frequency {{Ratio}} of {{Al}}{\textsuperscript{+}} and {{Hg}}{\textsuperscript{+}} {{Single-Ion Optical Clocks}}; {{Metrology}} at the 17{\textsuperscript{th}} {{Decimal Place}}},\ }\href {https://doi.org/10.1126/science.1154622} {\bibfield  {journal} {\bibinfo  {journal} {Science}\ }\textbf {\bibinfo {volume} {319}},\ \bibinfo {pages} {1808} (\bibinfo {year} {2008})}\BibitemShut {NoStop}%
\bibitem [{\citenamefont {Chou}\ \emph {et~al.}(2010)\citenamefont {Chou}, \citenamefont {Hume}, \citenamefont {Koelemeij}, \citenamefont {Wineland},\ and\ \citenamefont {Rosenband}}]{chou_frequency_2010}%
  \BibitemOpen
  \bibfield  {author} {\bibinfo {author} {\bibfnamefont {C.~W.}\ \bibnamefont {Chou}}, \bibinfo {author} {\bibfnamefont {D.~B.}\ \bibnamefont {Hume}}, \bibinfo {author} {\bibfnamefont {J.~C.~J.}\ \bibnamefont {Koelemeij}}, \bibinfo {author} {\bibfnamefont {D.~J.}\ \bibnamefont {Wineland}},\ and\ \bibinfo {author} {\bibfnamefont {T.}~\bibnamefont {Rosenband}},\ }\bibfield  {title} {\bibinfo {title} {Frequency {{Comparison}} of {{Two High-Accuracy Al}}{\textsuperscript{+}} {{Optical Clocks}}},\ }\href {https://doi.org/10.1103/PhysRevLett.104.070802} {\bibfield  {journal} {\bibinfo  {journal} {Physical Review Letters}\ }\textbf {\bibinfo {volume} {104}},\ \bibinfo {pages} {070802} (\bibinfo {year} {2010})}\BibitemShut {NoStop}%
\bibitem [{\citenamefont {Cui}\ \emph {et~al.}(2022)\citenamefont {Cui}, \citenamefont {Chao}, \citenamefont {Sun}, \citenamefont {Wang}, \citenamefont {Zhang}, \citenamefont {Wei}, \citenamefont {Yuan}, \citenamefont {Cao}, \citenamefont {Shu},\ and\ \citenamefont {Huang}}]{cui_evaluation_2022}%
  \BibitemOpen
  \bibfield  {author} {\bibinfo {author} {\bibfnamefont {K.}~\bibnamefont {Cui}}, \bibinfo {author} {\bibfnamefont {S.}~\bibnamefont {Chao}}, \bibinfo {author} {\bibfnamefont {C.}~\bibnamefont {Sun}}, \bibinfo {author} {\bibfnamefont {S.}~\bibnamefont {Wang}}, \bibinfo {author} {\bibfnamefont {P.}~\bibnamefont {Zhang}}, \bibinfo {author} {\bibfnamefont {Y.}~\bibnamefont {Wei}}, \bibinfo {author} {\bibfnamefont {J.}~\bibnamefont {Yuan}}, \bibinfo {author} {\bibfnamefont {J.}~\bibnamefont {Cao}}, \bibinfo {author} {\bibfnamefont {H.}~\bibnamefont {Shu}},\ and\ \bibinfo {author} {\bibfnamefont {X.}~\bibnamefont {Huang}},\ }\bibfield  {title} {\bibinfo {title} {Evaluation of the systematic shifts of a {\textsuperscript{40}}{{Ca}}{\textsuperscript{+}}-{\textsuperscript{27}}{{Al}}{\textsuperscript{+}} optical clock},\ }\href {https://doi.org/10.1140/epjd/s10053-022-00451-1} {\bibfield  {journal} {\bibinfo  {journal} {The European Physical Journal D}\ }\textbf {\bibinfo {volume} {76}},\ \bibinfo {pages} {140}
  (\bibinfo {year} {2022})}\BibitemShut {NoStop}%
\bibitem [{\citenamefont {Roos}\ \emph {et~al.}(2000)\citenamefont {Roos}, \citenamefont {Leibfried}, \citenamefont {Mundt}, \citenamefont {{Schmidt-Kaler}}, \citenamefont {Eschner},\ and\ \citenamefont {Blatt}}]{roos_experimental_2000}%
  \BibitemOpen
  \bibfield  {author} {\bibinfo {author} {\bibfnamefont {C.~F.}\ \bibnamefont {Roos}}, \bibinfo {author} {\bibfnamefont {D.}~\bibnamefont {Leibfried}}, \bibinfo {author} {\bibfnamefont {A.}~\bibnamefont {Mundt}}, \bibinfo {author} {\bibfnamefont {F.}~\bibnamefont {{Schmidt-Kaler}}}, \bibinfo {author} {\bibfnamefont {J.}~\bibnamefont {Eschner}},\ and\ \bibinfo {author} {\bibfnamefont {R.}~\bibnamefont {Blatt}},\ }\bibfield  {title} {\bibinfo {title} {Experimental demonstration of ground state laser cooling with electromagnetically induced transparency},\ }\href@noop {} {\bibfield  {journal} {\bibinfo  {journal} {Physical Review Letters}\ }\textbf {\bibinfo {volume} {85}},\ \bibinfo {pages} {5547} (\bibinfo {year} {2000})}\BibitemShut {NoStop}%
\bibitem [{\citenamefont {Lechner}\ \emph {et~al.}(2016)\citenamefont {Lechner}, \citenamefont {Maier}, \citenamefont {Hempel}, \citenamefont {Jurcevic}, \citenamefont {Lanyon}, \citenamefont {Monz}, \citenamefont {Brownnutt}, \citenamefont {Blatt},\ and\ \citenamefont {Roos}}]{lechner_electromagneticallyinducedtransparency_2016}%
  \BibitemOpen
  \bibfield  {author} {\bibinfo {author} {\bibfnamefont {R.}~\bibnamefont {Lechner}}, \bibinfo {author} {\bibfnamefont {C.}~\bibnamefont {Maier}}, \bibinfo {author} {\bibfnamefont {C.}~\bibnamefont {Hempel}}, \bibinfo {author} {\bibfnamefont {P.}~\bibnamefont {Jurcevic}}, \bibinfo {author} {\bibfnamefont {B.~P.}\ \bibnamefont {Lanyon}}, \bibinfo {author} {\bibfnamefont {T.}~\bibnamefont {Monz}}, \bibinfo {author} {\bibfnamefont {M.}~\bibnamefont {Brownnutt}}, \bibinfo {author} {\bibfnamefont {R.}~\bibnamefont {Blatt}},\ and\ \bibinfo {author} {\bibfnamefont {C.~F.}\ \bibnamefont {Roos}},\ }\bibfield  {title} {\bibinfo {title} {Electromagnetically-induced-transparency ground-state cooling of long ion strings},\ }\bibfield  {journal} {\bibinfo  {journal} {Physical Review A}\ }\textbf {\bibinfo {volume} {93}},\ \href {https://doi.org/10.1103/PhysRevA.93.053401} {10.1103/PhysRevA.93.053401} (\bibinfo {year} {2016})\BibitemShut {NoStop}%
\bibitem [{\citenamefont {Scharnhorst}\ \emph {et~al.}(2018)\citenamefont {Scharnhorst}, \citenamefont {Cerrillo}, \citenamefont {Kramer}, \citenamefont {Leroux}, \citenamefont {W{\"u}bbena}, \citenamefont {Retzker},\ and\ \citenamefont {Schmidt}}]{scharnhorst_experimental_2018}%
  \BibitemOpen
  \bibfield  {author} {\bibinfo {author} {\bibfnamefont {N.}~\bibnamefont {Scharnhorst}}, \bibinfo {author} {\bibfnamefont {J.}~\bibnamefont {Cerrillo}}, \bibinfo {author} {\bibfnamefont {J.}~\bibnamefont {Kramer}}, \bibinfo {author} {\bibfnamefont {I.~D.}\ \bibnamefont {Leroux}}, \bibinfo {author} {\bibfnamefont {J.~B.}\ \bibnamefont {W{\"u}bbena}}, \bibinfo {author} {\bibfnamefont {A.}~\bibnamefont {Retzker}},\ and\ \bibinfo {author} {\bibfnamefont {P.~O.}\ \bibnamefont {Schmidt}},\ }\bibfield  {title} {\bibinfo {title} {Experimental and theoretical investigation of a multimode cooling scheme using multiple electromagnetically-induced-transparency resonances},\ }\href {https://doi.org/10.1103/PhysRevA.98.023424} {\bibfield  {journal} {\bibinfo  {journal} {Physical Review A}\ }\textbf {\bibinfo {volume} {98}},\ \bibinfo {pages} {023424} (\bibinfo {year} {2018})}\BibitemShut {NoStop}%
\bibitem [{\citenamefont {Sun}\ \emph {et~al.}(2023)\citenamefont {Sun}, \citenamefont {Cui}, \citenamefont {Chao}, \citenamefont {Wei}, \citenamefont {Yuan}, \citenamefont {Cao}, \citenamefont {Shu},\ and\ \citenamefont {Huang}}]{sun_sympathetic_2023}%
  \BibitemOpen
  \bibfield  {author} {\bibinfo {author} {\bibfnamefont {C.}~\bibnamefont {Sun}}, \bibinfo {author} {\bibfnamefont {K.}~\bibnamefont {Cui}}, \bibinfo {author} {\bibfnamefont {S.}~\bibnamefont {Chao}}, \bibinfo {author} {\bibfnamefont {Y.}~\bibnamefont {Wei}}, \bibinfo {author} {\bibfnamefont {J.}~\bibnamefont {Yuan}}, \bibinfo {author} {\bibfnamefont {J.}~\bibnamefont {Cao}}, \bibinfo {author} {\bibfnamefont {H.}~\bibnamefont {Shu}},\ and\ \bibinfo {author} {\bibfnamefont {X.}~\bibnamefont {Huang}},\ }\bibfield  {title} {\bibinfo {title} {Sympathetic electromagnetically induced transparency ground state cooling of a {{40Ca}}+--{{27Al}}+ pair in an {{27Al}}+ clock},\ }\href {https://doi.org/10.1088/1674-1056/aca39d} {\bibfield  {journal} {\bibinfo  {journal} {Chinese Physics B}\ }\textbf {\bibinfo {volume} {32}},\ \bibinfo {pages} {050601} (\bibinfo {year} {2023})}\BibitemShut {NoStop}%
\bibitem [{\citenamefont {Chen}\ \emph {et~al.}(2017)\citenamefont {Chen}, \citenamefont {Brewer}, \citenamefont {Chou}, \citenamefont {Wineland}, \citenamefont {Leibrandt},\ and\ \citenamefont {Hume}}]{chen_sympathetic_2017}%
  \BibitemOpen
  \bibfield  {author} {\bibinfo {author} {\bibfnamefont {J.-S.}\ \bibnamefont {Chen}}, \bibinfo {author} {\bibfnamefont {S.~M.}\ \bibnamefont {Brewer}}, \bibinfo {author} {\bibfnamefont {C.~W.}\ \bibnamefont {Chou}}, \bibinfo {author} {\bibfnamefont {D.~J.}\ \bibnamefont {Wineland}}, \bibinfo {author} {\bibfnamefont {D.~R.}\ \bibnamefont {Leibrandt}},\ and\ \bibinfo {author} {\bibfnamefont {D.~B.}\ \bibnamefont {Hume}},\ }\bibfield  {title} {\bibinfo {title} {Sympathetic {{Ground State Cooling}} and {{Time-Dilation Shifts}} in an {\textsuperscript{27}}{{Al}}{\textsuperscript{+}} {{Optical Clock}}},\ }\href {https://doi.org/10.1103/PhysRevLett.118.053002} {\bibfield  {journal} {\bibinfo  {journal} {Physical Review Letters}\ }\textbf {\bibinfo {volume} {118}},\ \bibinfo {pages} {053002} (\bibinfo {year} {2017})}\BibitemShut {NoStop}%
\bibitem [{\citenamefont {Kramer}(2023)}]{kramer_aluminum_2023}%
  \BibitemOpen
  \bibfield  {author} {\bibinfo {author} {\bibfnamefont {J.~A.}\ \bibnamefont {Kramer}},\ }\emph {\bibinfo {title} {An Aluminum Optical Clock Setup and Its Evaluation Using {{Ca}}{\textsuperscript{+}}}},\ \href {https://doi.org/10.15488/13300} {Ph.D. thesis},\ \bibinfo  {school} {Leibniz Universit{\"a}t Hannover} (\bibinfo {year} {2023})\BibitemShut {NoStop}%
\bibitem [{\citenamefont {Hankin}\ \emph {et~al.}(2019)\citenamefont {Hankin}, \citenamefont {Clements}, \citenamefont {Huang}, \citenamefont {Brewer}, \citenamefont {Chen}, \citenamefont {Chou}, \citenamefont {Hume},\ and\ \citenamefont {Leibrandt}}]{hankin_systematic_2019}%
  \BibitemOpen
  \bibfield  {author} {\bibinfo {author} {\bibfnamefont {A.~M.}\ \bibnamefont {Hankin}}, \bibinfo {author} {\bibfnamefont {E.~R.}\ \bibnamefont {Clements}}, \bibinfo {author} {\bibfnamefont {Y.}~\bibnamefont {Huang}}, \bibinfo {author} {\bibfnamefont {S.~M.}\ \bibnamefont {Brewer}}, \bibinfo {author} {\bibfnamefont {J.-S.}\ \bibnamefont {Chen}}, \bibinfo {author} {\bibfnamefont {C.~W.}\ \bibnamefont {Chou}}, \bibinfo {author} {\bibfnamefont {D.~B.}\ \bibnamefont {Hume}},\ and\ \bibinfo {author} {\bibfnamefont {D.~R.}\ \bibnamefont {Leibrandt}},\ }\bibfield  {title} {\bibinfo {title} {Systematic uncertainty due to background-gas collisions in trapped-ion optical clocks},\ }\href {https://doi.org/10.1103/PhysRevA.100.033419} {\bibfield  {journal} {\bibinfo  {journal} {Physical Review A}\ }\textbf {\bibinfo {volume} {100}},\ \bibinfo {pages} {033419} (\bibinfo {year} {2019})}\BibitemShut {NoStop}%
\bibitem [{\citenamefont {Morigi}\ \emph {et~al.}(2000)\citenamefont {Morigi}, \citenamefont {Eschner},\ and\ \citenamefont {Keitel}}]{morigi_ground_2000}%
  \BibitemOpen
  \bibfield  {author} {\bibinfo {author} {\bibfnamefont {G.}~\bibnamefont {Morigi}}, \bibinfo {author} {\bibfnamefont {J.}~\bibnamefont {Eschner}},\ and\ \bibinfo {author} {\bibfnamefont {C.~H.}\ \bibnamefont {Keitel}},\ }\bibfield  {title} {\bibinfo {title} {Ground state laser cooling using electromagnetically induced transparency},\ }\href@noop {} {\bibfield  {journal} {\bibinfo  {journal} {Physical Review Letters}\ }\textbf {\bibinfo {volume} {85}},\ \bibinfo {pages} {4458} (\bibinfo {year} {2000})}\BibitemShut {NoStop}%
\bibitem [{\citenamefont {Hume}\ \emph {et~al.}(2007)\citenamefont {Hume}, \citenamefont {Rosenband},\ and\ \citenamefont {Wineland}}]{hume_highfidelity_2007}%
  \BibitemOpen
  \bibfield  {author} {\bibinfo {author} {\bibfnamefont {D.}~\bibnamefont {Hume}}, \bibinfo {author} {\bibfnamefont {T.}~\bibnamefont {Rosenband}},\ and\ \bibinfo {author} {\bibfnamefont {D.}~\bibnamefont {Wineland}},\ }\bibfield  {title} {\bibinfo {title} {High-{{Fidelity Adaptive Qubit Detection}} through {{Repetitive Quantum Nondemolition Measurements}}},\ }\href {https://doi.org/10.1103/PhysRevLett.99.120502} {\bibfield  {journal} {\bibinfo  {journal} {Physical Review Letters}\ }\textbf {\bibinfo {volume} {99}},\ \bibinfo {pages} {120502} (\bibinfo {year} {2007})}\BibitemShut {NoStop}%
\bibitem [{\citenamefont {Amairi}(2014)}]{amairi_long_2014}%
  \BibitemOpen
  \bibfield  {author} {\bibinfo {author} {\bibfnamefont {S.}~\bibnamefont {Amairi}},\ }\emph {\bibinfo {title} {A {{Long Optical Cavity For Sub-Hertz Laser Spectroscopy}}}},\ \href@noop {} {Ph.D. thesis},\ \bibinfo  {school} {Leibniz Universit{\"a}t Hannover} (\bibinfo {year} {2014})\BibitemShut {NoStop}%
\bibitem [{\citenamefont {Scharnhorst}\ \emph {et~al.}(2015)\citenamefont {Scharnhorst}, \citenamefont {W{\"u}bbena}, \citenamefont {Hannig}, \citenamefont {Jakobsen}, \citenamefont {Kramer}, \citenamefont {Leroux},\ and\ \citenamefont {Schmidt}}]{scharnhorst_highbandwidth_2015}%
  \BibitemOpen
  \bibfield  {author} {\bibinfo {author} {\bibfnamefont {N.}~\bibnamefont {Scharnhorst}}, \bibinfo {author} {\bibfnamefont {J.~B.}\ \bibnamefont {W{\"u}bbena}}, \bibinfo {author} {\bibfnamefont {S.}~\bibnamefont {Hannig}}, \bibinfo {author} {\bibfnamefont {K.}~\bibnamefont {Jakobsen}}, \bibinfo {author} {\bibfnamefont {J.}~\bibnamefont {Kramer}}, \bibinfo {author} {\bibfnamefont {I.~D.}\ \bibnamefont {Leroux}},\ and\ \bibinfo {author} {\bibfnamefont {P.~O.}\ \bibnamefont {Schmidt}},\ }\bibfield  {title} {\bibinfo {title} {High-bandwidth transfer of phase stability through a fiber frequency comb},\ }\href {https://doi.org/10.1364/OE.23.019771} {\bibfield  {journal} {\bibinfo  {journal} {Optics Express}\ }\textbf {\bibinfo {volume} {23}},\ \bibinfo {pages} {19771} (\bibinfo {year} {2015})}\BibitemShut {NoStop}%
\bibitem [{\citenamefont {Benkler}\ \emph {et~al.}(2019)\citenamefont {Benkler}, \citenamefont {Lipphardt}, \citenamefont {Puppe}, \citenamefont {Wilk}, \citenamefont {Rohde},\ and\ \citenamefont {Sterr}}]{benkler_endtoend_2019}%
  \BibitemOpen
  \bibfield  {author} {\bibinfo {author} {\bibfnamefont {E.}~\bibnamefont {Benkler}}, \bibinfo {author} {\bibfnamefont {B.}~\bibnamefont {Lipphardt}}, \bibinfo {author} {\bibfnamefont {T.}~\bibnamefont {Puppe}}, \bibinfo {author} {\bibfnamefont {R.}~\bibnamefont {Wilk}}, \bibinfo {author} {\bibfnamefont {F.}~\bibnamefont {Rohde}},\ and\ \bibinfo {author} {\bibfnamefont {U.}~\bibnamefont {Sterr}},\ }\bibfield  {title} {\bibinfo {title} {End-to-end topology for fiber comb based optical frequency transfer at the 10{\textsuperscript{-21}} level},\ }\href {https://doi.org/10.1364/OE.27.036886} {\bibfield  {journal} {\bibinfo  {journal} {Optics Express}\ }\textbf {\bibinfo {volume} {27}},\ \bibinfo {pages} {36886} (\bibinfo {year} {2019})}\BibitemShut {NoStop}%
\bibitem [{\citenamefont {Ye}\ \emph {et~al.}(2003)\citenamefont {Ye}, \citenamefont {Peng}, \citenamefont {Jones}, \citenamefont {Holman}, \citenamefont {Hall}, \citenamefont {Jones}, \citenamefont {Diddams}, \citenamefont {Kitching}, \citenamefont {Bize}, \citenamefont {Bergquist}, \citenamefont {Hollberg}, \citenamefont {Robertsson},\ and\ \citenamefont {Ma}}]{ye_delivery_2003}%
  \BibitemOpen
  \bibfield  {author} {\bibinfo {author} {\bibfnamefont {J.}~\bibnamefont {Ye}}, \bibinfo {author} {\bibfnamefont {J.-L.}\ \bibnamefont {Peng}}, \bibinfo {author} {\bibfnamefont {R.~J.}\ \bibnamefont {Jones}}, \bibinfo {author} {\bibfnamefont {K.~W.}\ \bibnamefont {Holman}}, \bibinfo {author} {\bibfnamefont {J.~L.}\ \bibnamefont {Hall}}, \bibinfo {author} {\bibfnamefont {D.~J.}\ \bibnamefont {Jones}}, \bibinfo {author} {\bibfnamefont {S.~A.}\ \bibnamefont {Diddams}}, \bibinfo {author} {\bibfnamefont {J.}~\bibnamefont {Kitching}}, \bibinfo {author} {\bibfnamefont {S.}~\bibnamefont {Bize}}, \bibinfo {author} {\bibfnamefont {J.~C.}\ \bibnamefont {Bergquist}}, \bibinfo {author} {\bibfnamefont {L.~W.}\ \bibnamefont {Hollberg}}, \bibinfo {author} {\bibfnamefont {L.}~\bibnamefont {Robertsson}},\ and\ \bibinfo {author} {\bibfnamefont {L.-S.}\ \bibnamefont {Ma}},\ }\bibfield  {title} {\bibinfo {title} {Delivery of high-stability optical and microwave frequency standards over an optical fiber network},\ }\href
  {https://doi.org/10.1364/JOSAB.20.001459} {\bibfield  {journal} {\bibinfo  {journal} {Journal of the Optical Society of America B}\ }\textbf {\bibinfo {volume} {20}},\ \bibinfo {pages} {1459} (\bibinfo {year} {2003})}\BibitemShut {NoStop}%
\bibitem [{\citenamefont {Kraus}\ \emph {et~al.}(2022)\citenamefont {Kraus}, \citenamefont {Dawel}, \citenamefont {Hannig}, \citenamefont {Kramer}, \citenamefont {Nauk},\ and\ \citenamefont {Schmidt}}]{kraus_phasestabilized_2022}%
  \BibitemOpen
  \bibfield  {author} {\bibinfo {author} {\bibfnamefont {B.}~\bibnamefont {Kraus}}, \bibinfo {author} {\bibfnamefont {F.}~\bibnamefont {Dawel}}, \bibinfo {author} {\bibfnamefont {S.}~\bibnamefont {Hannig}}, \bibinfo {author} {\bibfnamefont {J.}~\bibnamefont {Kramer}}, \bibinfo {author} {\bibfnamefont {C.}~\bibnamefont {Nauk}},\ and\ \bibinfo {author} {\bibfnamefont {P.~O.}\ \bibnamefont {Schmidt}},\ }\bibfield  {title} {\bibinfo {title} {Phase-stabilized {{UV}} light at 267 nm through twofold second harmonic generation},\ }\href {https://doi.org/10.1364/OE.471450} {\bibfield  {journal} {\bibinfo  {journal} {Optics Express}\ }\textbf {\bibinfo {volume} {30}},\ \bibinfo {pages} {44992} (\bibinfo {year} {2022})}\BibitemShut {NoStop}%
\bibitem [{\citenamefont {Colombe}\ \emph {et~al.}(2014)\citenamefont {Colombe}, \citenamefont {Slichter}, \citenamefont {Wilson}, \citenamefont {Leibfried},\ and\ \citenamefont {Wineland}}]{colombe_singlemode_2014}%
  \BibitemOpen
  \bibfield  {author} {\bibinfo {author} {\bibfnamefont {Y.}~\bibnamefont {Colombe}}, \bibinfo {author} {\bibfnamefont {D.~H.}\ \bibnamefont {Slichter}}, \bibinfo {author} {\bibfnamefont {A.~C.}\ \bibnamefont {Wilson}}, \bibinfo {author} {\bibfnamefont {D.}~\bibnamefont {Leibfried}},\ and\ \bibinfo {author} {\bibfnamefont {D.~J.}\ \bibnamefont {Wineland}},\ }\bibfield  {title} {\bibinfo {title} {Single-mode optical fiber for high-power, low-loss {{UV}} transmission},\ }\href {https://doi.org/10.1364/OE.22.019783} {\bibfield  {journal} {\bibinfo  {journal} {Optics Express}\ }\textbf {\bibinfo {volume} {22}},\ \bibinfo {pages} {19783} (\bibinfo {year} {2014})}\BibitemShut {NoStop}%
\bibitem [{\citenamefont {Marciniak}\ \emph {et~al.}(2017)\citenamefont {Marciniak}, \citenamefont {Ball}, \citenamefont {Hung},\ and\ \citenamefont {Biercuk}}]{marciniak_fully_2017}%
  \BibitemOpen
  \bibfield  {author} {\bibinfo {author} {\bibfnamefont {C.~D.}\ \bibnamefont {Marciniak}}, \bibinfo {author} {\bibfnamefont {H.~B.}\ \bibnamefont {Ball}}, \bibinfo {author} {\bibfnamefont {A.~T.-H.}\ \bibnamefont {Hung}},\ and\ \bibinfo {author} {\bibfnamefont {M.~J.}\ \bibnamefont {Biercuk}},\ }\bibfield  {title} {\bibinfo {title} {Towards fully commercial, {{UV-compatible}} fiber patch cords},\ }\href {https://doi.org/10.1364/OE.25.015643} {\bibfield  {journal} {\bibinfo  {journal} {Optics Express}\ }\textbf {\bibinfo {volume} {25}},\ \bibinfo {pages} {15643} (\bibinfo {year} {2017})}\BibitemShut {NoStop}%
\bibitem [{\citenamefont {Schwarz}(2022)}]{schwarz_cryogenic_2022}%
  \BibitemOpen
  \bibfield  {author} {\bibinfo {author} {\bibfnamefont {R.}~\bibnamefont {Schwarz}},\ }\emph {\bibinfo {title} {A Cryogenic {{Strontium}} Lattice Clock}},\ \href {https://doi.org/10.15488/11929} {\bibinfo {type} {{{DoctoralThesis}}}},\ \bibinfo  {school} {Hannover : Institutionelles Repositorium der Leibniz Universit{\"a}t Hannover} (\bibinfo {year} {2022})\BibitemShut {NoStop}%
\bibitem [{\citenamefont {Dawel}\ \emph {et~al.}(2025)\citenamefont {Dawel}, \citenamefont {Pelzer}, \citenamefont {Dietze}, \citenamefont {Kramer}, \citenamefont {Hild}, \citenamefont {King}, \citenamefont {Spethmann}, \citenamefont {Klose}, \citenamefont {Stahl}, \citenamefont {Dörscher}, \citenamefont {Benkler}, \citenamefont {Lisdat},\ and\ \citenamefont {Schmidt}}]{dawel_supplement_2025}%
  \BibitemOpen
  \bibfield  {author} {\bibinfo {author} {\bibfnamefont {F.}~\bibnamefont {Dawel}}, \bibinfo {author} {\bibfnamefont {L.}~\bibnamefont {Pelzer}}, \bibinfo {author} {\bibfnamefont {K.}~\bibnamefont {Dietze}}, \bibinfo {author} {\bibfnamefont {J.}~\bibnamefont {Kramer}}, \bibinfo {author} {\bibfnamefont {M.}~\bibnamefont {Hild}}, \bibinfo {author} {\bibfnamefont {S.}~\bibnamefont {King}}, \bibinfo {author} {\bibfnamefont {N.}~\bibnamefont {Spethmann}}, \bibinfo {author} {\bibfnamefont {J.}~\bibnamefont {Klose}}, \bibinfo {author} {\bibfnamefont {K.}~\bibnamefont {Stahl}}, \bibinfo {author} {\bibfnamefont {S.}~\bibnamefont {Dörscher}}, \bibinfo {author} {\bibfnamefont {E.}~\bibnamefont {Benkler}}, \bibinfo {author} {\bibfnamefont {C.}~\bibnamefont {Lisdat}},\ and\ \bibinfo {author} {\bibfnamefont {P.~O.}\ \bibnamefont {Schmidt}},\ }\bibfield  {title} {\bibinfo {title} {Supplement: A high-stability optical clock based on a continuously ground-state cooled al$^+$ ion without compromising its accuracy},\
  }\href@noop {} {\bibfield  {journal} {\bibinfo  {journal} {PR}\ } (\bibinfo {year} {2025})}\BibitemShut {NoStop}%
\bibitem [{\citenamefont {Rasmusson}\ \emph {et~al.}(2021)\citenamefont {Rasmusson}, \citenamefont {D'Onofrio}, \citenamefont {Xie}, \citenamefont {Cui},\ and\ \citenamefont {Richerme}}]{Rasmusson_optimized_2021}%
  \BibitemOpen
  \bibfield  {author} {\bibinfo {author} {\bibfnamefont {A.~J.}\ \bibnamefont {Rasmusson}}, \bibinfo {author} {\bibfnamefont {M.}~\bibnamefont {D'Onofrio}}, \bibinfo {author} {\bibfnamefont {Y.}~\bibnamefont {Xie}}, \bibinfo {author} {\bibfnamefont {J.}~\bibnamefont {Cui}},\ and\ \bibinfo {author} {\bibfnamefont {P.}~\bibnamefont {Richerme}},\ }\bibfield  {title} {\bibinfo {title} {Optimized pulsed sideband cooling and enhanced thermometry of trapped ions},\ }\href {https://doi.org/10.1103/PhysRevA.104.043108} {\bibfield  {journal} {\bibinfo  {journal} {Physical Review A}\ }\textbf {\bibinfo {volume} {104}},\ \bibinfo {pages} {043108} (\bibinfo {year} {2021})}\BibitemShut {NoStop}%
\bibitem [{\citenamefont {Steck}(2007)}]{steck_quantum_2007}%
  \BibitemOpen
  \bibfield  {author} {\bibinfo {author} {\bibfnamefont {D.~A.}\ \bibnamefont {Steck}},\ }\href@noop {} {\emph {\bibinfo {title} {Quantum and {{Atom Optics}}}}},\ \bibinfo {edition} {2025th}\ ed.\ (\bibinfo {year} {2007})\BibitemShut {NoStop}%
\bibitem [{\citenamefont {Bonin}\ and\ \citenamefont {Kresin}(1997)}]{bonin_electricdipole_1997}%
  \BibitemOpen
  \bibfield  {author} {\bibinfo {author} {\bibfnamefont {K.~D.}\ \bibnamefont {Bonin}}\ and\ \bibinfo {author} {\bibfnamefont {V.~V.}\ \bibnamefont {Kresin}},\ }\href@noop {} {\emph {\bibinfo {title} {Electric-Dipole Polarizabilities of Atoms, Molecules, and Clusters}}}\ (\bibinfo  {publisher} {World Scientific},\ \bibinfo {year} {1997})\BibitemShut {NoStop}%
\bibitem [{\citenamefont {{Dounas-Frazer}}\ \emph {et~al.}(2010)\citenamefont {{Dounas-Frazer}}, \citenamefont {Tsigutkin}, \citenamefont {Family},\ and\ \citenamefont {Budker}}]{dounas-frazer_measurement_2010}%
  \BibitemOpen
  \bibfield  {author} {\bibinfo {author} {\bibfnamefont {D.~R.}\ \bibnamefont {{Dounas-Frazer}}}, \bibinfo {author} {\bibfnamefont {K.}~\bibnamefont {Tsigutkin}}, \bibinfo {author} {\bibfnamefont {A.}~\bibnamefont {Family}},\ and\ \bibinfo {author} {\bibfnamefont {D.}~\bibnamefont {Budker}},\ }\bibfield  {title} {\bibinfo {title} {Measurement of dynamic {{Stark}} polarizabilities by analyzing spectral line shapes of forbidden transitions},\ }\href {https://doi.org/10.1103/PhysRevA.82.062507} {\bibfield  {journal} {\bibinfo  {journal} {Physical Review A}\ }\textbf {\bibinfo {volume} {82}},\ \bibinfo {pages} {062507} (\bibinfo {year} {2010})}\BibitemShut {NoStop}%
\bibitem [{\citenamefont {Shi}\ \emph {et~al.}(2015)\citenamefont {Shi}, \citenamefont {Robyr}, \citenamefont {Eismann}, \citenamefont {Zawada}, \citenamefont {Lorini}, \citenamefont {Targat},\ and\ \citenamefont {Lodewyck}}]{shi_polarizabilities_2015}%
  \BibitemOpen
  \bibfield  {author} {\bibinfo {author} {\bibfnamefont {C.}~\bibnamefont {Shi}}, \bibinfo {author} {\bibfnamefont {J.-L.}\ \bibnamefont {Robyr}}, \bibinfo {author} {\bibfnamefont {U.}~\bibnamefont {Eismann}}, \bibinfo {author} {\bibfnamefont {M.}~\bibnamefont {Zawada}}, \bibinfo {author} {\bibfnamefont {L.}~\bibnamefont {Lorini}}, \bibinfo {author} {\bibfnamefont {R.~L.}\ \bibnamefont {Targat}},\ and\ \bibinfo {author} {\bibfnamefont {J.}~\bibnamefont {Lodewyck}},\ }\bibfield  {title} {\bibinfo {title} {Polarizabilities of the {{87Sr Clock Transition}}},\ }\href@noop {} {\bibfield  {journal} {\bibinfo  {journal} {arXiv:1503.09167 [physics]}\ } (\bibinfo {year} {2015})},\ \Eprint {https://arxiv.org/abs/1503.09167} {arXiv:1503.09167 [physics]} \BibitemShut {NoStop}%
\bibitem [{\citenamefont {Hettrich}\ \emph {et~al.}(2015)\citenamefont {Hettrich}, \citenamefont {Ruster}, \citenamefont {Kaufmann}, \citenamefont {Roos}, \citenamefont {Schmiegelow}, \citenamefont {{Schmidt-Kaler}},\ and\ \citenamefont {Poschinger}}]{hettrich_measurement_2015}%
  \BibitemOpen
  \bibfield  {author} {\bibinfo {author} {\bibfnamefont {M.}~\bibnamefont {Hettrich}}, \bibinfo {author} {\bibfnamefont {T.}~\bibnamefont {Ruster}}, \bibinfo {author} {\bibfnamefont {H.}~\bibnamefont {Kaufmann}}, \bibinfo {author} {\bibfnamefont {C.~F.}\ \bibnamefont {Roos}}, \bibinfo {author} {\bibfnamefont {C.~T.}\ \bibnamefont {Schmiegelow}}, \bibinfo {author} {\bibfnamefont {F.}~\bibnamefont {{Schmidt-Kaler}}},\ and\ \bibinfo {author} {\bibfnamefont {U.~G.}\ \bibnamefont {Poschinger}},\ }\bibfield  {title} {\bibinfo {title} {Measurement of {{Dipole Matrix Elements}} with a {{Single Trapped Ion}}},\ }\href {https://doi.org/10.1103/PhysRevLett.115.143003} {\bibfield  {journal} {\bibinfo  {journal} {Physical Review Letters}\ }\textbf {\bibinfo {volume} {115}},\ \bibinfo {pages} {143003} (\bibinfo {year} {2015})}\BibitemShut {NoStop}%
\bibitem [{\citenamefont {Barakhshan}\ \emph {et~al.}(2022)\citenamefont {Barakhshan}, \citenamefont {Marrs}, \citenamefont {Bhosale}, \citenamefont {Arora}, \citenamefont {Eigenmann},\ and\ \citenamefont {Safronova}}]{UDportal}%
  \BibitemOpen
  \bibfield  {author} {\bibinfo {author} {\bibfnamefont {P.}~\bibnamefont {Barakhshan}}, \bibinfo {author} {\bibfnamefont {A.}~\bibnamefont {Marrs}}, \bibinfo {author} {\bibfnamefont {A.}~\bibnamefont {Bhosale}}, \bibinfo {author} {\bibfnamefont {B.}~\bibnamefont {Arora}}, \bibinfo {author} {\bibfnamefont {R.}~\bibnamefont {Eigenmann}},\ and\ \bibinfo {author} {\bibfnamefont {M.~S.}\ \bibnamefont {Safronova}},\ }\href@noop {} {\bibinfo {title} {Portal for {{High-Precision Atomic Data}} and {{Computation}} (version 2.0)}},\ \bibinfo {howpublished} {\emph{Portal for High-Precision Atomic Data and Computation} (version 2.0). University of Delaware, Newark, DE, USA. URL: https://www.udel.edu/atom[February 2022].} (\bibinfo {year} {2022})\BibitemShut {NoStop}%
\bibitem [{\citenamefont {Wei}\ \emph {et~al.}(2024{\natexlab{a}})\citenamefont {Wei}, \citenamefont {Chao}, \citenamefont {Cui}, \citenamefont {Li}, \citenamefont {Yu}, \citenamefont {Zhang}, \citenamefont {Shu}, \citenamefont {Cao},\ and\ \citenamefont {Huang}}]{wei_improved_2024}%
  \BibitemOpen
  \bibfield  {author} {\bibinfo {author} {\bibfnamefont {Y.-F.}\ \bibnamefont {Wei}}, \bibinfo {author} {\bibfnamefont {S.-J.}\ \bibnamefont {Chao}}, \bibinfo {author} {\bibfnamefont {K.-F.}\ \bibnamefont {Cui}}, \bibinfo {author} {\bibfnamefont {C.-B.}\ \bibnamefont {Li}}, \bibinfo {author} {\bibfnamefont {S.-C.}\ \bibnamefont {Yu}}, \bibinfo {author} {\bibfnamefont {H.}~\bibnamefont {Zhang}}, \bibinfo {author} {\bibfnamefont {H.-L.}\ \bibnamefont {Shu}}, \bibinfo {author} {\bibfnamefont {J.}~\bibnamefont {Cao}},\ and\ \bibinfo {author} {\bibfnamefont {X.-R.}\ \bibnamefont {Huang}},\ }\bibfield  {title} {\bibinfo {title} {Improved {{Measurement}} of the {{Differential Polarizability Using Co-Trapped Ions}}},\ }\href {https://doi.org/10.1103/PhysRevLett.133.033001} {\bibfield  {journal} {\bibinfo  {journal} {Physical Review Letters}\ }\textbf {\bibinfo {volume} {133}},\ \bibinfo {pages} {033001} (\bibinfo {year} {2024}{\natexlab{a}})}\BibitemShut {NoStop}%
\bibitem [{\citenamefont {Kotler}\ \emph {et~al.}(2011)\citenamefont {Kotler}, \citenamefont {Akerman}, \citenamefont {Glickman}, \citenamefont {Keselman},\ and\ \citenamefont {Ozeri}}]{kotler_singleion_2011}%
  \BibitemOpen
  \bibfield  {author} {\bibinfo {author} {\bibfnamefont {S.}~\bibnamefont {Kotler}}, \bibinfo {author} {\bibfnamefont {N.}~\bibnamefont {Akerman}}, \bibinfo {author} {\bibfnamefont {Y.}~\bibnamefont {Glickman}}, \bibinfo {author} {\bibfnamefont {A.}~\bibnamefont {Keselman}},\ and\ \bibinfo {author} {\bibfnamefont {R.}~\bibnamefont {Ozeri}},\ }\bibfield  {title} {\bibinfo {title} {Single-ion quantum lock-in amplifier},\ }\href {https://doi.org/10.1038/nature10010} {\bibfield  {journal} {\bibinfo  {journal} {Nature}\ }\textbf {\bibinfo {volume} {473}},\ \bibinfo {pages} {61} (\bibinfo {year} {2011})}\BibitemShut {NoStop}%
\bibitem [{\citenamefont {Dimarcq}\ \emph {et~al.}(2024)\citenamefont {Dimarcq}, \citenamefont {Gertsvolf}, \citenamefont {Mileti}, \citenamefont {Bize}, \citenamefont {Oates}, \citenamefont {Peik}, \citenamefont {Calonico}, \citenamefont {Ido}, \citenamefont {Tavella}, \citenamefont {Meynadier}, \citenamefont {Petit}, \citenamefont {Panfilo}, \citenamefont {Bartholomew}, \citenamefont {Defraigne}, \citenamefont {Donley}, \citenamefont {Hedekvist}, \citenamefont {Sesia}, \citenamefont {Wouters}, \citenamefont {Dub{\'e}}, \citenamefont {Fang}, \citenamefont {Levi}, \citenamefont {Lodewyck}, \citenamefont {Margolis}, \citenamefont {Newell}, \citenamefont {Slyusarev}, \citenamefont {Weyers}, \citenamefont {Uzan}, \citenamefont {Yasuda}, \citenamefont {Yu}, \citenamefont {Rieck}, \citenamefont {Schnatz}, \citenamefont {Hanado}, \citenamefont {Fujieda}, \citenamefont {Pottie}, \citenamefont {Hanssen}, \citenamefont {Malimon},\ and\ \citenamefont {Ashby}}]{dimarcq_roadmap_2024}%
  \BibitemOpen
  \bibfield  {author} {\bibinfo {author} {\bibfnamefont {N.}~\bibnamefont {Dimarcq}}, \bibinfo {author} {\bibfnamefont {M.}~\bibnamefont {Gertsvolf}}, \bibinfo {author} {\bibfnamefont {G.}~\bibnamefont {Mileti}}, \bibinfo {author} {\bibfnamefont {S.}~\bibnamefont {Bize}}, \bibinfo {author} {\bibfnamefont {C.~W.}\ \bibnamefont {Oates}}, \bibinfo {author} {\bibfnamefont {E.}~\bibnamefont {Peik}}, \bibinfo {author} {\bibfnamefont {D.}~\bibnamefont {Calonico}}, \bibinfo {author} {\bibfnamefont {T.}~\bibnamefont {Ido}}, \bibinfo {author} {\bibfnamefont {P.}~\bibnamefont {Tavella}}, \bibinfo {author} {\bibfnamefont {F.}~\bibnamefont {Meynadier}}, \bibinfo {author} {\bibfnamefont {G.}~\bibnamefont {Petit}}, \bibinfo {author} {\bibfnamefont {G.}~\bibnamefont {Panfilo}}, \bibinfo {author} {\bibfnamefont {J.}~\bibnamefont {Bartholomew}}, \bibinfo {author} {\bibfnamefont {P.}~\bibnamefont {Defraigne}}, \bibinfo {author} {\bibfnamefont {E.~A.}\ \bibnamefont {Donley}}, \bibinfo {author} {\bibfnamefont {P.~O.}\
  \bibnamefont {Hedekvist}}, \bibinfo {author} {\bibfnamefont {I.}~\bibnamefont {Sesia}}, \bibinfo {author} {\bibfnamefont {M.}~\bibnamefont {Wouters}}, \bibinfo {author} {\bibfnamefont {P.}~\bibnamefont {Dub{\'e}}}, \bibinfo {author} {\bibfnamefont {F.}~\bibnamefont {Fang}}, \bibinfo {author} {\bibfnamefont {F.}~\bibnamefont {Levi}}, \bibinfo {author} {\bibfnamefont {J.}~\bibnamefont {Lodewyck}}, \bibinfo {author} {\bibfnamefont {H.~S.}\ \bibnamefont {Margolis}}, \bibinfo {author} {\bibfnamefont {D.}~\bibnamefont {Newell}}, \bibinfo {author} {\bibfnamefont {S.}~\bibnamefont {Slyusarev}}, \bibinfo {author} {\bibfnamefont {S.}~\bibnamefont {Weyers}}, \bibinfo {author} {\bibfnamefont {J.-P.}\ \bibnamefont {Uzan}}, \bibinfo {author} {\bibfnamefont {M.}~\bibnamefont {Yasuda}}, \bibinfo {author} {\bibfnamefont {D.-H.}\ \bibnamefont {Yu}}, \bibinfo {author} {\bibfnamefont {C.}~\bibnamefont {Rieck}}, \bibinfo {author} {\bibfnamefont {H.}~\bibnamefont {Schnatz}}, \bibinfo {author} {\bibfnamefont {Y.}~\bibnamefont
  {Hanado}}, \bibinfo {author} {\bibfnamefont {M.}~\bibnamefont {Fujieda}}, \bibinfo {author} {\bibfnamefont {P.-E.}\ \bibnamefont {Pottie}}, \bibinfo {author} {\bibfnamefont {J.}~\bibnamefont {Hanssen}}, \bibinfo {author} {\bibfnamefont {A.}~\bibnamefont {Malimon}},\ and\ \bibinfo {author} {\bibfnamefont {N.}~\bibnamefont {Ashby}},\ }\bibfield  {title} {\bibinfo {title} {Roadmap towards the redefinition of the second},\ }\href {https://doi.org/10.1088/1681-7575/ad17d2} {\bibfield  {journal} {\bibinfo  {journal} {Metrologia}\ }\textbf {\bibinfo {volume} {61}},\ \bibinfo {pages} {012001} (\bibinfo {year} {2024})}\BibitemShut {NoStop}%
\bibitem [{\citenamefont {W{\"u}bbena}\ \emph {et~al.}(2012)\citenamefont {W{\"u}bbena}, \citenamefont {Amairi}, \citenamefont {Mandel},\ and\ \citenamefont {Schmidt}}]{wübbena_sympathetic_2012}%
  \BibitemOpen
  \bibfield  {author} {\bibinfo {author} {\bibfnamefont {J.~B.}\ \bibnamefont {W{\"u}bbena}}, \bibinfo {author} {\bibfnamefont {S.}~\bibnamefont {Amairi}}, \bibinfo {author} {\bibfnamefont {O.}~\bibnamefont {Mandel}},\ and\ \bibinfo {author} {\bibfnamefont {P.~O.}\ \bibnamefont {Schmidt}},\ }\bibfield  {title} {\bibinfo {title} {Sympathetic cooling of mixed-species two-ion crystals for precision spectroscopy},\ }\href {https://doi.org/10.1103/PhysRevA.85.043412} {\bibfield  {journal} {\bibinfo  {journal} {Physical Review A}\ }\textbf {\bibinfo {volume} {85}},\ \bibinfo {pages} {043412} (\bibinfo {year} {2012})}\BibitemShut {NoStop}%
\bibitem [{\citenamefont {Berkeland}\ \emph {et~al.}(1998)\citenamefont {Berkeland}, \citenamefont {Miller}, \citenamefont {Bergquist}, \citenamefont {Itano},\ and\ \citenamefont {Wineland}}]{berkeland_minimization_1998}%
  \BibitemOpen
  \bibfield  {author} {\bibinfo {author} {\bibfnamefont {D.~J.}\ \bibnamefont {Berkeland}}, \bibinfo {author} {\bibfnamefont {J.~D.}\ \bibnamefont {Miller}}, \bibinfo {author} {\bibfnamefont {J.~C.}\ \bibnamefont {Bergquist}}, \bibinfo {author} {\bibfnamefont {W.~M.}\ \bibnamefont {Itano}},\ and\ \bibinfo {author} {\bibfnamefont {D.~J.}\ \bibnamefont {Wineland}},\ }\bibfield  {title} {\bibinfo {title} {Minimization of ion micromotion in a {{Paul}} trap},\ }\href {https://doi.org/10.1063/1.367318} {\bibfield  {journal} {\bibinfo  {journal} {Journal of Applied Physics}\ }\textbf {\bibinfo {volume} {83}},\ \bibinfo {pages} {5025} (\bibinfo {year} {1998})}\BibitemShut {NoStop}%
\bibitem [{\citenamefont {Arora}\ \emph {et~al.}(2007)\citenamefont {Arora}, \citenamefont {Safronova},\ and\ \citenamefont {Clark}}]{Ca2007}%
  \BibitemOpen
  \bibfield  {author} {\bibinfo {author} {\bibfnamefont {B.}~\bibnamefont {Arora}}, \bibinfo {author} {\bibfnamefont {M.~S.}\ \bibnamefont {Safronova}},\ and\ \bibinfo {author} {\bibfnamefont {C.~W.}\ \bibnamefont {Clark}},\ }\bibfield  {title} {\bibinfo {title} {Blackbody‐radiation shift in a ${}^{43}\mathrm{Ca}^+$ ion optical frequency standard},\ }\href {https://doi.org/10.1103/PhysRevA.76.064501} {\bibfield  {journal} {\bibinfo  {journal} {Physical Review A}\ }\textbf {\bibinfo {volume} {76}},\ \bibinfo {pages} {064501} (\bibinfo {year} {2007})}\BibitemShut {NoStop}%
\bibitem [{\citenamefont {Kiruga}\ \emph {et~al.}(2025)\citenamefont {Kiruga}, \citenamefont {Cheung}, \citenamefont {Filin}, \citenamefont {Barakhshan}, \citenamefont {Bhosale}, \citenamefont {Badhan}, \citenamefont {Arora}, \citenamefont {Eigenmann},\ and\ \citenamefont {Safronova}}]{Portal}%
  \BibitemOpen
  \bibfield  {author} {\bibinfo {author} {\bibfnamefont {A.}~\bibnamefont {Kiruga}}, \bibinfo {author} {\bibfnamefont {C.}~\bibnamefont {Cheung}}, \bibinfo {author} {\bibfnamefont {D.}~\bibnamefont {Filin}}, \bibinfo {author} {\bibfnamefont {P.}~\bibnamefont {Barakhshan}}, \bibinfo {author} {\bibfnamefont {A.}~\bibnamefont {Bhosale}}, \bibinfo {author} {\bibfnamefont {V.}~\bibnamefont {Badhan}}, \bibinfo {author} {\bibfnamefont {B.}~\bibnamefont {Arora}}, \bibinfo {author} {\bibfnamefont {R.}~\bibnamefont {Eigenmann}},\ and\ \bibinfo {author} {\bibfnamefont {M.~S.}\ \bibnamefont {Safronova}},\ }\href {https://arxiv.org/abs/2506.08170} {\bibinfo {title} {Portal for high-precision atomic data and computation}} (\bibinfo {year} {2025}),\ \Eprint {https://arxiv.org/abs/2506.08170} {arXiv:2506.08170 [physics.atom-ph]} \BibitemShut {NoStop}%
\bibitem [{\citenamefont {Kozlov}\ \emph {et~al.}(1996)\citenamefont {Kozlov}, \citenamefont {Porsev},\ and\ \citenamefont {Flambaum}}]{KozPorFla96}%
  \BibitemOpen
  \bibfield  {author} {\bibinfo {author} {\bibfnamefont {M.~G.}\ \bibnamefont {Kozlov}}, \bibinfo {author} {\bibfnamefont {S.~G.}\ \bibnamefont {Porsev}},\ and\ \bibinfo {author} {\bibfnamefont {V.~V.}\ \bibnamefont {Flambaum}},\ }\bibfield  {title} {\bibinfo {title} {Manifestation of the nuclear anapole moment in the m1 transitions in bismuth},\ }\href@noop {} {\bibfield  {journal} {\bibinfo  {journal} {J. \ Phys. \ B}\ }\textbf {\bibinfo {volume} {29}},\ \bibinfo {pages} {689} (\bibinfo {year} {1996})}\BibitemShut {NoStop}%
\bibitem [{\citenamefont {Kozlov}\ \emph {et~al.}(2015)\citenamefont {Kozlov}, \citenamefont {Porsev}, \citenamefont {Safronova},\ and\ \citenamefont {Tupitsyn}}]{KozPorSaf15}%
  \BibitemOpen
  \bibfield  {author} {\bibinfo {author} {\bibfnamefont {M.~G.}\ \bibnamefont {Kozlov}}, \bibinfo {author} {\bibfnamefont {S.~G.}\ \bibnamefont {Porsev}}, \bibinfo {author} {\bibfnamefont {M.~S.}\ \bibnamefont {Safronova}},\ and\ \bibinfo {author} {\bibfnamefont {I.~I.}\ \bibnamefont {Tupitsyn}},\ }\bibfield  {title} {\bibinfo {title} {Ci-mbpt: A package of programs for relativistic atomic calculations based on a method combining configuration interaction and many-body perturbation theory},\ }\href@noop {} {\bibfield  {journal} {\bibinfo  {journal} {Comp. Phys. Comm.}\ }\textbf {\bibinfo {volume} {195}},\ \bibinfo {pages} {199} (\bibinfo {year} {2015})}\BibitemShut {NoStop}%
\bibitem [{\citenamefont {Dzuba}\ \emph {et~al.}(1996)\citenamefont {Dzuba}, \citenamefont {Flambaum},\ and\ \citenamefont {Kozlov}}]{DzuFlaKoz96}%
  \BibitemOpen
  \bibfield  {author} {\bibinfo {author} {\bibfnamefont {V.~A.}\ \bibnamefont {Dzuba}}, \bibinfo {author} {\bibfnamefont {V.~V.}\ \bibnamefont {Flambaum}},\ and\ \bibinfo {author} {\bibfnamefont {M.~G.}\ \bibnamefont {Kozlov}},\ }\bibfield  {title} {\bibinfo {title} {Combination of the many-body perturbation theory with the configuration-interaction method},\ }\href@noop {} {\bibfield  {journal} {\bibinfo  {journal} {Phys.\ Rev.\ A}\ }\textbf {\bibinfo {volume} {54}},\ \bibinfo {pages} {3948} (\bibinfo {year} {1996})}\BibitemShut {NoStop}%
\bibitem [{\citenamefont {{Safronova}}\ \emph {et~al.}(2009)\citenamefont {{Safronova}}, \citenamefont {{Kozlov}}, \citenamefont {{Johnson}},\ and\ \citenamefont {{Jiang}}}]{SafKozJoh09}%
  \BibitemOpen
  \bibfield  {author} {\bibinfo {author} {\bibfnamefont {M.~S.}\ \bibnamefont {{Safronova}}}, \bibinfo {author} {\bibfnamefont {M.~G.}\ \bibnamefont {{Kozlov}}}, \bibinfo {author} {\bibfnamefont {W.~R.}\ \bibnamefont {{Johnson}}},\ and\ \bibinfo {author} {\bibfnamefont {D.}~\bibnamefont {{Jiang}}},\ }\bibfield  {title} {\bibinfo {title} {{Development of a configuration-interaction plus all-order method for atomic calculations}},\ }\href@noop {} {\bibfield  {journal} {\bibinfo  {journal} {Phys. Rev. A}\ }\textbf {\bibinfo {volume} {80}},\ \bibinfo {pages} {012516} (\bibinfo {year} {2009})}\BibitemShut {NoStop}%
\bibitem [{\citenamefont {Dzuba}\ \emph {et~al.}(1998)\citenamefont {Dzuba}, \citenamefont {Kozlov}, \citenamefont {Porsev},\ and\ \citenamefont {Flambaum}}]{DzuKozPor98}%
  \BibitemOpen
  \bibfield  {author} {\bibinfo {author} {\bibfnamefont {V.~A.}\ \bibnamefont {Dzuba}}, \bibinfo {author} {\bibfnamefont {M.~G.}\ \bibnamefont {Kozlov}}, \bibinfo {author} {\bibfnamefont {S.~G.}\ \bibnamefont {Porsev}},\ and\ \bibinfo {author} {\bibfnamefont {V.~V.}\ \bibnamefont {Flambaum}},\ }\bibfield  {title} {\bibinfo {title} {Using effective operators in calculating the hyperfine structure of atoms},\ }\href@noop {} {\bibfield  {journal} {\bibinfo  {journal} {Zh. \ Eksp. \ Teor. \ Fiz.}\ }\textbf {\bibinfo {volume} {114}},\ \bibinfo {pages} {1636} (\bibinfo {year} {1998})},\ \bibinfo {note} {[Sov. \ Phys.--JETP {\bf 87}, 885 (1998)]}\BibitemShut {NoStop}%
\bibitem [{\citenamefont {{Dzuba}}\ \emph {et~al.}(1987)\citenamefont {{Dzuba}}, \citenamefont {{Flambaum}}, \citenamefont {{Silvestrov}},\ and\ \citenamefont {{Sushkov}}}]{DzuFlaSil87}%
  \BibitemOpen
  \bibfield  {author} {\bibinfo {author} {\bibfnamefont {V.~A.}\ \bibnamefont {{Dzuba}}}, \bibinfo {author} {\bibfnamefont {V.~V.}\ \bibnamefont {{Flambaum}}}, \bibinfo {author} {\bibfnamefont {P.~G.}\ \bibnamefont {{Silvestrov}}},\ and\ \bibinfo {author} {\bibfnamefont {O.~P.}\ \bibnamefont {{Sushkov}}},\ }\bibfield  {title} {\bibinfo {title} {{Correlation potential method for the calculation of energy levels, hyperfine structure and E1 transition amplitudes in atoms with one unpaired electron}},\ }\href@noop {} {\bibfield  {journal} {\bibinfo  {journal} {J. Phys. B}\ }\textbf {\bibinfo {volume} {20}},\ \bibinfo {pages} {1399} (\bibinfo {year} {1987})}\BibitemShut {NoStop}%
\bibitem [{\citenamefont {Blundell}\ \emph {et~al.}(1989)\citenamefont {Blundell}, \citenamefont {Johnson}, \citenamefont {Liu},\ and\ \citenamefont {Sapirstein}}]{BlaJohLiu89}%
  \BibitemOpen
  \bibfield  {author} {\bibinfo {author} {\bibfnamefont {S.~A.}\ \bibnamefont {Blundell}}, \bibinfo {author} {\bibfnamefont {W.~R.}\ \bibnamefont {Johnson}}, \bibinfo {author} {\bibfnamefont {Z.~W.}\ \bibnamefont {Liu}},\ and\ \bibinfo {author} {\bibfnamefont {J.}~\bibnamefont {Sapirstein}},\ }\bibfield  {title} {\bibinfo {title} {Relativistic all-order calculations of energies and matrix elements for li and ${\mathrm{be}}^{+}$},\ }\href@noop {} {\bibfield  {journal} {\bibinfo  {journal} {Phys. Rev. A}\ }\textbf {\bibinfo {volume} {40}},\ \bibinfo {pages} {2233} (\bibinfo {year} {1989})}\BibitemShut {NoStop}%
\bibitem [{\citenamefont {Brewer}\ \emph {et~al.}(2019{\natexlab{b}})\citenamefont {Brewer}, \citenamefont {Chen}, \citenamefont {Hankin}, \citenamefont {Clements}, \citenamefont {Chou}, \citenamefont {Wineland}, \citenamefont {Hume},\ and\ \citenamefont {Leibrandt}}]{BreCheHan19}%
  \BibitemOpen
  \bibfield  {author} {\bibinfo {author} {\bibfnamefont {S.~M.}\ \bibnamefont {Brewer}}, \bibinfo {author} {\bibfnamefont {J.-S.}\ \bibnamefont {Chen}}, \bibinfo {author} {\bibfnamefont {A.~M.}\ \bibnamefont {Hankin}}, \bibinfo {author} {\bibfnamefont {E.~R.}\ \bibnamefont {Clements}}, \bibinfo {author} {\bibfnamefont {C.~W.}\ \bibnamefont {Chou}}, \bibinfo {author} {\bibfnamefont {D.~J.}\ \bibnamefont {Wineland}}, \bibinfo {author} {\bibfnamefont {D.~B.}\ \bibnamefont {Hume}},\ and\ \bibinfo {author} {\bibfnamefont {D.~R.}\ \bibnamefont {Leibrandt}},\ }\bibfield  {title} {\bibinfo {title} {$^{27}{\mathrm{al}}^{+}$ quantum-logic clock with a systematic uncertainty below ${10}^{\ensuremath{-}18}$},\ }\href@noop {} {\bibfield  {journal} {\bibinfo  {journal} {Phys. Rev. Lett.}\ }\textbf {\bibinfo {volume} {123}},\ \bibinfo {pages} {033201} (\bibinfo {year} {2019}{\natexlab{b}})}\BibitemShut {NoStop}%
\bibitem [{\citenamefont {Wei}\ \emph {et~al.}(2024{\natexlab{b}})\citenamefont {Wei}, \citenamefont {Chao}, \citenamefont {Cui}, \citenamefont {Li}, \citenamefont {Yu}, \citenamefont {Zhang}, \citenamefont {Shu}, \citenamefont {Cao},\ and\ \citenamefont {Huang}}]{WeiChaCui24}%
  \BibitemOpen
  \bibfield  {author} {\bibinfo {author} {\bibfnamefont {Y.-F.}\ \bibnamefont {Wei}}, \bibinfo {author} {\bibfnamefont {S.-J.}\ \bibnamefont {Chao}}, \bibinfo {author} {\bibfnamefont {K.-F.}\ \bibnamefont {Cui}}, \bibinfo {author} {\bibfnamefont {C.-B.}\ \bibnamefont {Li}}, \bibinfo {author} {\bibfnamefont {S.-C.}\ \bibnamefont {Yu}}, \bibinfo {author} {\bibfnamefont {H.}~\bibnamefont {Zhang}}, \bibinfo {author} {\bibfnamefont {H.-L.}\ \bibnamefont {Shu}}, \bibinfo {author} {\bibfnamefont {J.}~\bibnamefont {Cao}},\ and\ \bibinfo {author} {\bibfnamefont {X.-R.}\ \bibnamefont {Huang}},\ }\bibfield  {title} {\bibinfo {title} {Improved measurement of the differential polarizability using co-trapped ions},\ }\href@noop {} {\bibfield  {journal} {\bibinfo  {journal} {Phys. Rev. Lett.}\ }\textbf {\bibinfo {volume} {133}},\ \bibinfo {pages} {033001} (\bibinfo {year} {2024}{\natexlab{b}})}\BibitemShut {NoStop}%
\end{thebibliography}%

\end{document}